\documentclass{jfm}

\usepackage{graphicx}
\usepackage{natbib}

\usepackage{amsmath}
\usepackage{bm}

\ifCUPmtlplainloaded \else
  \checkfont{eurm10}
  \iffontfound
    \IfFileExists{upmath.sty}
      {\typeout{^^JFound AMS Euler Roman fonts on the system,
                   using the 'upmath' package.^^J}%
       \usepackage{upmath}}
      {\typeout{^^JFound AMS Euler Roman fonts on the system, but you
                   dont seem to have the}%
       \typeout{'upmath' package installed. JFM.cls can take advantage
                 of these fonts,^^Jif you use 'upmath' package.^^J}%
      }
  \else
  \fi
\fi

\ifCUPmtlplainloaded \else
  \checkfont{msam10}
  \iffontfound
    \IfFileExists{amssymb.sty}
      {\typeout{^^JFound AMS Symbol fonts on the system, using the
                'amssymb' package.^^J}%
       \usepackage{amssymb}%
       \let\le=\leqslant  
       \let\ge=\geqslant  \let\geq=\geqslant
      }{}
  \fi
\fi

\ifCUPmtlplainloaded \else
  \IfFileExists{amsbsy.sty}
    {\typeout{^^JFound the 'amsbsy' package on the system, using it.^^J}%
     \usepackage{amsbsy}}
    {\providecommand\boldsymbol[1]{\mbox{\boldmath $##1$}}}
\fi

\providecommand\bnabla{\boldsymbol{\nabla}}

\newcommand\Real{\mbox{Re}} 
\newcommand\phihat{\hat{\bm{\phi}}}
\newcommand\curl{\boldsymbol{\nabla}  \times}

\newsavebox{\astrutbox}
\sbox{\astrutbox}{\rule[-5pt]{0pt}{20pt}}

\usepackage{color}
\definecolor{light-gray}{gray}{0.5}
\definecolor{blue}{rgb}{0.0,0.0,1.0}
\definecolor{green}{rgb}{0.0,0.5,0.0}
\definecolor{red}{rgb}{1.0,0.0,0.0}
\definecolor{cyan}{rgb}{0.0,0.75,0.75}
\definecolor{magenta}{rgb}{0.75,0.0,0.75}
\definecolor{yellow}{rgb}{0.75,0.75,0.0}

\newcommand{\mbB}{{\mathbf B}}
\newcommand{\mbu}{{\mathbf u}}
\newcommand{\mbA}{{\mathbf A}}
\newcommand{\mbj}{{\mathbf j}}
\newcommand{\mbF}{{\mathbf F}}
\newcommand{\mbS}{{\mathbf S}}

\title[The Turbulent Dynamo]{The Turbulent Dynamo}
\author[S.~M.~Tobias]%
{S.~M.~Tobias}
\shorttitle{The Turbulent Dynamo} 
\shortauthor{ S.~M.~Tobias} 

\affiliation{Department of Applied Mathematics, University of Leeds, Leeds LS2 9JT, UK}

\pubyear{2010}
\volume{650}
\pagerange{119--126}
\date{?; revised ?; accepted ?. - To be entered by editorial office}
\begin{document}

\maketitle
%
\begin{abstract}
The generation of magnetic field in an electrically conducting fluid generally involves the complicated nonlinear interaction of flow turbulence, rotation and field. This \textit{dynamo} process is of great importance in  geophysics, planetary science and astrophysics, since magnetic fields are known to play a key role in the dynamics of these systems. This paper gives an introduction to dynamo theory for the fluid dynamicist. It proceeds by laying the groundwork, introducing the equations and techniques that are at the heart of dynamo theory, before presenting some simple dynamo solutions. The problems currently exercising dynamo theorists  are then introduced, along with the attempts to make progress. The paper concludes with the  argument that progress in dynamo theory will be made in the future by utilising and advancing some of the current breakthroughs in neutral fluid turbulence such as those in transition, self-sustaining processes, turbulence/mean-flow interaction, statistical methods and maintenance and loss of balance.
\end{abstract}


\section{\label{sec:intro}Introduction}

\subsection{\label{dynthefluid}Dynamo Theory for the Fluid Dynamicist}

It's really just a matter of perspective. To the fluid dynamicist, dynamo theory may appear as a rather esoteric and niche branch of fluid mechanics --- in dynamo theory much attention has focused on seeking solutions to the induction equation rather than those for the Navier-Stokes equation. Conversely to a practitioner dynamo theory is a field with myriad subtleties; in a severe interpretation the Navier-Stokes equations and the whole of neutral fluid mechanics may be regarded as  forming a useful invariant subspace of the dynamo problem, with --- it has to be said --- non-trivial dynamics. In this Perspective, I shall attempt to present the important and interesting developments in dynamo theory from the point of view of a fluid dynamicist, pointing out common themes. I shall focus on explaining how recent developments in fluid mechanics can  contribute to future breakthroughs for magnetohydrodynamic dynamos. 

This is not a review in the typical sense. Although I shall present the key results and features of dynamo theory, I shall not be exhaustive by any means. This perspective is focused on those areas of dynamo theory that I believe are both accessible and of interest to fluid dynamicists, drawing analogies with other areas of fluids where necessary. It is also concentrated on those areas that I believe offer the greatest scope for imminent breakthroughs. These are not necessarily those areas of research that may lead to the most accurate modelling of any given astrophysical object; though it is undoubtedly the case (as described in the next section) that the generation of magnetic fields in these objects nearly always forms the motivation for dynamo investigations. Further details of dynamo theory are contained in  myriad reviews and monographs, for example those of \citet{Moffatt:1978,krauraed:1980,bransub:2005,JONES:2008} and the recently published monograph of  \citet{MoffattDormy:2019} and review of \citet{rincon:2019}.

We begin, however, by giving motivation for the study of dynamos--- much of which arises from observations of cosmical magnetic fields, including those of planets, stars, galaxies and disks.

\subsection{\label{motivation}Motivation}

\subsubsection{\label{geo}The Geomagnetic Field}

\begin{figure}
\centerline{\includegraphics[angle=0,width=0.99\textwidth]{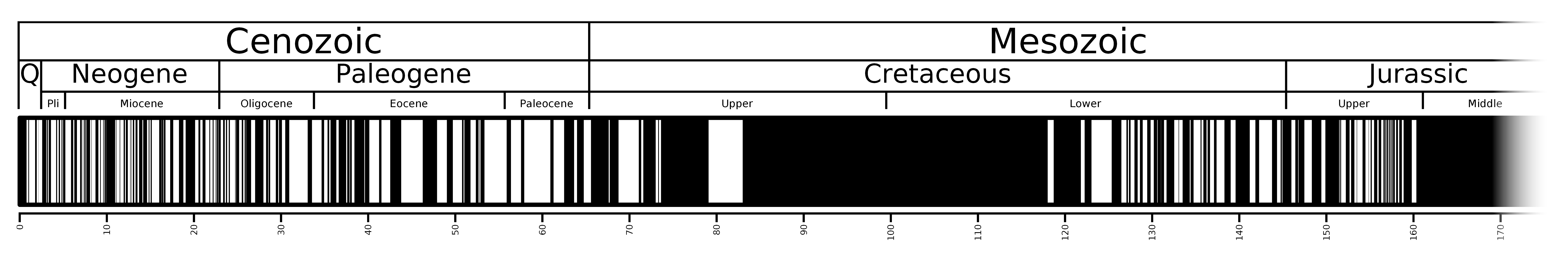}}
\caption{The Geomagnetic record showing the history of magnetic field reversals (source Wikipedia). The black white bars relate to magnetic fields of opposite polarities.}
\label{Geo_record}
\end{figure}

The Earth's magnetic field is presently predominantly dipolar with a  mean surface strength of approximately $40 \mu T$. Currently the dipole axis is offset by about $10^\circ$ from the rotation axis. Paleomagnetic records indicate that the magnetic field has persisted for greater than three billion years and has always had a significant dipole component \citep[see e.g.][]{dvc:2000,aubertetal:2010}. Figure \ref{Geo_record} shows how the record is punctuated by episodic reversals of the polarity of the dipole component; whilst the field reverses (in a time of about $10^4$ years) the field energy decreases and the field becomes smaller scale and more multipolar. The figure clearly shows the timescale for reversals of the field is exceptionally long, with some events (superchrons) lasting over ten million  years. The Earth's magnetic field can currently be measured at the surface to spherical harmonic degree 13, with higher harmonics being screened by remnant magnetism of the crust. In addition to long term variation of the field, it is also subject to `secular variation', which is seen in current and historical observations \citep{Jackson:2000}. Here the spatial structure and strength of the field varies with timescales ranging from years to centuries, with significant features including the `Westward Drift' of magnetic flux patches and changes in the length of day. These changes arise because of the interaction of magnetic fields with flows in the electrically conducting fluid outer core of the Earth.

\subsubsection{\label{ssplan}Solar System Planets}

Most, if not all, solar system planets (including terrestrial, gas giants and ice giants) currently possess or have possessed  dynamo-generated magnetic fields. Even smaller satellites such as Ganymede and the Moon show signs of current or historic dynamo action. It seems as though magnetic field generation is possible whether the planet is terrestrial, a gas giant, or an ice giant. It is therefore expected, and there is some observational evidence to support the theory, that many exoplanets should also be capable of dynamo action \citep{shkolnik:2008}. 

As is the case for the Earth, dynamo action usually takes place in the interior of planets but the magnetic fields are measured as a potential field having diffused through poorly conducting regions. These potential fields are usually decomposed into spherical harmonics with amplitudes given by the so-called gauss coefficients \citep[see e.g.][]{schsoder:2011}.  Briefly, the gas giants Jupiter and Saturn possess strong dipole dominated magnetic fields. Jupiter's field, confirmed by the Pioneer 10 flyby has a largely axial dipole and, with a mean surface strength of $550 \mu T$, has the strongest field in the solar system. Before the recent Juno mission the observations of the magnetic field had a poor resolution (up to spherical harmonic degree three), though recent flybys by the Juno satellite are beginning to establish that the magnetic field has much more structure at smaller scales and a distinct hemispheric asymmetry \citep{mooreetal:2017,jonesholme:2017,Cetal:2018}. It is worth mentioning at this point that the Juno mission will probably give us the closest direct observation of a naturally occurring dynamo generated magnetic field and so it will be fascinating to follow the progress of the mission. Saturn's magnetic field, first revealed by Pioneer 11, has  been measured to spherical harmonic degree three and has a mean surface strength of $30 \mu T$. The axis of the dipole is remarkably aligned with the rotation axis (with an offset of less than $1^\circ$) meaning that the results are consistent with the field being axisymmetric. As we shall see in section~\ref{antidynamo}, it is not possible for such a field configuration to be generated by dynamo action \citep{cowling:1933}; this prompts the widely held belief that Saturn's magnetic field exists solely to annoy Cowling.

The terrestrial planets --- possessing iron-alloy central cores, silicate mantles and rocky crusts --- also largely exhibit dynamo generated magnetic fields. Mercury's magnetic field, the weakest in the solar system at $0.3 \mu T$ is dipole dominated, with the dipole largely aligned with the rotation axis; similarly the magnetic field of the Jovian satellite Ganymede is an axially aligned dipole of about $1 \mu T$. Meanwhile the Moon and Mars do not have magnetic fields that are currently maintained by dynamo action, but there is evidence of extinct dynamos in both cases (see \citet{schsoder:2011} and the references therein). It is unclear in both cases when or why the dynamo switched off, though theories for the cessation of such dynamos usually involve the cooling of the body leading to the unsustainability of thermal convection or the inner core of the body growing to such a size to make dynamo action inefficient. Venus has no measurable intrinsic magnetic field. It may be that the absence of an inner core in Venus means that dynamo action is not possible or that the core is not convective at all. Finally the ice giants Neptune and Uranus have magnetic fields as discovered by the Voyager 2 flybys. These planets possess fundamentally different {\it multipolar} magnetic fields, with mean surface field strengths of $30 \mu T$, and with  the dipole component exhibiting a significant tilt from the rotation axis.

\subsubsection{\label{sunstars}The Sun and Stars}

\begin{figure}
\includegraphics[width=0.99\textwidth]{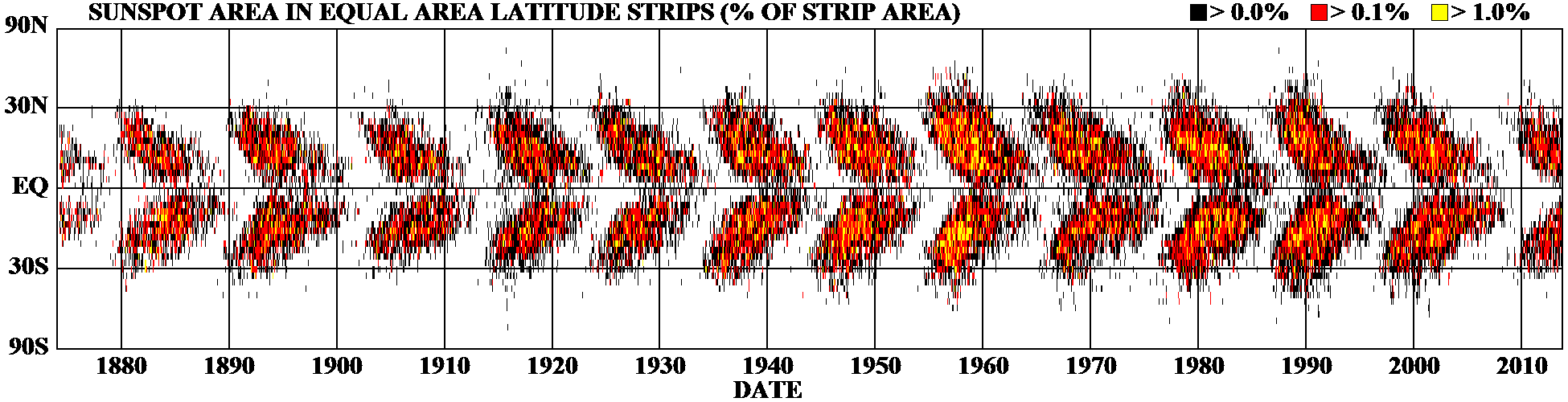}
\caption{The butterfly diagram of the solar cycle. This shows the positions of the spots for each rotation of the sun since 1874. Magnetic activity  first appears at mid-latitudes, widen, and then moves towards the equator as each cycle progresses (image courtesy of D. Hathaway).}
\label{butterfly}
\end{figure}

Arguably the most remarkable example of natural dynamo action is the solar activity cycle \citep{brunbrowning:2017,usoskin:2017,hathaway:2015}. Solar activity has been observed for many centuries, with both ancient Chinese and Athenian observations of sunspots being recorded. Sunspots have been systematically observed since the early seventeenth century, when Galileo turned the newly invented telescope towards our nearest star. It is now clear that solar magnetic activity exhibits an astonishingly systematic spatio-temporal behaviour. Sunspots appear in flux belts
confined between the equator and latitudes of about $\pm30^\circ$. They show cyclic activity with a period of approximately eleven years; as the cycle progresses the latitude at which activity is found moves in a wave from mid-latitude to the equator, before dying out as shown in the solar butterfly diagram of figure~\ref{butterfly}. Note that the activity in this diagram is basically symmetric about the equator; this sunspot activity has been linked to magnetic fields though the Zeeman splitting of spectral lines. The sun has an azimuthally averaged field of a few hundred $\mu T$ and the large-scale 
radial magnetic field changes sign across equator and changes parity,  giving a 22-year magnetic cycle. The sunspot field reverses at sunspot minimum --- giving a magnetic field that is currently dominated by its dipole component ---  whilst the  coronal field reverses out of phase at sunspot maximum. The sun also possesses small-scale magnetic fields, which are only weakly coupled with the solar cycle.

Indications of the long-term dynamics of solar activity come from both direct and indirect observations \citep{wt:2007}. Telescopic observations demonstrate the modulation of the basic eleven year cycle and also reveal the presence of a period in the seventeenth century, known as the Maunder Minimum, when sunspots were almost completely absent. Interestingly as the Sun emerged from this minimum sunspots were to be found solely in the Southern hemisphere \citep{ribesnesmeribes:1993} and sunspots were found at the equator in the nineteenth century \citep{aw:2015}.
Indirect obervations of terrestrial isotopes (see \citet{usoskin:2017}), whose production rate is anti-correlated with magnetic activity, show that the solar cycle persisted through the Maunder Minimum \citep{btw:1998} and that that minima such as the Maunder Minima are recurrent events, that appear in clusters \citep{wt:2016,btw:2018}.

Our understanding of the origin of the solar magnetic field owes much to observations of magnetic field generation in other stars. Stellar observations can obviously be used to calibrate dynamo theories, but will also give a clue as to past and future behaviour. It is difficult to measure stellar fields directly and proxy observations are usually required. Such techniques include measuring photometric variability \citep{strassmeier:2009}, atmospheric heating via emission in Ca II H and K lines \citep{baliunasetal:1995}, spectrapolarimetry \citep{donland:2009} and seismic proxies for magnetic activity as for example measured by {\it Kepler} and COROT \citep{stelloetal:2016}.

The results from such observations are varied and bewildering to the non-expert (and sometimes to the expert). For our purposes it is sufficient to state the main results. For solar-like main-sequence stars, broadly speaking magnetic activity increases with rotation rate. There is a systematic increase in magnetic field strength with decreasing Rossby number of the star (here defined as $Ro^{-1}= 2 \Omega \ell / u_{rms}$ where $\ell$ is a characteristic lengthscale given by mixing length theory, until activity levels off past a threshold in Rossby number (typically Ro $\approx 0.1-0.3$). Interestingly this threshold appears to be independent of the mass of the star. It is also the case that many of these stars exhibit activity cycles similar the solar cycle. The period of the cycle is also a function of Rossby number with the cycle period decreasing with Rossby number (i.e.\ faster rotators have shorter period). The precise form of this relationship is difficult to determine and is complicated by the possible presence of active and inactive branches, though there is now some evidence that these branches are an artefact of limited observations. Finally for these stars the morphology of the generated field also appears to depend on rotation rate, with faster rotators being more dominated by their zonal fields \citep{seeetal:2016}. Other stars show a wide variety of behaviour, depending on their mass, age and spectral type (for example whether they exhibit convection in a core or in an envelope plays a major role). Young stars (e.g. fully convective T-Tauri stars) have strong fields (average field strengths of around $0.2$T)

Suffice to say that stellar magnetic fields are ubiquitous and the properties of the generated field depend on the nature of the turbulence and the rotation rate of the star. The interested reader is directed to the excellent review by \citet{brunbrowning:2017} for more details.

\subsubsection{\label{galaxies}Galaxies and Galaxy Clusters}


The properties of magnetic field in galaxies (including our own galaxy) have been extensively measured, with a variety of observational techniques --- see \citet{bransub:2005,beck:2015} and the reviews referenced therein. Our galaxy has an estimated local field strength of about $5 \times 10^{-10}$\,T, which is a typical amplitude for fields in galaxies. The magnetic field in galaxies usually has an ordered \textit{and} a tangled component; in spiral galaxies the large-scale order manifests itself over several kpc. Usually the random component is strongest in the spiral arms, whilst the regular field extends into the interarm regions. Galaxy clusters are the largest bound systems in the universe and are found to have magnetic fields; for these the total magnetic fields are estimated to be of the order of $10^{-11}$--$10^{-9}$T.
For most astrophysical objects it is now widely accepted that dynamo action is the {\it only} mechanism that can explain their magnetic  properties and the persistence of the field for many ohmic decay times. For the galactic dynamo there appears still to be some debate (as the decay times for such vast objects are enormous), though there are a number of very significant arguments pointing to the dynamo origin of such fields \citep[see e.g.][]{kulsrudzweibel:2008,bransub:2005}.

\subsubsection{\label{discs}Accretion Discs}

Accretion discs are gaseous discs of material spinning around a central object (for example young stars, white dwarfs, neutron stars and black holes). Although direct observations of dynamo generated magnetic fields in such objects is difficult (though see \cite{donatietal:2005}), proxies such as emission or the magnetism of meteorites formed in the disc around the young sun give indirect evidence. Direct observations may place upper limits on the strength of any generated magnetic fields. However it is believed that magnetic fields play an important role in the collimation of jets and the sustainment of accretion (as we shall see later). The dynamo in such discs is very interesting theoretically --- being {\it essentially nonlinear} \citep{tcb:2011} --- with many of the characteristics of the transition to dynamo action being similar to those for the transition to turbulence in wall-bounded flows (see e.g. \citet{waleffe:1997,barkley:2016}). We shall discuss in details these \textit{essentially nonlinear} dynamos in section~\ref{ess:non}.

\section{\label{sec:fundamentals}Fundamentals}

In this section we introduce the fundamentals of dynamo theory, pointing out important analogies for fluid dynamicists. More details can be found for example in the excellent expositions and reviews referred to below.

\subsection{\label{eqnsbcs}Equations and Boundary Conditions}

\subsubsection{\label{indmom}The induction and momentum equation}

Dynamo theory typically (minimally) involves the construction of solutions to the pair of coupled partial differential equations for the evolution of the velocity field, ${\mathbf u}$ and the magnetic field ${\mathbf B}$ in an electrically conducting fluid under the magnetohydrodynamic (MHD) formalism. A complete derivation of the equations and a discussion of their applicability in any given physical situation is beyond the scope of this perspective; the interested reader is directed to \citet{Moffatt:1978,cowling:1978,roberts:1994,robsow:1992} and \cite{JONES:2008} for details of the derivation. Briefly, under the MHD approximation, which combines the non-relativistic Maxwell equations of electrodynamics with Ohm's law for a moving conductor, these take the form
\begin{equation}
\dfrac{\partial \mbu}{\partial t}
+ \mbu \cdot \boldsymbol{\nabla}     \mbu = - \dfrac{1}{\rho}\nabla p + \dfrac{1}{\rho}\, \mbj \times \mbB+\nu \nabla^2 \mbu + \dfrac{\mathbf F}{\rho},
\label{NS}
\end{equation}
\begin{equation}
\dfrac{\partial \mbB}{\partial t}
 = \bnabla \times \left( \mbu \times \mbB \right) -\bnabla \times \left(\eta \bnabla \times \mbB \right),
\label{Induction}
\end{equation}
\begin{equation}
\bnabla \cdot \mbB
 = 0.
\label{divB}
\end{equation}
Here the standard fluid parameters are $\rho$, which is the density of the fluid, $p$ which is the pressure and $\nu$ is the kinematic viscosity of the fluid, whilst $\mbF$ represents all the body forces (such as buoyancy or mechanical driving) acting to drive the fluid motion. The additional force term in the Navier-Stokes equations is termed the Lorentz force ${\mbF}_{Lor} = \mbj \times \mbB $ and arises owing to the interaction of the magnetic field with the current $\mbj$ flowing through the conductor. The current is itself given by the pre-Maxwell version of  Amp\`ere's Law, i.e.\
\begin{equation}
\mbj = \dfrac{1}{\mu} \bnabla \times \mbB,
\label{ampere}
\end{equation}
and hence 
\begin{equation}
{\mbF}_{Lor} = \dfrac{1}{\mu \rho} (\bnabla \times \mbB) \times \mbB.
\label{Lorentz}
\end{equation}
Here, and henceforth in this paper,  $\mu$ is the permeability. 
The evolution equation for the magnetic field given by equation~(\ref{Induction}) is termed the {\it induction equation}. Here $\eta = (\mu \sigma)^{-1}$ is the magnetic diffusivity which is a property of the conducting fluid, with electrical conductivity $\sigma$. Magnetic diffusivity is large for poor conductors.  In circumstances in which the magnetic diffusivity is constant  equation~(\ref{Induction}) simplifies (utilising equation~(\ref{divB})) to 
\begin{equation}
\dfrac{\partial \mbB}{\partial t}
 = \bnabla \times \left( \mbu \times \mbB \right) + \eta \nabla^2 \mbB.
\label{Inductionce}
\end{equation}

\subsubsection{\label{bcs}Magnetic Boundary Conditions}

The induction and momentum equations are usually (though not always) solved in a finite domain subject to the imposition of boundary conditions on the fluid velocity and magnetic field. The boundary conditions for the fluid velocity are straightforward and standard; usually one considers impenetrable boundaries with either stress-free or no-slip conditions on the tangential component of the velocity.

The magnetic boundary conditions are more problematic and usually involve some degree of simplification. Typically this involves considering the fluid in a domain $V$ with an exterior bounding surface $S$, external to which is either an insulator or a perfect conductor.
If the region outside of the evolution of the dynamo fluid is perfectly conducting, surface charges, $\rho_S$, and surface currents, ${\bf j}_S$, are allowed on the boundary. If we define $[.]$ as
the jump across the surface $S$ and ${\bf n}$ as an outward pointing normal to that surface, then integrating the pre-Maxwell equations across the surface gives
\begin{equation}
[ {\bf n} \cdot {\bf E} ] = \frac{\rho_S}{\epsilon}, \qquad
[ {\bf n} \cdot {\bf B} ] = 0, \qquad
[ {\bf n} \times {\bf B} ] = \mu {\bf j}_S , \qquad
[ {\bf n} \times {\bf E} ] = 0,
\end{equation}
where $\epsilon$ is the permittivity.
If there are no surface currents or charges (i.e.\ for non-perfectly conducting boundaries) then  ${\bf B}$ is continuous, provided $\mu$
is constant. If the region external to the solution domain does not allow currents (i.e.\ if this region is an insulator) then this continuity of ${\bf B}$ defines the problem.\footnote{If the outside region allows currents then the normal derivative of ${\bf n} \cdot {\bf B}$ is also continuous, though
the normal derivatives of the tangential components of ${\bf B}$ are not 
necessarily continuous. The boundary conditions on these depend on the velocity boundary conditions and can be found via the continuity of ${\bf n} \times {\bf E}$ and Ohm's law \citep[see the detailed discussion in][]{JONES:2008}.} 

However, if the outside of the domain is a static {perfect conductor} it is normal to assume that there is no trapped magnetic field there and so (for no normal flow conditions)
\begin{equation}
{\bf n} \cdot {\bf B} =0, \quad \quad
{\bf n} \times {\bf j} =0.
\end{equation} 
This gives
$$B_z = \dfrac{\partial B_x}{\partial z} = \dfrac{\partial B_y}{\partial z} = 0, \,\,\quad \rm{or}\,\, \quad
 B_r =  \dfrac{\partial (rB_{\theta})}{\partial r} =  \dfrac{\partial(rB_{\phi})}{\partial r} = 0,$$
at a Cartesian boundary $z=$const or  
at a spherical boundary $r=$const respectively.

\subsection{\label{dynamodef}What is a dynamo?}
Simply put, a self-exciting hydromagnetic dynamo is a self-consistent solution of the coupled Navier-Stokes and induction equation for which the magnetic energy, 
\begin{equation}
M(t) = \int_V \dfrac{B^2}{2 \mu}\, dV,
\end{equation}
remains finite as $t \rightarrow \infty$. Here $V$ is the volume over which the dynamo equations are solved, which could in principle be finite and bounded by a surface $S$ or (in an idealisation) taken to be all space. If the volume is finite it is traditional to take the region external to the domain as being an insulator (${\bf j}_{\rm ext.}=0$) so that the magnetic field is  maintained entirely by the current distribution within the domain and so ${\bf B} = {\mathcal O}(|{\bf x}|^{-3})$ as $|{\bf x}| \rightarrow \infty$ \citep[see e.g][]{Moffatt:1978}.

\subsection{\label{en:cons}Energetics and Conservation Laws}

As in hydrodynamics, much can be understood by deriving global conservation laws, valid in the absence of driving and dissipation. For  inviscid  hydrodynamic flows in the absence of body forces (such as gravity) and 
boundary forces the total kinetic energy and kinetic helicity 
defined as 
\begin{equation}
E_k = \dfrac{1}{2}\int_V \rho \mbu \cdot \mbu \, dV , \quad
H_k = \int_V \rho \mbu \cdot \bm{\omega} \, dV ,
\end{equation}
are conserved \citep[][]{moreau1961,moffatt1969}\footnote{though see earlier unpublished work by Feynman.}.

In the presence of a magnetic field these quantities are no longer conserved. Transfer of energy may take place between kinetic and magnetic energies via the action of induction and the Lorentz force respectively. The magnetic and kinetic energy equations take the form after a little vector calculus and ignoring surface terms,
\begin{equation}
\dfrac{d}{dt}\int_V \dfrac{\mbB^2}{2 \mu} \, dV = 
-\int_V \mbu\cdot (\mbj \times \mbB) \, dV -  \eta \mu \int_V {\mbj^2} \, dV,
\label{ME_eqn}
\end{equation}
and for an incompressible flow
\begin{equation}
\dfrac{d}{dt}\int_V \dfrac{\rho \mbu^2}{2} \, dV  = 
+\int_V \mbu\cdot (\mbj \times \mbB) \, dV  -  \int_V 2 \nu \rho {\mbS^2} \, dV ,
\label{KE_eqn}
\end{equation}
where the dissipative terms have now been included and $S_{ij} = \frac{1}{2}\left(\frac{\partial u_i}{\partial x_j}+\frac{\partial u_j}{\partial x_i}\right)$ is the rate of strain tensor for an incompressible fluid.

Clearly adding equation~(\ref{ME_eqn}) and (\ref{KE_eqn}) for an ideal (inviscid and perfectly conducting) fluid immediately reveals that the total (kinetic plus magnetic) energy is conserved, with magnetic energy only being created at the expense of kinetic energy. The $\mbu\cdot (\mbj \times \mbB)$ term is therefore responsible for this transfer of energy and arises from inductive effects in the induction equation and the Lorentz force in the momentum equation. 

In addition to the total energy there are two other quadratic invariants of the ideal system, namely the magnetic helicity and the cross helicity. The cross helicity, given by 
\begin{equation}
H^c = \int_V \mbu \cdot \mbB \, dV ,
\label{def_cross}
\end{equation}
is clearly not sign definite; moreover cross-helicity dissipation is not 
sign-definite. Cross-helicity may be either amplified or damped locally and so less attention has focussed on the implications of its conservation in the non-dissipative case  \citep[though see][]{biskamp:2003,yokoi:2013}.

However the implications of the conservation of magnetic helicity and its role in the saturation of nonlinear dynamos has received much attention and for this reason we devote some time to this here.

\subsubsection{\label{consmh}Conservation of Magnetic Helicity}

Because $\bnabla \cdot \mbB = 0$, it is often useful to write $\mbB = \bnabla \times \mbA$. Here $\mbA$ is termed the vector potential for the magnetic field. Clearly $\mbA$ is only defined up to a choice of gauge, so that transforming $\mbA \rightarrow \mbA + \nabla \psi$ for any scalar $\psi$ leaves the magnetic field unchanged. 
It is also then convenient to `uncurl' the induction equation to give
\begin{equation}
\dfrac{\partial {\mbA}}{\partial t} =({\mbu}\times {\mbB})-\eta \,\bnabla \times {\mbB} + \nabla \phi,
\label{eq:uncurled}
\end{equation}
where $\phi$ is related to the choice of gauge.

Various choices of gauge are possible; the \textit{Coulomb gauge} has 
\begin{equation}
\bnabla \cdot \mbA  = 0; \quad \nabla^2 \phi = -\bnabla \cdot ({\mbu}\times {\mbB}).
\end{equation}
The \textit{winding gauge} (suitable for calculations in Cartesian geometries) has 
\begin{equation}
\nabla_H \cdot \mbA = 0; \quad \nabla_H^2 \phi^\prime = -\bnabla \cdot ({\mbu}\times {\mbB}),
\end{equation} 
where $\nabla_H = (\partial_x,\partial_y,0)$ and $\phi^\prime = \phi- \eta \frac{\partial A_z}{\partial z}$ \citep{PriorYeates:2014}.
A numerically convenient gauge involves  setting
\begin{equation}
\phi = \dfrac{\partial \psi}{\partial t},
\end{equation}
as described in \citet{bransub:2005}

Now, magnetic helicity is defined as 
\begin{equation}
H = \int_V \mbA \cdot \mbB \, dV ,
\label{def_maghel}
\end{equation}
and its evolution can be easily shown to be given by
\begin{equation}
\dfrac{dH}{dt}=-2 \eta \mu \int_V \mbj \cdot \mbB \, dV  + F_s,
\label{hel_evolution}
\end{equation}
where $F_s$ represents the surface flux of magnetic helicity.
Hence for a perfectly conducting fluids, with no loss or gain of magnetic helicity through the boundaries (i.e. $F_s = 0$), magnetic helicity is conserved. This has implications for dynamo action, which requires the generation or destruction of magnetic helicity, as we shall see. Magnetic helicity is a measure of the topological complexity and linkage of field lines, and in the absence of diffusive processes (by which the field can reconnect) this complexity is maintained. Furthermore magnetic helicity appears to be a more robust invariant than say total energy in the presence of small diffusion. The dissipative term for total energy, $-\eta \mu \int_V \,{\mbj^2}\, dV $, may remain finite as $\eta \rightarrow 0$, because $\mbj^2$ gets large in this limit. However the dissipation of magnetic helicity, $-2 \eta \mu \int_V \mbj \cdot \mbB \, dV $, appears to tend to zero as $\eta \rightarrow 0$ so magnetic helicity is well conserved. Of course the situation changes if magnetic helicity is allowed to enter or leave the domain of interest; it may do so via either advective or diffusive processes, as we shall see in section~\ref{fluxes}.

In hydrodynamic turbulence the presence of quadratic invariants has consequences for the nature of the cascades. Briefly, the same reasoning applies to MHD. As argued above for small dissipation, energy decays faster than magnetic helicity and cross-helicity \citep{biskamp:2003}.  Therefore, in a turbulent state, energy cascades
towards small scales (analogous to the energy cascade in 3D hydrodynamics). However, the magnetic 
helicity cascades toward large scales (analogous to the energy cascade in 2D hydrodynamics); the  inverse cascade of magnetic helicity may lead to the formation of large-scale magnetic 
fields --- this fact may prove to be important for the generation of systematic fields by nonlinear dynamos.

\subsection{\label{ie:kin_dyn}The induction Equation and Kinematic Dynamos: The basics}

 For fluid dynamicists an obvious useful analogy can be made between the induction equation~(\ref{Induction}) and the incompressible vorticity equation given by
\begin{equation}
\dfrac{\partial \bm{\omega}}{\partial t}
 = \bnabla \times \left( \mbu \times \bm{\omega} \right) +\nu \nabla^2 \bm{\omega}.
 \label{vorticity}
 \end{equation}
For experts on vorticity dynamics, this analogy can lead to significant insight into the dynamics of the magnetic field. Magnetic flux tubes may in certain circumstances have similar dynamics to those of vortex tubes; this analogy becomes a formal correspondence when the magnetic field and vorticity are weak. However it is important not to push the analogy too far. Equation~(\ref{vorticity}) is a nonlinear evolution equation for the vorticity (since the advecting velocity is related to the vorticity), whereas equation~(\ref{Inductionce}) \textit{is} formally linear in the magnetic field, with the system only becoming nonlinear when coupled to the momentum equation via the Lorentz force. As a rule of thumb, if the magnetic field is weak compared with the velocity it behaves analogously to the vorticity; if it is of a similar strength it 
behaves more like like the velocity. 

Important limits of the induction equation are the so-called diffusive and perfectly conducting  limits. If ${\bf u} = 0$, equation~(\ref{Inductionce}) reduces to the vector diffusion equation,
\begin{equation}
\dfrac{\partial \mbB}{\partial t}
 = \eta \nabla^2 \mbB,
\end{equation}
so that for no fluid motion the field must diffuse away (assuming there are no fields at infinity). Hence motion is needed to maintain magnetic field. The timescale for diffusion of field with a typical lengthscale $\ell_B$ is given by  $\tau_D=\ell_B^2 /\eta$. To give some idea of some typical diffusive timescales, we note that $\tau_d \sim 2$\,seconds for $\ell_B = 1$\,metre and $\eta=0.04$\,metre$^2 s^{-1}$, which may be appropriate for a liquid sodium experiment. Whereas for a magnetic field of the scale of the Earth's core and a relevant diffusivity $\tau_d \sim 10^4$ years; the Sun has a diffusion time of $10^9$ years. For galaxies the diffusion time of magnetic field is significantly longer than the age of the universe! 

In the absence of diffusion (the so-called perfectly conducting limit where $\sigma \rightarrow \infty$ and $\eta=0$), equation~(\ref{Inductionce})  becomes 
\begin{equation}
\dfrac{\partial \mbB}{\partial t}
 = \bnabla \times \left( \mbu \times \mbB \right).
\end{equation}
Sometimes this is called the frozen flux limit; the magnetic flux $\int_{\cal{S}} \mbB \cdot d{\bf S}$ through the surface \cal{S} bounded by any closed curve \cal{C} moving with the fluid, remains fixed (Alfv\'en's theorem). 
Hence we can think of magnetic field as being
frozen into the fluid, in a similar manner to vortex lines in an inviscid fluid,  Alfv\'en's theorem is the magnetic counterpart of Kelvin's circulation theorem.

\subsubsection{\label{Rmrole}Importance of the magnetic Reynolds number $Rm$}

Clearly from the above discussion, magnetic field will decay away unless the advective term in the induction equation is large enough to overcome diffusive effects. The relative importance of the two terms $\curl  ({\bf u} \times \mbB)$
and $\eta \nabla^2  \mbB$ can be estabilshed by  non-dimensionalising. 
We choose a typical length scale $L$ which is the size of the object or region
under consideration and a typical fluid velocity $U$. On introducing  
scaled   variables
$ t = ({L}/{U}) \tilde{t} $, $ {\bf x} = L {\tilde {\bf x}}$,
${\bf u} = U {\bf \tilde u}$, 
 and dropping tildes the induction equation becomes
\begin{equation} 
\dfrac{\partial \mbB}{\partial t} = {\bf \nabla} {\bf \times}  
({\bf  u} \times \mbB) + Rm^{-1} 
\nabla^2  \mbB, 
\label{ndinduction}
\end{equation}
where $Rm = {U L}/{\eta}$ is the dimensionless magnetic Reynolds number. 

In general, large $Rm$ means induction
dominates over diffusion, whilst small $Rm$ means diffusion
wins out over induction, and as we shall see minimum values of $Rm$ for dynamo action (so-called dynamo bounds) can sometimes be found.

It is worth noting at this point, however, that $Rm$ should only be used as a guide to determine the relative importance of advection and diffusion. In defining $Rm$ it has explicitly been assumed that $L$ is a typical lengthscale for both the magnetic field and the velocity (i.e. that $\ell_B = \ell_U = L$). This makes sense if $L$ is the size of the astrophysical object (i.e. the largest lengthscale available). Of course this may not be the case, with the potential for $\ell_B$ to be very different from $\ell_U$. For example if $\ell_B \gg \ell_U$ then the relative importance of advection and diffusion is given by $Rm_T = U \ell_B^2 / \eta \ell_U \sim \omega \ell_B^2/\eta$, where $\omega$ is a typical vorticity amplitude. This may be large even if $\ell_U$ and $U$ are small. Such a basic misunderstanding of the limitations of the information encoded in the  magnetic Reynolds number may lead to the incorrect dismissal of certain small-scale flows as the possible origin of large-scale fields. Large-scale fields, as they are weakly diffusive, require very little induction for their maintenance.

\subsubsection{\label{axisymm}A useful technique: axisymmetric field decomposition}
It is clear from the form of the induction equation that 
a non-axisymmetric flow immediately creates a non-axisymmetric field --- the converse is not true. If both the flow and field are axisymmetric then one can decompose the flow and field by setting (in cylindrical polars $(s,\phi,z)$)
\begin{equation}
{\bf u}=s \Omega \phihat + {\bf u}_P = s \Omega \phihat + \curl \left(\frac{\psi}{s}\right) \phihat,
\end{equation}
\begin{equation}
{\bf B} =  B \phihat + {\bf B}_P =B \phihat + \curl (A \phihat),
\end{equation}
where $s = r \sin \theta$ (where $r$ and $\theta$ relate to spherical polars $(r,\theta,\phi)$). Here $\Omega$ is the differential rotation, $\psi$ is the streamfunction, $B$ is the zonal field (sometimes termed toroidal field in an axisymmetric setting) and $A$ is the scalar potential. 
The induction equation then simplifies to 
\begin{equation}
 \frac{\partial A}{\partial t} + \frac{1}{s} ({\bf u}_P \cdot \nabla) (sA) = \eta \left(\nabla^2 - \frac{1}{s^2}\right) A,
\label{pol_eqn}
\end{equation}
\begin{equation}
\frac{\partial B}{\partial t} + s ({\bf u}_P \cdot \nabla) \left(\frac{B}{s}\right) = \eta \left(\nabla^2 - \frac{1}{s^2}\right) B
+ s{\bf B}_P \cdot \nabla \Omega.
\label{tor_eqn}
\end{equation}
The form of these equations reveals the limitation of an axisymmetric representation for dynamo action. This will be exploited in the next section to prove so-called anti-dynamo theorems. Both the $A$ and $B$ equations have advective and diffusive terms. The field stretching term, $s{\bf B}_P  \cdot \nabla \Omega$, only appears in equation~(\ref{tor_eqn}) if gradients in angular velocity are present, and equation~(\ref{pol_eqn}) shows that there is no corresponding source term for $A$.

\subsubsection{\label{antidynamo}Anti-dynamo theorems}

 It is fair to say that the psychology of the dynamo practitioner has been strongly shaped by the early results of dynamo theory --- results that showed the complexity and difficulty of achieving dynamo-generated fields. Primary of these results are the so-called \textit{anti-dynamo theorems}. These show the impossibility of dynamo action for large classes of magnetic fields and velocity fields with certain symmetries. We shall not reproduce all of the demonstrations and proofs here, since they are readily available from many sources \citep[for example][]{JONES:2008,dormysoward:2007,MoffattDormy:2019}, but the importance of these for the subsequent direction of the development of the field can not be overstated.

\medskip
\noindent{\textbf{Cowling's Theorem:}}  \textit{An axisymmetric field vanishing at infinity can not be maintained by dynamo action} \citep{cowling:1933}.

\medskip
\noindent{\textit{Proof:}} As noted above, a non-axisymmetric velocity field immediately generates non-axisymmetric magnetic field and so it is necessary to consider only axisymmetric flows and fields. The evolution of the field is therefore given by equations~(\ref{pol_eqn}-\ref{tor_eqn}). Multiplying equation~(\ref{pol_eqn}) by $s^2 A$ and integrating over all space gives
\begin{equation}
\frac{d}{dt} \int \frac{1}{2} s^2 A^2 \, dV = - \eta \int | \nabla (sA) |^2 \, dV .
\label{A_decay}
\end{equation}
Derivation of equation~(\ref{A_decay}) has utilised the divergence theorem and the vanishing of the surface terms at infinity. This equation clearly shows that $sA$ decays to zero as $t \rightarrow \infty$\footnote{Note $sA$ can not decay to a constant since this would imply $A\rightarrow \infty$ as $s\rightarrow 0$.}. Once $A$ has decayed there is no source term in the toroidal equation~(\ref{tor_eqn}). Similar arguments then ensure the subsequent decay of $B/s$ and therefore the impossibility of the creation of an axisymmetric magnetic field by dynamo action. 
It is important to note that it is the absence of source terms in equation~(\ref{pol_eqn}) that causes the problems for dynamo action. Relaxation of the constraint of axisymmetry allows for the re-inclusion of such a source term.

As noted above, the effect of Cowling's Theorem ruling out such simple symmetric solutions made the search for any dynamo solutions (which were necessarily three dimensional!) seem a formidable task. Indeed it was not clear that any such dynamo solutions existed for a long time. The situation is encapsulated in the following story taken from \citet{krause:1993} (which attributes the source as Paul Roberts.) \textit{`Walter Elsasser and Einstein were friends in Germany before they both emigrated to the US in the 1930s. Several years after Elsasser had settled there (in
the late 1930s in fact), he became interested in the origin of the geomagnetic field. Einstein paid him a visit, and (as people do) asked ``What are you working on these
days?". Elsasser told him, and Einstein invited him to explain dynamo theory to him. Elsasser set up the problem and then told Einstein about Cowling's theorem.
Einstein's response was, ``If such simple solutions are impossible, self-excited fluid
dynamos cannot exist". For once, the great man's craving for simplicity seems to
have misled him.'}

\noindent{\textbf{Other Antidynamo Theorems:}}
There are many extensions of Cowling's antidynamo theorem to other geometries and for slightly different setups. For example, in Cartesian coordinates $(x,y,z)$, no magnetic field that vanishes at infinity and is independent of $z$ can be generated
by dynamo action; the proof proceeds along similar lines to that given above \citep[see e.g.][]{JONES:2008}. Note again that this is a restriction on the form of a dynamo-generated magnetic field.

Anti-dynamo theorems placing constraints on the form of the velocities that can lead to dynamo action have been proven by \citet{bg:1954,backus:1958} and in Cartesian coordinates by \citet{zel:1957}. These results essentially show that a velocity field must have all three components in order to be capable of acting as a dynamo \citep[see the long discussion and derivation in][]{Moffatt:1978}. 

In conclusion, both the field and the flow must be sufficiently complicated for dynamo action to occur. A minimal requirement is that the field must be three-dimensional and the flow must not be purely poloidal (i.e. can not be planar).

\subsubsection{\label{bounds}Bounds on dynamo action}

Even for flows and fields that are not ruled out as dynamos on symmetry grounds, there are bounds that constrain dynamo action in finite domains. We noted earlier that a crude measure of the expected efficiency of a dynamo is given by the magnetic Reynolds number $Rm$, which gives the ratio of advection to diffusion at a particular scale in the flow (usually taken to be the integral or system scale). This argument can be formalised in the following ways

\noindent{\bf The Backus Bound \citep{backus:1958}:} In order to derive rigorous bounds on dynamo action, it is necessary to look for solutions for which the magnetic energy does not decay. The evolution equation for the magnetic energy is given by (see equation~(\ref{ME_eqn}))
\begin{equation} 
\frac{\partial}{\partial t} \int \frac{{\bf B}^2}{2 \mu} 
\, dV = \int {\bf j} \cdot ({\bf u} \times {\bf B}) \, dV 
-  {\mu \eta} \int {\bf j}^2 \, dV , 
\end{equation}
or
\begin{equation}
\mu \frac{\partial E_M}{\partial t} = 
- \eta \int | \nabla \times  {\bf B}|^2 \, dV  
+ \int (\nabla \times {\bf B}) \cdot ({\bf u} 
 \times {\bf B}) \, dV . 
 \label{mag_en}
\end{equation}
Now
\begin{equation}
\int (\nabla \times {\bf B}) \cdot ({\bf u} 
 \times {\bf B}) \, dV  = \int  {\bf B} \cdot \bnabla \times({\mbu} \times {\bf B}) \,dV  = \int {\bf B}\cdot ({\bf B} \cdot \nabla) {\mbu} \,dV  \le e_{max} \int | {\bf B}|^2 \, dV ,   \end{equation}
where $e_{max}$ is the maximum of the rate of strain tensor \citep[see e.g.][]{MoffattDormy:2019}. In order to make
progress we shall consider the case of field generation in a sphere of radius $a$ matching to a decaying potential outside. For this case (as $\bm{\nabla}\cdot {\bf B}=0$)
\begin{equation}
\int |\nabla \times {\bf B}|^2 \, dV \ge \frac{\pi^2}{a^2}
\int | {\bf B}|^2 \, dV,
\label{ineq1}
\end{equation}
and so 
\begin{equation}
\mu \frac{\partial E_M}{\partial t} \le \left( e_{max}
- \frac{\eta \pi^2}{a^2} \right) \int |  {\bf B}|^2 \, dV. 
\end{equation}
Hence dynamo action requires 
\begin{equation}
Rm = \frac{a^2 e_{max}}{\eta} \ge \pi^2.
\end{equation}
Note here that $Rm$ is defined in terms of the maximum strain (and not a typical velocity amplitude), which seems natural.\\ 

\noindent{\textbf{The Childress bound \citep{childress:1969}}} A slightly different bound can be used by taking equation~(\ref{mag_en}) and noting that 
\begin{equation}
\int (\nabla \times {\bf B}) \cdot ({\bf u} 
 \times {\bf B}) \, dV
\le u_{max} \left( \int |\nabla \times {\bf B}|^2 \, dV \right)^{1/2}
\left( \int | {\bf B}|^2 \, dV \right)^{1/2} 
\end{equation}
and so, utilising equation~(\ref{ineq1}), this yields
\begin{equation}
\int (\nabla \times {\bf B}) \cdot ({\bf u} 
 \times {\bf B}) \, dV
\le \frac{a u_{max}}{\pi} \int |\nabla \times {\bf B}|^2 \, dV, 
\end{equation}
and hence
\begin{equation}
\mu \frac{\partial E_M}{\partial t} \le \left( \frac{a u_{max}}{\pi}
-\eta \right) \int | \nabla \times  {\bf B}|^2 \, dV . 
\end{equation}
Therefore dynamo action is only possible if 
\begin{equation}
{Rm = \frac{a u_{max}}{\eta} \ge \pi}
\end{equation}

Other bounds on dynamo action are also possible. However it can be shown that a steady or periodic dynamo can exist in a bounded conductor with an arbitrarily small value of the kinetic energy. Hence  there is no lower bound on dynamo action when $Rm$ is defined using the root mean square velocity, 
without placing limitations on the rate of strain \citep{proctor:2015}.

Motivated by the desire to construct experimental dynamos in the laboratory, there is now an effort to optimise dynamo flows given certain constraints. This involves taking the machinery developed for understanding the transition to turbulence  and using them to maximise growth-rates in the induction equation \citep{willis:2012}.

\subsection{\label{kd:work}Kinematic dynamos: some simple flows that work}

The kinematic problem considers the induction equation in isolation, for a prescribed velocity field $\mbu$. It is then natural to consider flows that are either steady, periodic in time or statistically steady. Because the induction equation is linear in the magnetic field, solutions take the form of magnetic fields that grow or decay exponentially on average and the task for the dynamo theorist is to determine flows that lead to growing solutions (i.e. that can act as dynamos) and perhaps examine the form of the solutions as a function of $Rm$. 

\subsubsection{\label{kinematic}Kinematic dynamos}

Having convinced ourselves that overly symmetric fields and flows are not good for dynamo action and that sufficient stretching is required, it is time to discuss some flows that do actually work as dynamos. These are not presented in chronological order, as we shall start with the simpler case of flows in an infinite domain, before moving onto flows in spheres and spherical shells. 

\medskip
\noindent{\textbf{The Ponomarenko Flow \citep{Ponomarenko:1973}:}}
\medskip

Here we consider the simplest possible flow that leads to dynamo action. It takes the form of a localised discontinuous ``screw" or vortex flow that in cylindrical co-ordinates $(s, \phi, z)$ is given by 
\begin{equation}
{\bf u} = 
\begin{cases}
s \Omega {\phihat} 
+ U {\bf \hat z}, \ s<a,\\  
0, \ s>a,
\end{cases}
\end{equation}
where $\Omega$ and $U$ are constants.
Note that this flow is not planar (owing to the presence of the throughflow $U {\bf \hat z}$) and has kinetic helicity, 
$
H = {\bf u} \cdot \bm{\omega} = 2 U \Omega.
$
We shall see the importance of kinetic helicity for large-scale dynamos in section~\ref{sec:org}.
Strong shear naturally occurs at (the physically unrealistic) discontinuity at $s=a$. In principle there are two independent parameters defining the flow $U$ and $\Omega$. These can be re-expressed in terms of an overall amplitude of the flow given by
$U_{amp} = (U^2 + a^2 \Omega^2)^{1/2}$ and the pitch angle $\chi = U / a \Omega$.

The trick for elegant solution for the kinematic dynamo modes in such a configuration is to note first that
the flow is steady and, together with the linearity of the induction equation, this implies that magnetic field will either grow or decay exponentially in time, with potentially a complex growth-rate $\lambda = \sigma + i \omega$. Secondly, 
the flow is independent of $z$ and $\phi$ and therefore, again making use of the linearity of the induction equation, monochromatic magnetic fields can be sought in those directions.

{It is therefore advantageous to seek solutions of the form 
\begin{equation}
{\bf B} = {\bf b}(s) \exp \left( \lambda t + i m \phi + ikz \right),
\end{equation}}
where $m$ and $k$ are the azimuthal and vertical wavenumbers, which must be non-zero to avoid the anti-dynamo theorems discussed earlier.

This model is extremely illuminating  as it can be solved pseudo-analytically. Non-dimensionalising length with  $a$ and time with the diffusive timescale $a^2/\eta$, the dimensionless input parameters are then the  the pitch of the spiral $\chi$, the magnetic Reynolds number defined as $Rm =  a U_{amp}/ \eta$ and the non-dimensional wavenumbers
$ka$ and $m$. The problem now constitutes an eigenvalue problem for the non-dimensional growth-rate  $a^2 \sigma / \eta$ and frequency $a^2 \omega/\eta$. 
This eigenvalue problem is solved separately for $s \le a$ (ensuring the magnetic field is regular at $s=0$) and $s > a$ (ensuring solutions decay as $s \rightarrow \infty$), with matching achieved by setting the magnetic field and the $z$-component of the electric field continuous at $s=a$.

This eigenvalue problem can be solved \citep[details are given in][]{JONES:2008} and marginally stable solutions can be found by setting $\Real{\lambda} = 0$ (for a given $Rm$, $\chi$, $ka$ and $m$). The critical solutions can then be ascertained by minimising $Rm$ over the other three input variables. 
The Ponomarenko flow is indeed a dynamo! It can be thought of as a prototype dynamo that is a model of vortical plume. Dynamo action sets in 
at $Rm_{crit}=17.72$, for 
$ka_{crit} = -0.3875$, $m=1$, $a^2 \omega / \eta= -0.41$
and $\chi= 1.31$, so the optimal pitch is $\mathcal{O}(1)$. 
The magnetic field is strongest near $s=a$, where it is generated by (the unphysical) shear.
Although it is important to examine how dynamo action onsets,  it is also of great interest to determine the behaviour at large $Rm$. Asymptotic solutions of the Ponomarenko dynamo (in the form of Bessel functions) show that \citep{Gilbert:1988}
the fastest growing modes are given by
\begin{equation}
|m| = (6(1 + \chi^{-2}))^{-3/4} \left( \frac{a^2 \Omega}{2 \eta}
\right)^{1/2},\quad
\sigma = 6^{-3/2} \Omega \,(1 + \chi^{-2})^{-1/2}. 
\end{equation}
In the presence of viscosity,  velocities with discontinuities are not  realistic and so the Ponomarenko dynamo has been extended to the continuous case where $ {\bf u} = s \Omega(s) \phihat + U(s) {\bf \hat z}$, where $\Omega(s)$ and $U(s)$ are smooth functions. In general such a problem must be solved numerically, using a two-point boundary eigenvalue solver. However at large $Rm$ some  elegant analysis \citep{Gilbert:1988} localises the field at the point
$s=a$ where
$m \Omega'(a) + k U'(a) = 0$.
Satisfyingly, it can be shown in this case that in order to achieve  dynamo action 
\begin{equation}
\left| \frac{\Omega''(a)}{\Omega'(a)} - \frac{U''(a)}{U'(a)} \right| < \frac{4}{a}.
\end{equation}

\medskip
\noindent{\textbf{The Roberts Flow \citep{Roberts:1972}:}}
\medskip

Perhaps the most illuminating, kinematic dynamo flow is the so-called G.O.~Roberts flow. This flow, like the Ponomarenko flow, is specially crafted to circumvent both the Cartesian version of Cowling's anti-dynamo theorem and the planar velocity anti-dynamo theorem, whilst still remaining fairly tractable.

The G.O.~Roberts flow is the special case of the well-studied 
ABC (Arnol'd, Beltrami and Childress) flow
in Cartesian co-ordinates $(x,y,z)$ in an infinite domain given by
\begin{equation}
{\bf u} = (C \sin z + B \cos y, A \sin x + C \cos z, B \sin y + A \cos x).
\end{equation}
Here $A=B=1$, $C=0$, so that
\begin{equation}
{\bf u}(x,y) = (\cos y, \sin x, \sin y + \cos x).
\end{equation}
This flow is two-dimensional (in the sense that it only depends on two coordinates), but it has all three components (thus not falling foul of the planar flow anti-dynamo theorem).

Before discussing the dynamo properties of the Roberts flow, it is useful to describe some of its basic hydrodynamic properties. The Roberts flow is integrable in the sense that it can be written in terms of a single steady streamfunction $\psi(x,y)$, so that
\begin{equation}
{\bf u} = \left( \frac{\partial \psi}{\partial y},
-\frac{\partial \psi}{\partial x}, \psi \right),
\end{equation}
where
\begin{equation}
\qquad \psi = \sin y + \cos x . 
\end{equation}

\begin{figure}
\begin{center} 
\includegraphics[width=5.5cm]{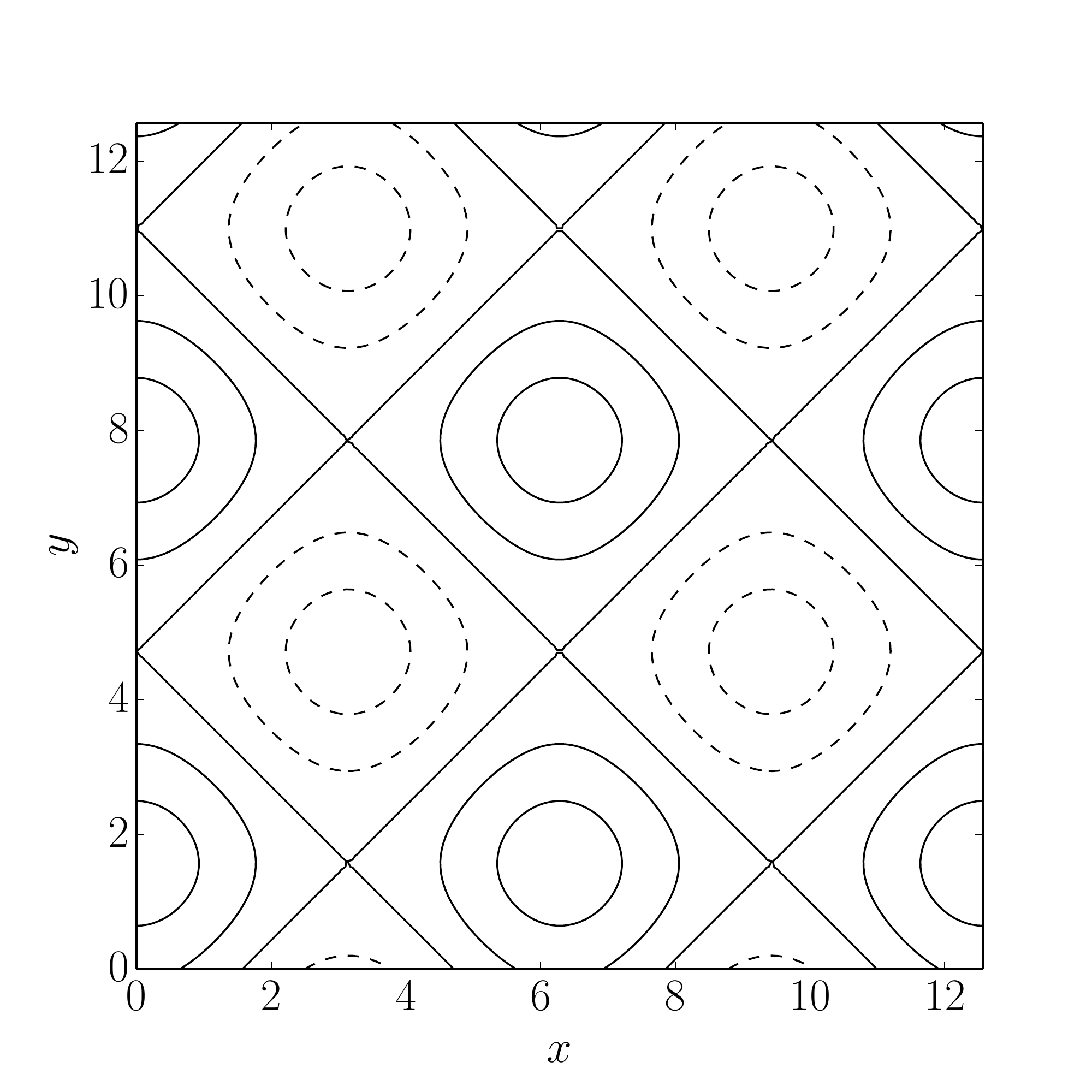}
\includegraphics[height=5.5cm]{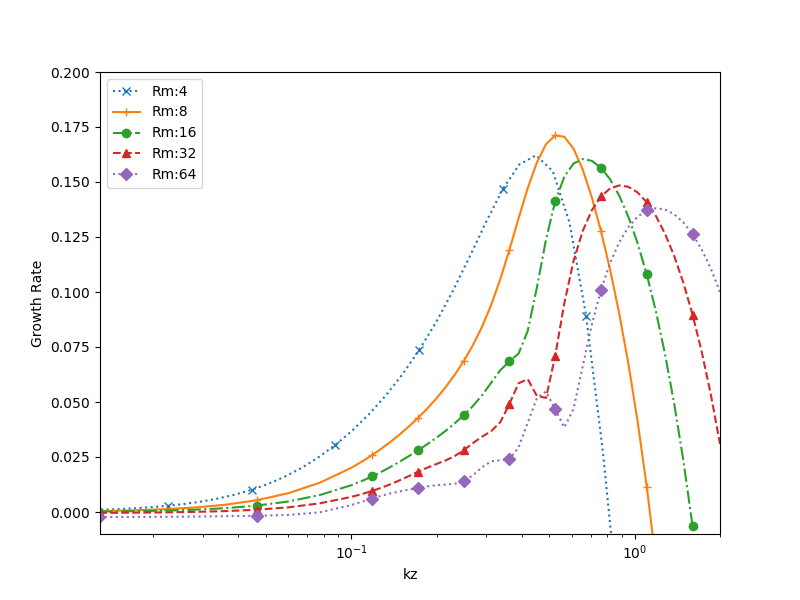}
\caption{(a) Contours of the streamfunction $\psi$ for the G.O.~Roberts flow. Positive (and zero) contours are solid and negative contours are dashed. (b)  Growth rate $\sigma$ as a function of wavenumber, $k_z$
for various $Rm$ \citep[after][]{Roberts:1972}. 
Courtesy of Andrew Clarke.}
\label{roberts_spatial_flow}
\end{center}
\end{figure}
The flow therefore takes the form of an array of helical cells with throughflow (see Figure~\ref{roberts_spatial_flow}). It has a typical horizontal spatial scale of $2 \pi$ and an infinite vertical scale. Moreover 
the winding sense of each helix in the array is the same. This ensures that the normalised relative kinetic helicity defined to be
\begin{equation}
\mathcal{H}_{rel} = \int \frac{\left| {\bf u}\cdot \bm{\omega}\right| }{\left| {\bf u}\cdot {\bf u} \right| ^{1/2}\left| \bm{\omega}\cdot \bm{\omega}\right| ^{1/2}}\,dV 
\end{equation}
is unity (a so-called \textit{maximally helical} flow); the importance of helicity for the large-scale dynamo properties of the flow will be discussed at some length in section~\ref{lin_ded}.

Roberts utilised the same considerations as Ponomarenko (though a year previous) to search for magnetic field solutions to the induction equation of the form
\begin{equation}
{\bf B} = {\bf b}(x,y) \exp ( \lambda t + ik_z z), \end{equation}
where ${\bf b}(x,y)$ is periodic in $x$ and $y$ (including the possibility that ${\bf b}(x,y)$ has a mean part).
The solution of the two-dimensional problem requires spectral methods yielding a matrix eigenvalue problem for the growth-rate $\sigma = \Real(\lambda)$ as a function of the wavenumber $k_z$ and $Rm$.  
For each value of $Rm$
there is an optimal value of $k_z$ that 
maximises the growth rate, as shown in Figure~\ref{roberts_spatial_flow}(b).

\begin{figure}
\begin{center} 
\includegraphics[width=6.50cm]{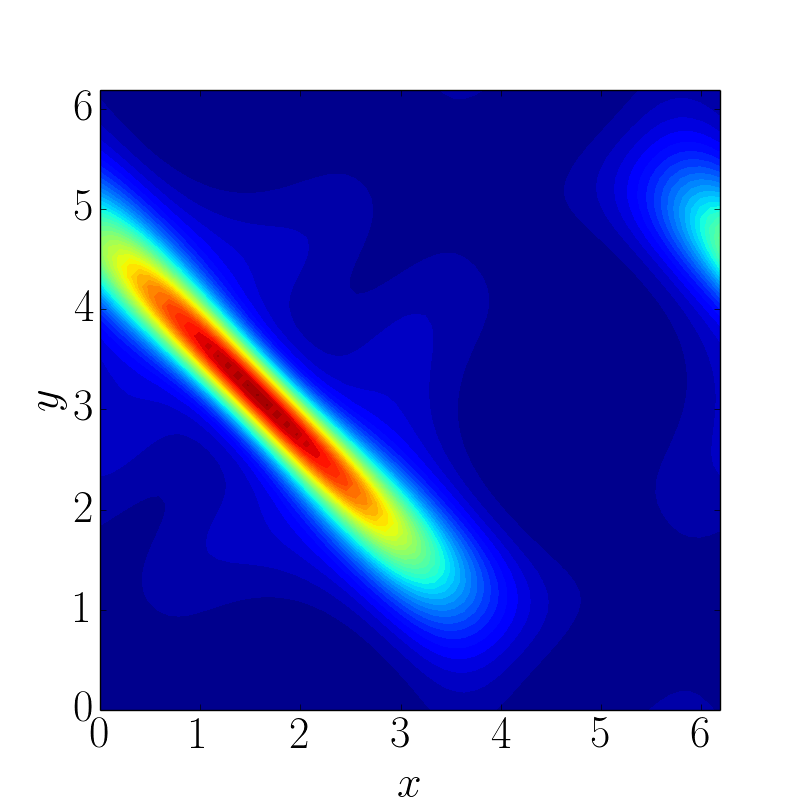}
\includegraphics[width=6.50cm]{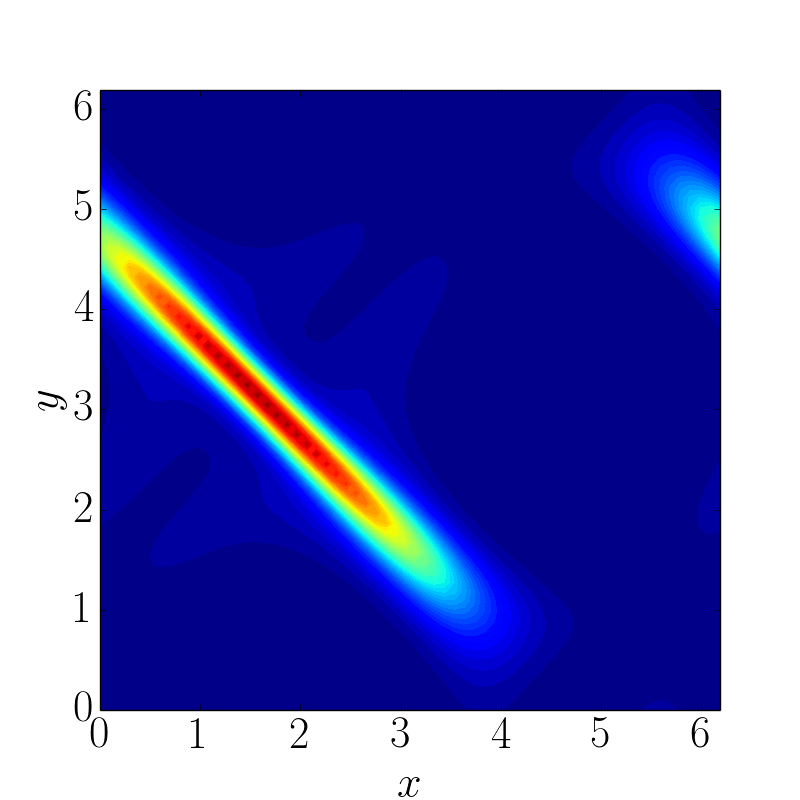}
\includegraphics[width=6.50cm]{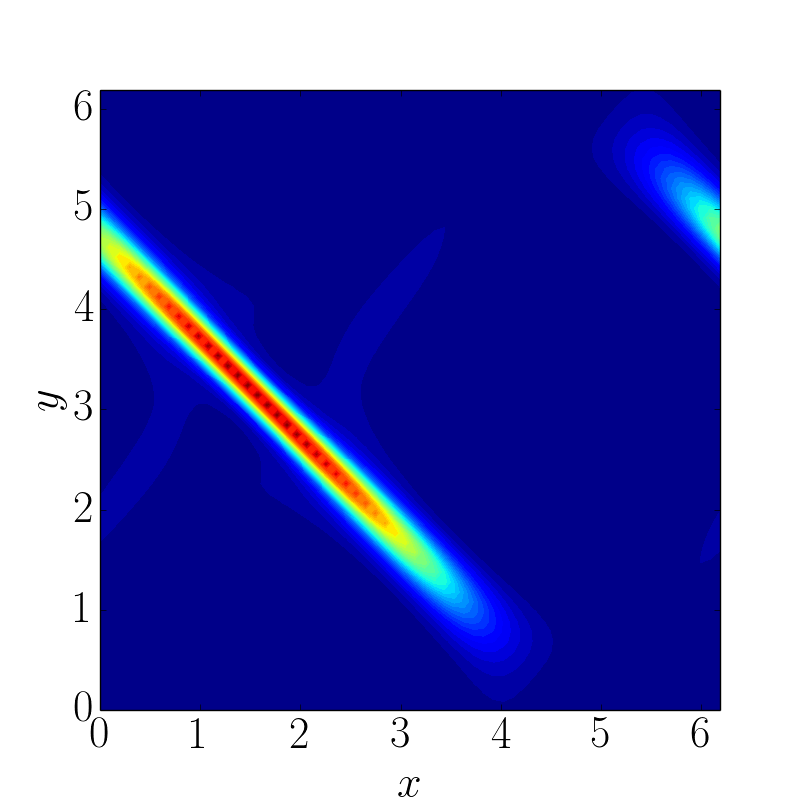}
\includegraphics[width=6.50cm]{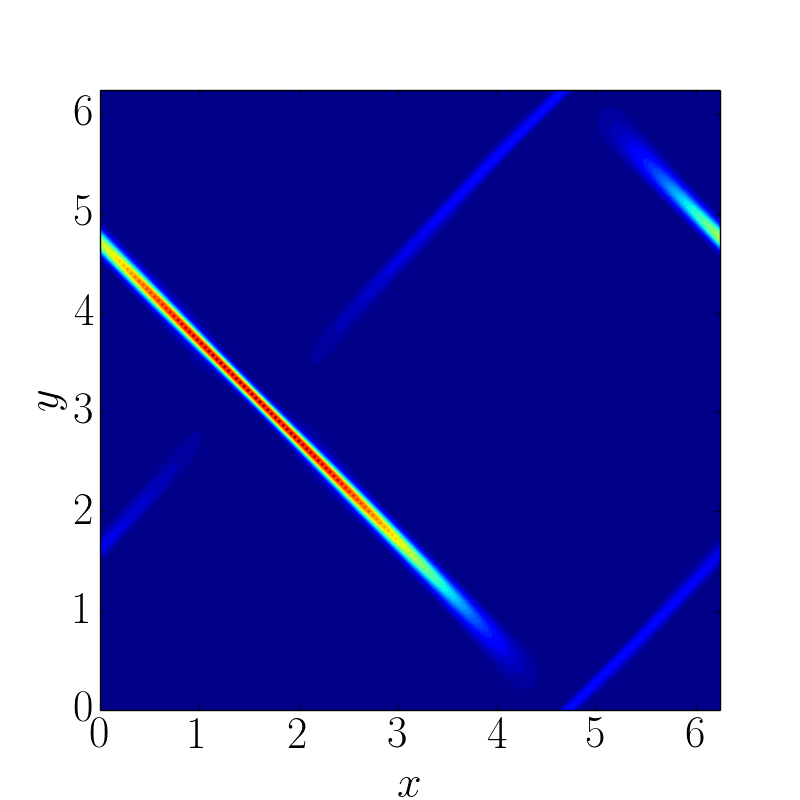}
\caption{Scaled magnetic energy in the plane $z=0$ for the Roberts flow for four different $Rm=16$, $32$, $64$ and $512$. As $Rm$ is increased the field is expelled into magnetic boundary layers of width $\mathcal{O}(Rm^{-1/2})$. Note only the domain between $0$ and $2 \pi$ is shown. The magnetic energy is scaled between $0$ and $1$. Figure courtesy of Andrew Clarke.}
\label{fig:robenergy}
\end{center}
\end{figure}

The form of the magnetic field for this flow is very illuminating. Figure~\ref{fig:robenergy} shows the magnetic energy in the plane $z=0$. For this steady flow the field is generated by the flow between the stagnation points. As $Rm$ is increased the field is expelled into magnetic boundary layers of \textit{width} $\mathcal{O}(Rm^{-1/2})$ enhancing diffusion (though the \textit{length} of the filamentary field structures is determined by the geometry of the flow). This is therefore an example of a \textit {small-scale dynamo} --- the field generated is dominated by structure at a scale smaller than a typical scale in the velocity; such a dynamo relies on the stretching overcoming the dissipative effects of diffusion.

In a tour-de-force paper \citet{Sow87} analysed the behaviour of the Roberts dynamo at high $Rm$ utilising asymptotic methods. He showed that the maximal growth-rate for this dynamo $\sigma \rightarrow 0$ as $Rm \rightarrow \infty$ (though very slowly --- indeed 
\begin{equation}
\sigma \sim \frac{\log \log Rm}{\log Rm}
\end{equation}
for large $Rm$). In the parlance of dynamo theory this makes the Roberts flow a \textit{slow dynamo} --- we shall discuss \textit{slow}, \textit{fast} and \textit{quick} dynamos later.

Finally for this section we stress again that the Roberts flow is a small-scale dynamo that generates field via stretching. Although the flow is helical, this is not its defining characteristic here. Indeed Roberts also considered a flow with no net helicity, \textit{viz.}
${\bf u}~=~(\sin 2y, \sin 2x, \sin(x+y))$. Although this is a less efficient dynamo than the helical Roberts flow, it is nonetheless a (slow) dynamo.\footnote{\textit{ 
\lq Helicity is not essential for
dynamo action, but it helps'} H.K.~Moffatt.}

\subsubsection{\label{kinsph}Kinematic dynamos in a spherical domain}

The examples above are instructive (and similar types of flows will be utilised later to illustrate other dynamo properties). These dynamos do have the drawback that, unlike astrophysical objects, both the flow and the magnetic field occupy all space and do not decay at large distances. A natural question to pose is whether similar types of flow can lead to dynamos in a bounded domain --- the most natural examples of which are spheres or spherical shells.

In spherical geometry one may decompose an incompressible velocity in terms of two scalar fields 
\citep{bg:1954}, i.e.
\begin{equation}
{\bf u} = \nabla \times \left( T(r,\theta,\phi,t)\,{\bf \hat r} \right) + \nabla \times \nabla \times \left( S(r,\theta,\phi,t)\, {\bf \hat r} \right).
\label{pol_tor_expansion}
\end{equation}
Here $T$ and $S$ are the toroidal and poloidal components. Alternatively we write
\begin{equation}
{\bf u} = \sum_{l,m} {\bf t}_l^m + {\bf s}_l^m,  
\end{equation}
where ${\bf t}_l^m$ and ${\bf s}_l^m$ are given by
\begin{eqnarray}
{\bf t}_l^m &=& \nabla \times \left( t_l^m (r,t) Y_l^m 
(\theta, \phi)\,{\bf \hat r}\right), \\
{\bf s}_l^m &=& \nabla \times \nabla \times \left(s_l^m (r,t) Y_l^m 
(\theta, \phi)\, {\bf \hat r}\right)
\end{eqnarray}
where $-l \le m \le l$ (and $Y_l^m$ is the spherical harmonic).

The flow is defined by choosing the values of $l$ and $m$  and the corresponding radial functional form for the scalar fields $t_l^m(r,t)$ and $s_l^m(r,t)$. In their landmark paper, 
Bullard and Gellman, nearly twenty years before the flows discussed in the last section, chose ${\bf u} = \epsilon {\bf t}_1^0 + {\bf s}_2^2$ and set $t_1^0 = r^2 (1-r)$,   and $s_2^2 = r^3 (1-r)^2$. Having made a similar expansion to equation~(\ref{pol_tor_expansion}) for the magnetic field, they used spectral interaction rules to determine the growth-rate of the field.
They reported dynamo action for this flow for high enough $Rm$. Unfortunately the dynamo growth reported here was spurious (as shown by subsequent higher resolution calculations). We shall return to this unfortunate property of dynamo calculations in section~\ref{caution}.
Although the reported dynamo action was incorrect, the work of Bullard \& Gellman revitalised the field, after the depressing wilderness years of anti-dynamo theorems, giving hope that self-excited dynamos were indeed possible.

Generalisations and extensions of the Bullard \& Gellman type dynamos have been studied by \citet{kumrob75} and \citet{dud89}.
The Kumar-Roberts flow, designed to model convective flows is non-axisymmetric and takes the form 
\begin{equation}
{\bf u} = \epsilon_0{\bf t}_1^0 + \epsilon_1 {\bf s}_2^0
+ \epsilon_2 {\bf s}_2^{2c} + \epsilon_3 {\bf s}_2^{2s},
\end{equation}
where $2c$ indicates $\cos 2 \phi$  and $2s$ indicates $\sin 2 \phi$. These dynamos were studied in detail by \citet{gub2000} for a range of choices of flow. Interestingly, although dynamos are typically found for large enough $Rm$, there are open parameter sets where  no dynamo occurs
at any $Rm$. The reason for this seems to be the presence of enhanced dissipation owing to the expulsion of magnetic flux into steady boundary layers. This once again shows the sensitivity of the dynamo process, especially in steady flows.

If the non-axisymmetric components are small ($\mathcal{O}(\epsilon)$) then it is possible to perform an asymptotic expansion with the axisymmetric components of flow and field dominating over the non-axisymmetric parts. This is the nearly axisymmetric dynamo of \citet{bra:1975}, which is an example of a self-consistent mean field model (see section~\ref{sec:org}).

The simpler steady spherical axisymmetric flows of \citet{dud89}, take the form
\begin{equation}
{\bf u} = {\bf t}_2^0 + \epsilon {\bf s}_2^0, \quad
 {\bf u} = {\bf t}_1^0 + \epsilon {\bf s}_2^0, \quad
 {\bf u} = {\bf t}_1^0 + \epsilon {\bf s}_1^0 
\end{equation}
with
$ t_1^0 = s_1^0 = r \sin \pi r$, and $t_2^0 = s_2^0 = r^2 \sin \pi r.$ 
These all give dynamo action, with the magnetic field having an $\exp (i m \phi)$
dependence, with $m=1$ being preferred.

\subsubsection{\label{caution}A note of caution}

The evolution equation for the magnetic field as described by the induction equation takes the form of a competition between magnetic field stretching (advection) and diffusion. Often these are both exponential processes; whether magnetic field grows or decays is then determined by the small difference between the efficiency of these processes. It is therefore of \textit{vital importance} that both of these processes are accurately represented by the numerical scheme utilised to solve the induction equation. Failure to do so will inevitably lead to the incorrect determination of the dynamo properties of the flow. This was first demonstrated  by the spurious solutions to the Bullard-Gellman dynamo \citep{bg:1954}. Owing to computational limitations, the resolution chosen for the spectral scheme was not sufficient to resolve the dissipative structures (current sheets) in the magnetic field. Hence the dissipation was underestimated and dynamo action was claimed when no such sustained magnetic field generation was possible. It is always this way. Insufficient resolution in a dynamo calculation will lead to the system appearing to be a dynamo when in fact it is not. This is troubling to dynamo theorists, though not as troubling to some as it should be.

It should be clear from the above discussion that any misrepresentation of the magnetic diffusion in a dynamo calculation is to be avoided at all costs. This includes not only misrepresentations that arise owing to lack of resolution, but also those that arise owing to the numerical scheme. It is often the case in fluids calculations that sub-grid processes are modelled (say by the inclusion of a hyperdiffusion, a numerical fix or by nominally solving the diffusionless problem with a stable scheme and using numerical errors to take the form of the dissipative processes 
\citep[see e.g.][]{miesch:2015}). In hydrodynamics, where there may be many competing terms in the nonlinear evolution equation for the velocity and dissipation may be small, such schemes \textit{may} not be too damaging --- but for dynamo calculations \textit{extreme care} must be taken in interpreting the results from such schemes. Let me stress again that if one is examining the competition between two processes in the linear induction equation, one of which is known to be incorrectly represented, there is a strong chance that the results are incorrect.

\section{\label{sec:wtp}So what's the problem then?}

The previous sections demonstrated that dynamo action is possible in simple steady flows in the kinematic regime, i.e.\ for a prescribed velocity field, in both infinite and finite domains. Despite the restrictions placed on dynamo fields and the velocities that generate them, growing solutions for the magnetic field are possible.

The rest of this Perspective will focus on the current issues that are troubling dynamo theorists. In this section we shall introduce each issue, indicating its importance for our understanding, before describing in subsequent sections the past and current attempts at a resolution of each issue and concluding in section~\ref{Future} with possible future lines of research  (most of which are based on current techniques utilised in hydrodynamics).

\subsection{\label{turb}Turbulence --- high and low $Pm$}

The flows considered above are defined at a single spatial scale and are steady. Of course naturally occurring flows in geophysics and astrophysics are neither. In general, owing to the vast lengthscale of such flows, the Reynolds numbers of the dynamo flows are enormous  and the flows are extremely turbulent --- hence the title of this Perspective. Moreover similar arguments may pertain to the magnetic Reynolds number $Rm$. The ratio of these two non-dimensional numbers is given by the magnetic Prandtl number $Pm = \nu/\eta$, which is a property of the fluid/plasma. In geophysics and astrophysics $Pm$ is usually either very large or very small; in numerical experiments it is usually $\mathcal{O}(1)$. The combination of turbulence, with its large range of spatial and temporal scales to be resolved and the naturally occurring extreme values of $Pm$ presents a formidable problem to the theorist and the numericist.

When $Pm$ is large ($Rm \gg Re$) the magnetic field can be generated on scales much smaller than the viscous dissipation scale. How dynamos (at least kinematically) behave in this regime is the preserve of \textit{Fast Dynamo Theory}, which is discussed in section~\ref{fast_dynamo}. The situation is reversed (and more complicated) when $Pm$ is small ($Re \gg Rm$); in this case the magnetic field dissipates in the inertial range of the turbulence; with substantial implications for the dynamo --- this case is discussed in sections~\ref{kazantsev} and ~\ref{coherent}.

\subsection{\label{organisation}Organisation}

The dynamos described so far tend to be small-scale dynamos in the sense that they generate field kinematically on a scale smaller than the typical velocity scale (sometimes on the diffusive scale, which gets very small as $Rm$ increases). However the observed geophysical and astrophysical magnetic fields described earlier display organisation on the scale of the astrophysical object (and are sometimes called large-scale dynamos). A natural question, therefore, is 
``What is the origin of this organisation and how does the mechanism leading to organised flows compete with that leading the production of small-scale fields?'' Can this competition be understood within a kinematic framework, or are the nonlinear effects from the momentum equation required? Is it the case that the organisation arises as a consequence of turbulent interactions or despite them? These questions are discussed in section~\ref{sec:org}, where the concept of mean-field electrodynamics will be introduced and critiqued.

\subsection{\label{saturation}Saturation}

From the point of view of a fluid dynamicist, the focus of dynamo theorists on the induction equation is somewhat mysterious, though, to be fair, the difficulties in finding dynamo solutions described above go some way to explaining this preoccupation. However, in the present day it makes no sense to make the kinematic assumption and take the velocity as prescribed. What is required is to solve the coupled Navier-Stokes and induction system. Questions that can be addressed within this framework include: How does a turbulent small-scale dynamo saturate? Do organised fields saturate in a similar manner? Are there dynamos that operate through instabilities of a magnetic field driving a flow --- if so can these lead to subcritical dynamo action? These questions are addressed in section~\ref{sec:saturation}

\subsection{\label{rotation}The Role of Rotation --- rapid or otherwise}
As we shall see in section~\ref{sec:org}, the dynamo generation of organised magnetic field requires a breaking of reflexional symmetry of the system \citep{Moffatt:1978}. In geophysical and astrophysical systems this naturally occurs via the effects of rotation (often in combination with stratification). Rotation is responsible for providing  correlations that lead to the generation of a net electromotive force, in much the same way as it can drive large-scale flows via the introduction of correlations and non-trivial Reynolds stresses \citep[see e.g.][]{vallis:2006}. For dynamos, rotation plays a key role in determining the solutions of the induction equation. Moreover, once the field is generated it becomes dynamic in the momentum equation. In the case of rapid rotation, such as is the case in the Earth's core and in rapidly rotating stars, this is a particularly interesting and delicate issue. Magnetic fields may break leading-order geostrophic balances (leading to so-called magnetostrophic balance) or, even if the primary balance
is still geostrophic, the fields may play a crucial role in the prognostic equation for unbalanced motions, as discussed in section~\ref{balances}. The presence of strong rotation can lead to dynamo-generated magnetic field acting as a conduit to turbulence via nonlinear processes. The magnetic field can act so as to relax the constraints engendered by strong rotation, leading to more efficient (convective) driving of turbulence and strongly subcritical behaviour. All these issues are discussed in section~\ref{balances}.

\section{\label{sec:small}Small-scale magnetic field generation}

In this section we describe current research into small-scale dynamos. One problem that appears to have been solved to the satisfaction of dynamo theorists is the \textit{fast dynamo problem}.

\subsection{\label{fast_dynamo}One-scale velocity fields and fast dynamo theory}

\begin{figure}
\begin{center} 
\includegraphics[width=7.0cm]{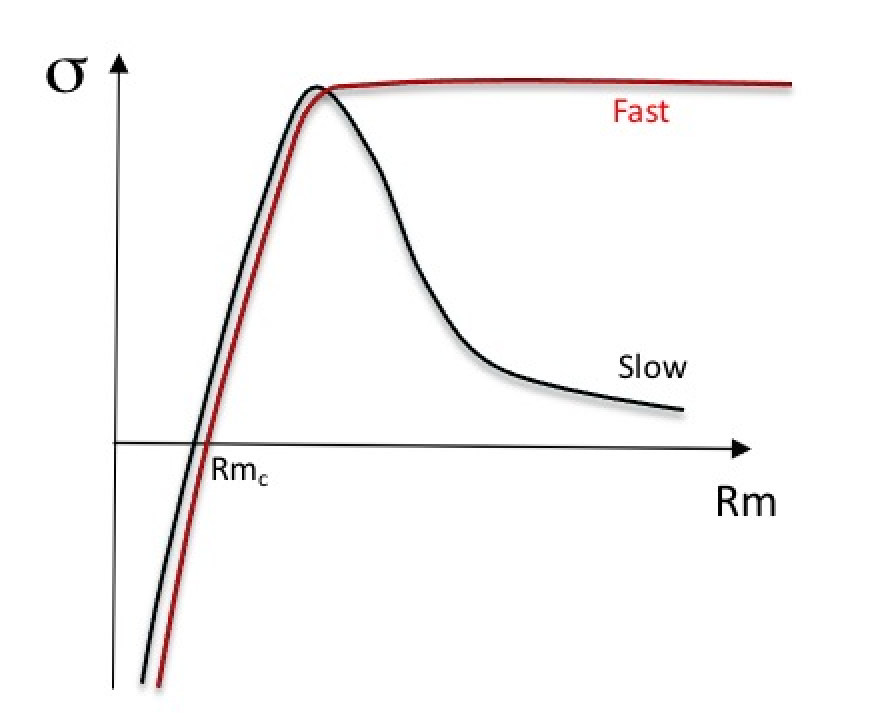}
\end{center}
\caption{Growth rate $\sigma$ as a function of $Rm$ for a slow dynamo (black curve) and a fast dynamo (red curve). }
\label{fastgr}
\end{figure}

Simply put, fast dynamo theory is concerned with the kinematic generation of magnetic field (on any scale) at high $Rm$ (such as pertains in virtually all astrophysical objects). Consider a velocity field defined at a single scale $\ell_0$ with a characteristc velocity $u_0$. Then, defining $Rm = u_0 \ell_0 /\eta$ in the usual way, the fast dynamo problem is concerned with the behaviour of the growth-rate $\sigma(Rm)$ of the dynamo at high $Rm$. There are two possibilities, either $\sigma \rightarrow 0$ as $Rm \rightarrow \infty$  in which case the dynamo is described as `slow'. Alternatively $\sigma$ may $\rightarrow {\rm const} > 0$ as $Rm \rightarrow \infty$ ---  a so-called \textit{fast dynamo}. The two possible options for the growth-rate curve are illustrated in Figure~\ref{fastgr}.

\citet{moffatproctor:1985} demonstrated that the eigenmodes associated with fast dynamo action may exist, providing that they have a scale of variation $\mathcal{O}(Rm^{-1/2})$ as $Rm \rightarrow \infty$, nearly everywhere in the fluid domain.
Much of our understanding of the behaviour of fast dynamos arises from the field of dynamical systems and mixing; progress has been made by examining the simpler problem where the flow is modelled as a discontinuous (in time) map. Indeed there are strong parallels between the two problems. I will not go into details here, but the interested reader should consult the excellent \cite{chilgil:1995}.

A central result arising from dynamical systems approaches to fast dynamo action is that which bounds the asymptotic growth-rate by the topological entropy of the flow \citep{Klapper:1995}; see also \citet{finnott:1988}. This is important as it immediately rules out the possibility of fast dynamo action for integrable flows, such as the steady $2 \frac{1}{2}$-dimensional flows discussed earlier. Chaotic particle paths are therefore required for a flow to be a fast dynamo --- these may be introduced in two obvious ways, either by making the flow fully three-dimensional or by introducing time dependence to the $2 \frac{1}{2}$-dimensional flows (whilst for simplicity still keeping the flows at a single spatial scale).
\begin{figure}
\begin{center} 
\includegraphics[height=5.45cm]{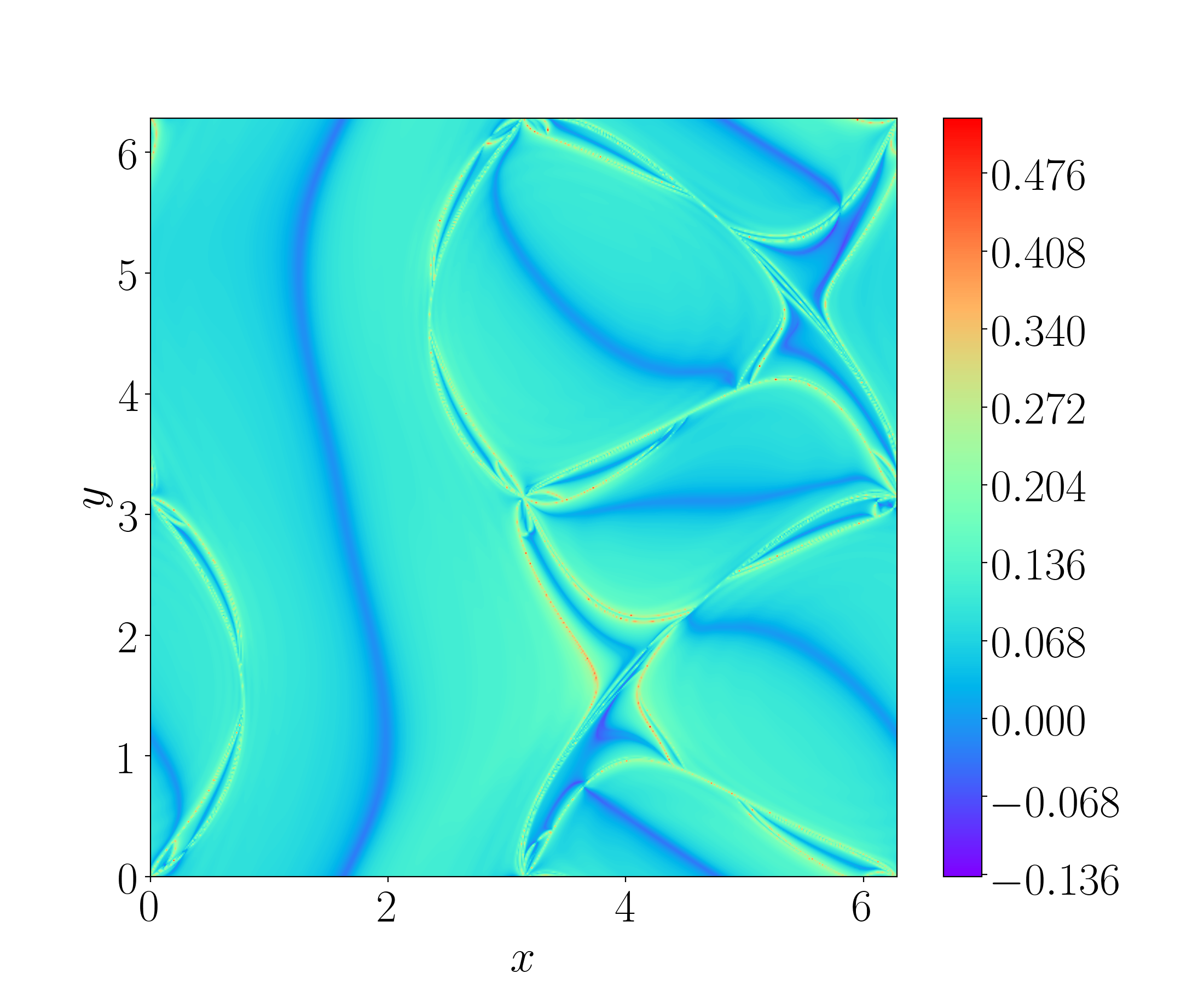}
\includegraphics[height=5.2cm]{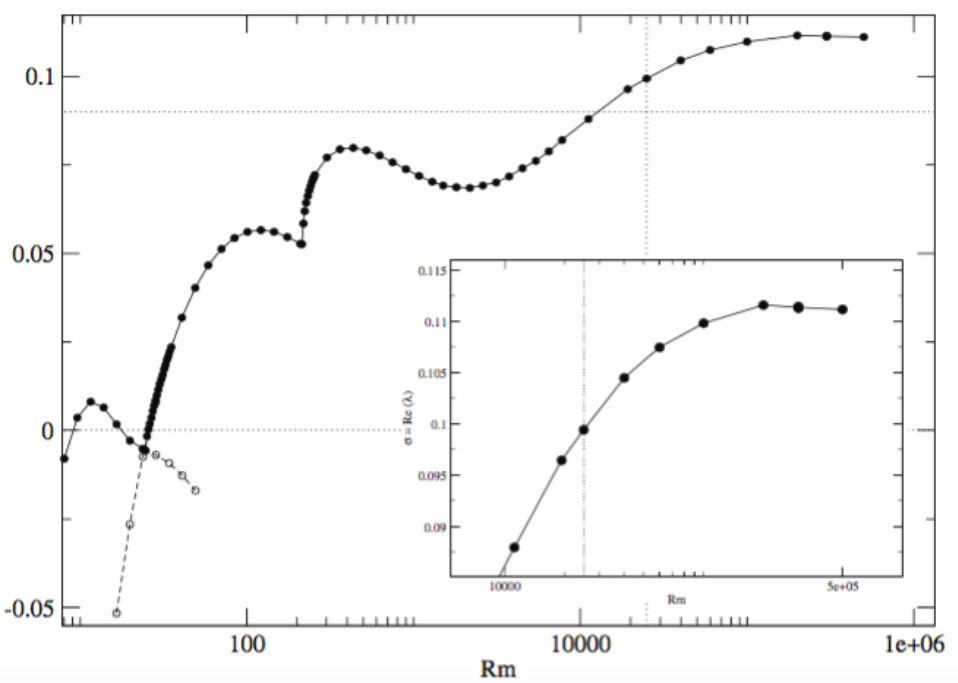}
\caption{ Finite time lyapunov exponents for the $ABC=1$ flow \citep[after][]{BrummCattTob:2001}. The $ABC=1$ flow is
unusual in having rather large integrable KAM regions and small chaotic regions; Poincar\'e sections for this flow can be found in \cite{dom:1986}. (b) growth-rate as a function of $Rm$ for the ABC dynamo \citep[after][]{bouyadormy}}
\label{ABC}
\end{center}
\end{figure}
Moving to three dimensions  makes computations at high $Rm$ extremely challenging, owing to the severe constraints imposed by the requirement to resolve structures on the scale $Rm^{-1/2}$ in three dimensions. Nonetheless, progress has recently been made in investigating the dynamo properties of the ABC flow given earlier as
\begin{equation}
{\bf u}_{ABC} = (C \sin z + B \cos y, A \sin x + C \cos z, B \sin y + A \cos x).
\end{equation}
This flow is chaotic, as shown by the Poincar\'e sections of the particle paths in  \citet{dom:1986} and the finite-time Lyapunov exponents shown in Figure~\ref{ABC}(a). Calculations of the growth-rate as a function of $Rm$ have periodically been made, since the earliest calculations \citep[see e.g.][]{arnkork:1983,gallfrisch:1984} extending to higher and higher $Rm$.
Figure~\ref{ABC}(b) taken from \citet{bouyadormy} shows that with current computing facilities $Rm \sim 10^5$ is possible for this flow.
The figure shows that even at this value of $Rm$ the dynamo is presumably not in its asymptotic regime, as the growth-rate is still above the theoretical bound provided by the topological entropy.

\begin{figure}
\begin{center} 
\includegraphics[height=5.45cm]{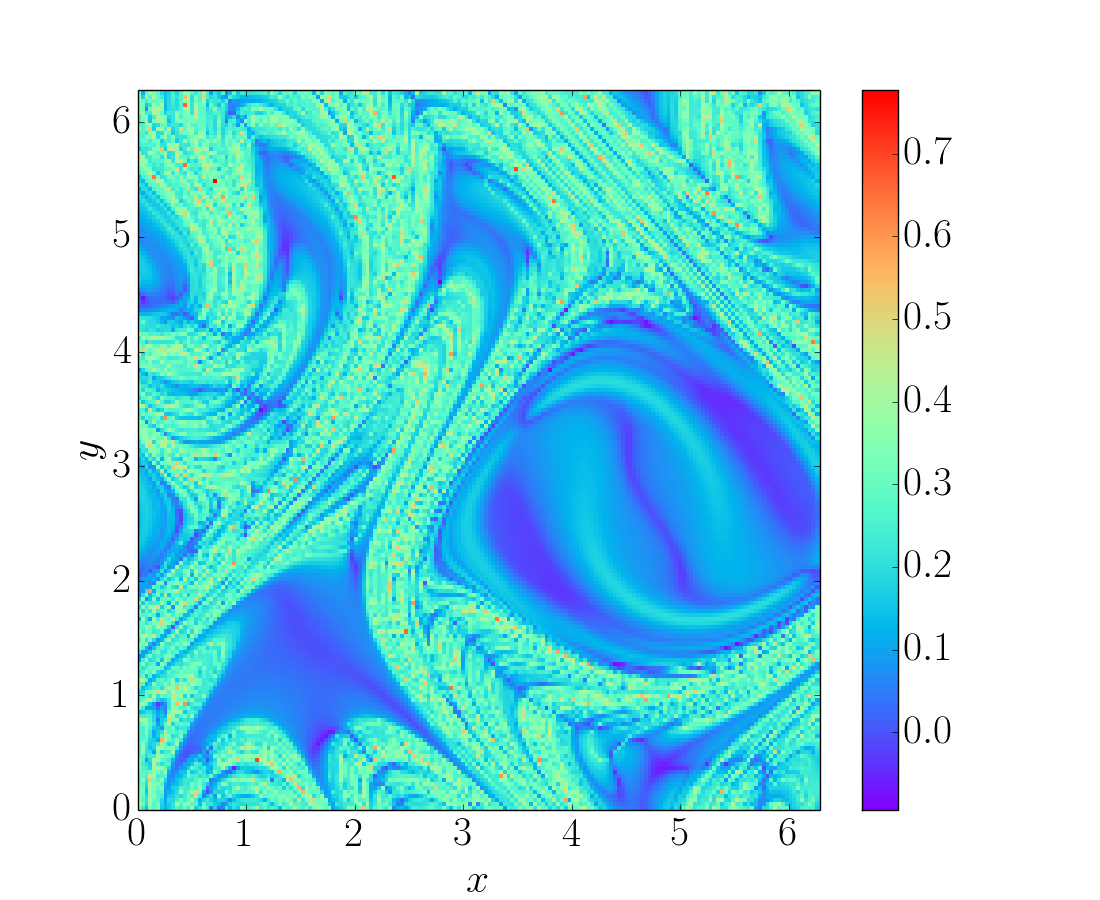}
\includegraphics[height=5.0cm]{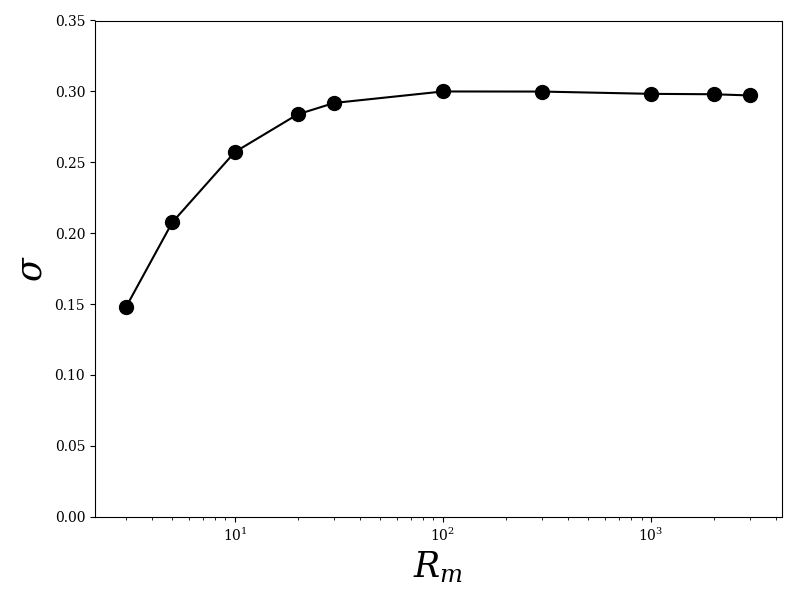}
\caption{(a) Lyapunov exponents as a function of starting position in the plane $z=0$ for the GPCP flow. (b) Growth-rate as a function of $Rm$ for fixed $k_z = 0.57$ for the Galloway-Proctor CP flow. Courtesy of Andrew Clarke (after \cite{Galloway:1992}).
}
\label{lyap_gp}
\end{center}
\end{figure}

A much more promising way to investigate fast dynamo action is to introduce chaos via time-dependence in a $2 \frac{1}{2}$-dimensional flow. This of course has the benefit of allowing computations to proceed in 2 dimensions. This approach was pioneered by \citet{otani:1988,otani:1993} and \cite{Galloway:1992}, who constructed similar flows. Here I give details dynamo action in the Galloway-Proctor circularly polarised (GPCP) flow, which takes the form
\begin{equation}
{\bf u} = \nabla \times (\psi (x,y,t) {\bf \hat z}) + \psi (x,y,t){\bf \hat z},
\end{equation}
where
\begin{equation}
\psi = \sin ( y + \sin t) + 
\cos (x + \cos t).
\end{equation}
This is based on the Roberts flow, with the time-dependence being introduced via a rotation of the cellular pattern around a circle. This introduces a significant region of chaos (though regions of integrability still remain) as shown in the finite-time Lyapunov exponents of Figure~\ref{lyap_gp}).

The magnetic field for the GPCP-flow is again generated on the small-diffusive lengthscale $\ell_b \sim Rm^{-1/2}$ as $Rm \rightarrow \infty$. The growth-rate (which is  a function of vertical wavenumber $k_z$ and $Rm$) has the form shown in Figure~\ref{lyap_gp}(b) for fixed $k_z$. Interestingly, for large enough $Rm$ the preferred wavenumber becomes independent of $Rm$, as does the optimised growth-rate. Notice also that this dynamo reaches its asymptotic growth-rate for moderate $Rm$ and so is an example of a \textit{quick} dynamo \citep{tobcatt:2008a}, discussed below. Although it is impossible to prove numerically that these time-dependent flows are fast dynamos, all the evidence certainly points in this direction. It is now widely accepted that sufficiently chaotic flows at a single scale will act as fast dynamos. Of course it is possible to introduce time-dependence to three-dimensional, single scale cellular flows such as the ABC-flow, which has the tendency to increase the regions of chaos and hence their dynamo efficiency. Such dynamos have been investigated in both the linear and nonlinear regimes \citep{BrummCattTob:2001}.

One aspect of fast dynamo theory that is not widely appreciated is that it is only applicable for high $Pm$ fluids, as noted by \cite{tobcatt:2008a}. The paradigm of holding the flow fixed and increasing $Rm$ is equivalent to increasing the $Pm$ of the fluid. Clearly increasing $Rm$ whilst holding $Pm$ fixed requires an equivalent increase in $Re$, which will lead to a change in the form of the flow for any realistic forcing mechanism. In particular, increasing $Re$ almost inevitably leads to turbulence with a wide range of spatial and temporal scales.

\subsection{\label{multi-scale}Multi-scale velocity fields}

In this section we examine kinematic dynamos where the underlying flow has a spectrum of spatial scales. As discussed above one can think of two cases; at high $Pm$ the magnetic field dissipates at scales much smaller than the  smallest eddy and one can rely somewhat on fast dynamo theory based on considering the smallest eddy alone (since the smallest eddy is the one with the fastest turnover time). At low $Pm$ the magnetic field dissipates in the inertial range of the turbulence and life becomes more complicated. We shall start by considering the simplified (some might say over-simplified) case of random velocity fields.

\subsubsection{\label{kazantsev}Random dynamos - the Kraichnan-Kazantsev formulation}

\begin{figure}
\begin{center}
\includegraphics[height=8cm]{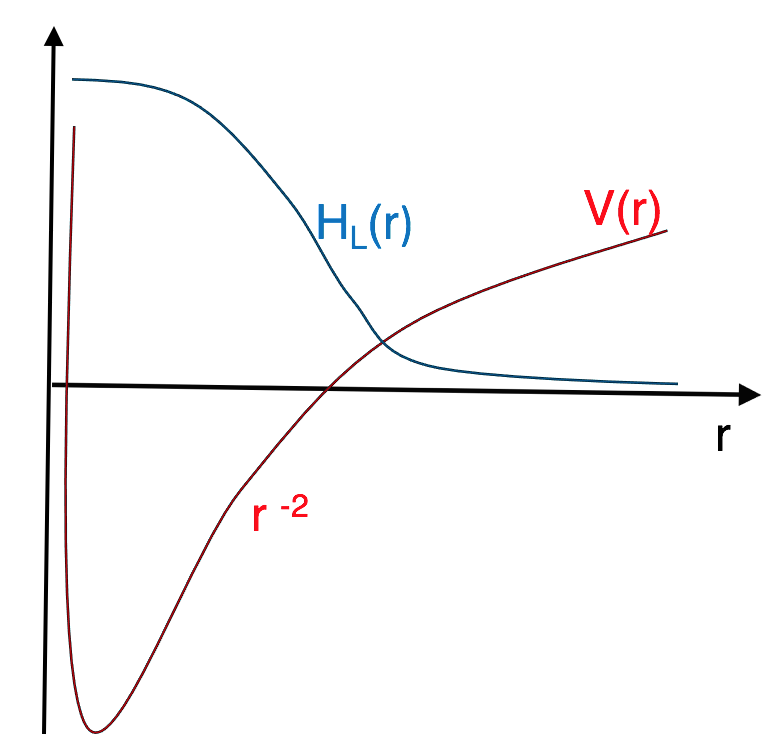}
\end{center}
\caption{Schematic representation of the potential $V(r)$ and the spatial part of the longitudinal magnetic correlator
$H_L(r)$ in the Kazantsev model. The potential has a $1/r^2$ behaviour in the inertial range. Its overall height is
 determined by the velocity roughness, and its large $r$ behaviour by the boundary conditions. The magnetic correlator is 
 peaked at small scales and decays exponentially for large values of $r$. After \citet{tcb11}
}
\label{fig_kaz}
\end{figure}



The discussion developed in this section summarises that of \citet{tcb11}. Turbulent velocity fields have a coherent and random component, both of which contribute to the dynamo properties. The simplest model of (kinematic) dynamo action driven by a purely random flow on a range of scales is the so-called 
Kazantsev model \citep{kaz:1968}. It is an example of a solvable model for the statistics of the magnetic field based on those for a prescribed velocity and as such may be viewed as an early example of Direct Statistical Simulation (see section~\ref{Future}). 

The prescribed velocity takes the form of  a Gaussian, $\delta$-correlated 
(in time) velocity field, with zero mean and covariance given by
\begin{eqnarray} 
\langle u_i({\bf x+r}, t)u_j({\bf x}, \tau) \rangle
=\kappa_{ij}({\bf x},{\bf r})\delta(t-\tau).
\label{correlator}
\end{eqnarray}
Geophysical and astrophysical   flows that lead to dynamo action are in general anisotropic and inhomogeneous (and this should be a feature of any description of dynamo action). However analytic progress can be made for the dynamo problem by assuming that the underlying flow is isotropic and homogeneous, in which case the velocity correlation 
function has the form
\begin{eqnarray}
\kappa_{ij}({\bf r})=\kappa_N(r)\left(\delta_{ij}-\frac{r_ir_j}{r^2} 
\right)+\kappa_L(r)\frac{r_ir_j}{r^2},
\label{kappa}
\end{eqnarray}
where $r = |{\bf r}|$. In addition, for incompressible velocity fields we have that 
$\kappa_N=\kappa_L+(r\kappa_L^\prime)/2$, and so the
velocity statistics can be characterized by the single scalar function 
$\kappa_L(r)$. 
Progress is made by defining a corresponding expression for the magnetic covariance
\begin{eqnarray}
\langle B_i({\bf x+r}, t)B_j({\bf x}, t) \rangle=H_{ij}({\bf x},{\bf r}, t),
\label{bcorrelator}
\end{eqnarray}
where for similar reasons to above
\begin{eqnarray}
H_{ij}=H_N(r,t)\left(\delta_{ij}-\frac{r_ir_j}{r^2} 
\right)+H_L(r,t)\frac{r_ir_j}{r^2}.
\label{hdef}
\end{eqnarray}
Similarly, $\nabla\cdot {\bf B}=0$ gives $H_N=H_L+(rH_L^\prime)/2$, and so the correlator is completely determined by $H_L(r,t)$.
The evolution equation for $H_L$, in terms of the input function
$\kappa_L(r)$, is then given by
\begin{eqnarray}
{\partial_t H}_L=\kappa H_L^{\prime \prime} + \left(\frac{4}{r}\kappa+\kappa' \right)H_L^{\prime}
+\left( \kappa''+\frac{4}{r}\kappa' \right)H_L,
\label{kequation}
\end{eqnarray}
where $\kappa(r)=2\eta+\kappa_L(0)-\kappa_L(r)$ is the 
the `renormalized' velocity correlation 
function.
Remarkably,  changing variable via $H_L=\psi(r,t) r^{-2}\kappa(r)^{-1/2}$ leads to  a related equation
that formally coincides with the Schr\"odinger equation in imaginary time, i.e.
\begin{eqnarray}
\partial_t \psi=\kappa(r)\psi''-V(r)\psi.
\label{sequation}
\end{eqnarray}
Here $\psi$ describes the wave function of a quantum particle with variable 
mass, $m(r)=1/[2\kappa(r)]$, in a one-dimensional potential~($r>0$) given by
\begin{eqnarray}
V(r)=\frac{2}{r^2}\kappa(r)-\frac{1}{2}\kappa''(r)-\frac{2}{r}\kappa'(r)-
\frac{(\kappa'(r))^2}{4\kappa(r)}.
\end{eqnarray}

Equation~(\ref{sequation}) has been investigated in depth for various choices of prescribed $\kappa(r)$ (see \cite{zelruzsok:1990, arphorvai:2007,chertetal:1999}).
For a more thorough review see \citet{tcb11,rincon:2019}. 

Briefly, progress can be made by considering  homogeneous, isotropic turbulent flow with a wide-range of scales, and
a well-defined inertial range that extends from the integral scale $\ell_0$ to the dissipative scale $\ell_\nu$, with a 
dissipative sub-range for the scales $\ell < \ell_\nu$.
The flow can then be characterised by the second order structure function $\Delta_2(r) = \langle |({\bf u}({\bf x},t) - {\bf u}({\bf x}+{\bf 
r},t)).{\bf e}_r|^2  \rangle$, where $r = |{\bf r}|$ and ${\bf e}_r={\bf r}/r$. 
The inertial and dissipative ranges are then described  by the scaling exponents of $\Delta_2(r)$ with 
$\Delta_2(r) \sim r^{2 \alpha}$, where $\alpha$ is termed the \textit{roughness exponent} of the flow. In the dissipative sub-range we expect $\alpha=1$ as the velocity is a smooth function of position, whilst for the inertial range the velocity is {\it rough} and  $\alpha < 1$\footnote{
for Kolmogorov turbulence $\alpha = 1/3$.}.
If the slope of the energy spectrum in the inertial range is given by $E_k \sim k^{-p}$ then  $p$ is related 
to the  roughness exponent by $p = 2 \alpha+1$. 

We shall briefly describe dynamo action in the 
$Pm \gg 1$ case, where the  the velocity is smooth on the dissipative scale of the field, and the $Pm \ll 1$ case where the velocity is rough there.
For the smooth case, exponentially growing solutions of equation~(\ref{kequation}) can be found  
with the magnetic energy spectrum $E_M$ peaked at the magnetic dissipation scale, just as for the single-scale flows considered earlier. The spectrum for the magnetic energy in the range $1/l_\eta < k < 
1/l_\nu$ has  $E_M \sim k^{3/2}$, independent of the  spectral index for the velocity in the 
inertial range \citep{kuland:1992,schek:2002}. This regime for a smooth velocity
is also referred to as the Batchelor regime, as it corresponds to that of \citet{batchelor:1959} for passive-scalar advection.

The low $Pm$ case is more complicated, and we shall not go into much detail here. Briefly,  when $Pm \ll 1$, the magnetic field dissipates in the inertial range, where $\kappa(r) \sim r^{1+\alpha}$ \citep[see e.g][]{tcb11}.   Hence in the inertial range the 
Schr\"odinger equation~(\ref{sequation})  has an effective 
potential $U_{eff}(r)=V(r)/\kappa(r)=A(\alpha)/r^2$,  
where $A(\alpha)=2-3(1+\alpha)/2-3(1+\alpha)^2/4 $.
An example of the form of this potential is given in 
Figure~\ref{fig_kaz}.
At small scales ($\ell < \ell_\eta$) this effective potential is regularised 
by magnetic diffusion. Growing dynamo solutions
correspond to bound states for the wave-function $\psi$. These are guaranteed to exist if $A(\alpha) < -1/4$, which is always the case for the physically realistic range $0 < \alpha < 1$. Hence, dynamo action is always possible even in the case of a rough 
velocity at low $Pm$ \citep{bolcatt:2004}. Interestingly, the corresponding wave-function is  concentrated around the minimum of the potential 
at $r \sim \ell_\eta$ and decays exponentially for $r > \ell_\eta$. It is important to note that 
the effective potential always remains $\propto 1/r^2$ in the inertial range;  its depth decreases as $\alpha \rightarrow 0$. Hence, it is harder to drive dynamo action the 
rougher the velocity --- this has serious consequences for our ability to generate magnetic fields in liquid metal experiments, as discussed in section~\ref{expts}.

Equation~(\ref{sequation}) can be solved asymptotically and solutions demonstrate that the growth-time of the dynamo is of 
the order of the turnover time of the eddies at the diffusive scale ($\ell_\eta$).
Moreover it shows \citep{bolcatt:2004} that there is a dramatic increase in the effort (computational or experimental) 
as the 
velocity becomes rougher (1+$\alpha$ gets smaller). Hence the effort required to drive a dynamo in a rough velocity also increases. 

For these random flows we may now describe the behaviour of the critical magnetic Reynolds number $Rm_c$ as one moves from large $Pm$ to small $Pm$.
At high $Pm$ the 
effort necessary to drive a dynamo is modest. As $Pm$ decreases through unity the viscous scale becomes smaller than the resistive scale and the dynamo begins to operate in the inertial 
range. There is then an increase in the effort required to sustain dynamo action. However, once 
the dynamo is in the inertial range, further decreases in $Pm$ do not make any difference to the effort 
required (as measured by $Rm_c$).
This picture is largely consistent with numerical models of dynamo action in random flows \citep{christensenetal:1999, 
yousefetal:2003,
schekochihietal:2004c,schekochihinetal:2005c}
as we shall see in the next section.

The Kazantsev model as proposed is extremely restrictive, though it can easily be extended to cases where the correlation time of the turbulence is finite \citep{vainkit:1986,kleerogsok:2002}, providing the growth time of the dynamo is long compared with this correlation time. Anisotropic versions of the Kazantsev model have also 
been constructed by Schekochihin {\it et al.} (2004b).
These solvable models will also prove useful in understanding the generation of organised field (as discussed in section~\ref{sec:orgkaz})

\begin{figure}
\centering
\includegraphics[width=1.0\textwidth]{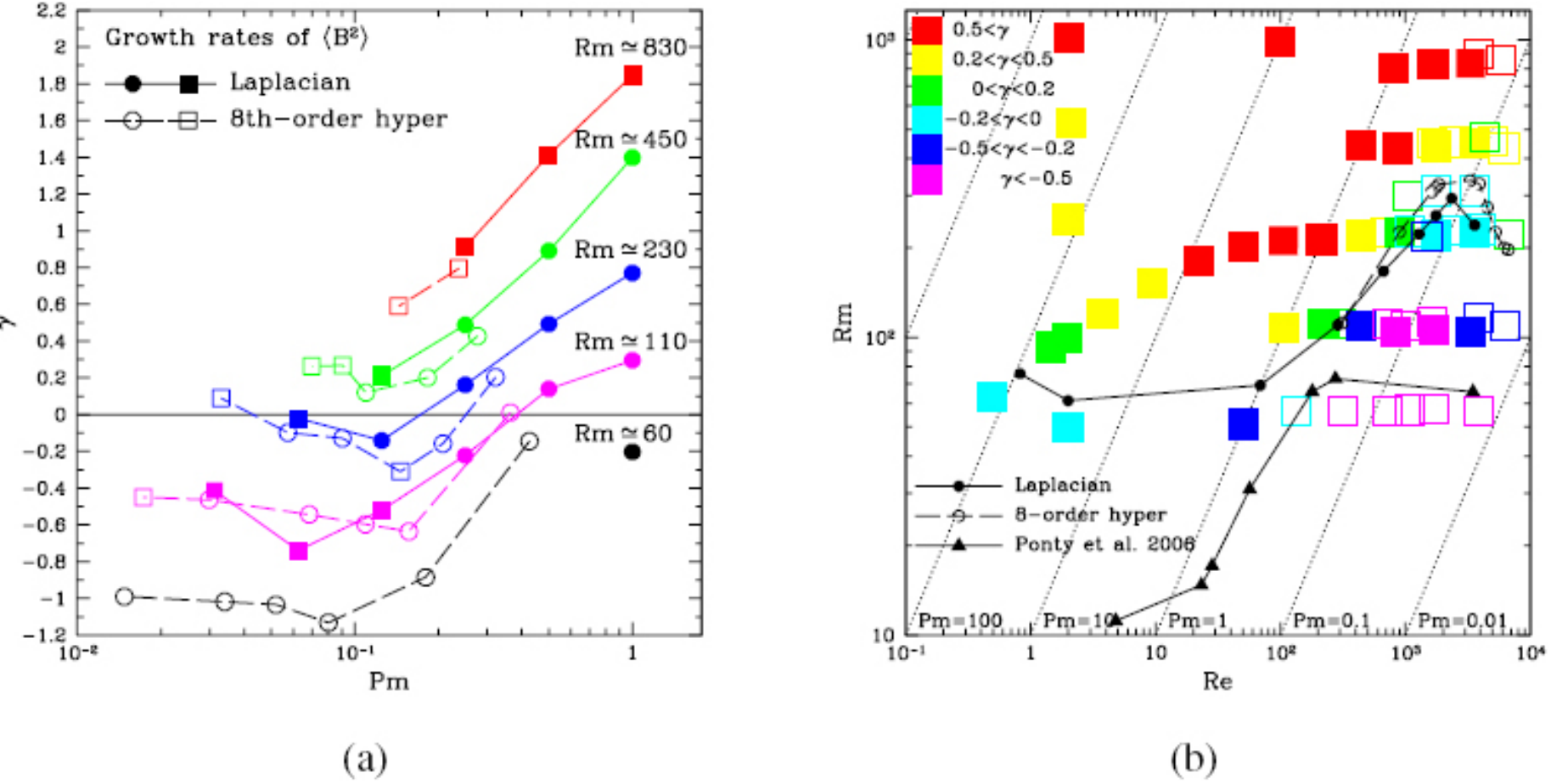}
\caption{Onset of dynamo action at moderate $Pm$, from \citet{scheketal:2007}. (a) Growth rate of 
magnetic energy as a function of Pm for five values of Rm. (b) Growth/decay rates in the parameter space (Re, Rm). Also 
shown are the interpolated stability curves $Rm^{crit}$ as a function of $Re$ based
on the Laplacian. Hyperviscous runs are shown separately. }
\label{crit_rm_low_pm}
\end{figure}

\subsubsection{\label{lowpmkin}Numerical Solutions of Random Dynamos --- the low $Pm$ problem}

Owing the the importance of understanding how dynamos onset in turbulent flows at low $Pm$ for experiments (see section~\ref{expts}) there has been much numerical effort in this direction. These numerical calculations are fraught with difficulty as extremely large calculations are required; \citet{tcb11} calculate that the size of the numerical calculation required to answer the question of the behaviour of the critical value of $Rm$ for dynamo action at  low $Pm$ is beyond the reach of current computational resources (requiring a calculation of size $10,000^3$ points).

In order to relate the Kazantsev models to numerical kinematic numerical simulations, where the Navier-Stokes equations are solved (with no Lorentz force) to provide the input to the induction equation, the Kazantsev models have been extended take into account departures from Gaussianity in the flow, with similar conclusions being drawn as for Gaussian flows. The results are summarised in Figure~\ref{crit_rm_low_pm}, which shows $Rm_c$ as a function of $Re$ for a collection of such calculations \citep{scheketal:2007}. 
At large and moderate $Pm$ these results are consistent with the predictions of the Kazantsev model, as shown in Figure~\ref{crit_rm_low_pm}. As noted above, calculations rapidly becomes prohibitive at small $Pm$, and so this  regime is not really accessible to direct numerical 
simulation (DNS), unless 
large eddy simulations (LES) are utilised 
\citep{pontyetal:2007}. However care must be taken here ---  in this regime the dynamo growth is completely controlled by the roughness exponent of the flow and so one would require an LES scheme with exceptional representation of the roughness. 

Turbulence, however, is  significantly more complicated than random flows that are $\delta$-correlated in time. In addition geophysical and astrophysical turbulence often has a substantial non-random 
component --- usually taking the form of coherent structures or vortex tubes, outside of the range of applicability of the Kazantsev model. For these flows it should be the case that characteristics other than the
spectral slope of the velocity (and hence the roughness exponent) control the dynamo growth. We discuss this case in the next section.


\subsubsection{\label{coherent}Flows with coherence}

\begin{figure}
\centering
\includegraphics[width=1.0\textwidth]{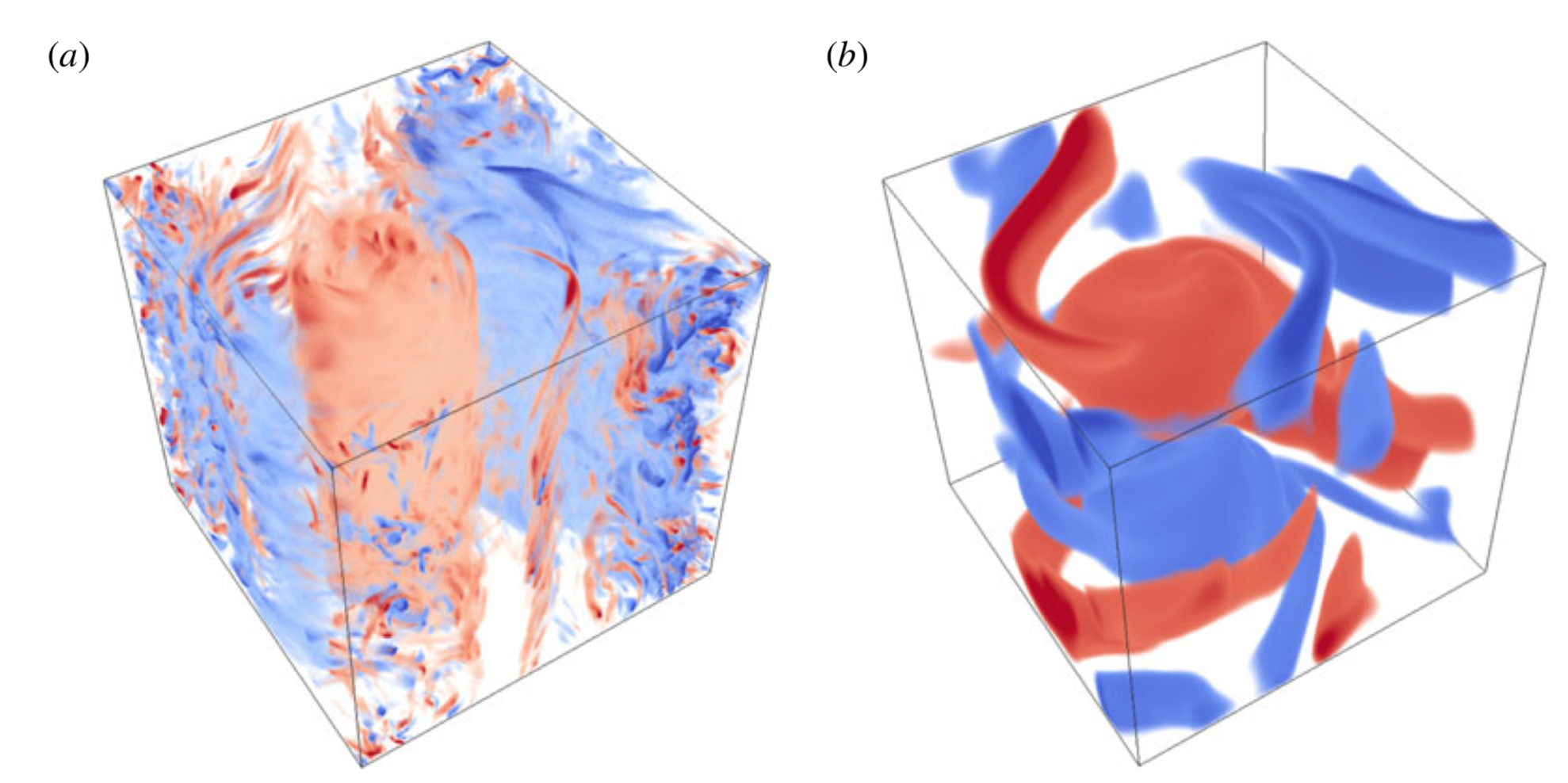}
\caption{(a) Colour plot of the vertical vorticity $\omega_z$ and (b) the vertical current $j_z$ in a rotating dynamo simulation at moderately low $Pm=0.0375$ \citep[from][]{sesh:2017}. Note the effect of rotation is found in the formation of long-lived coherent vortex tubes that make the dynamo more efficient.}
\label{sesha}
\end{figure}

In this section we consider the case more relevant to geophysical and astrophysical flows, where 
 the flow has two components, a random component as described above and a systematic 
component whose coherence time is long compared with its turnover time. Such coherent structures are ubiquitous in flows where rotation and stratification are important. The structures also exist on a wide range of spatial scales and so it is natural to ask what
determines the kinetic dynamo growth-rate in a multi-scale flow with coherent structures. 

The theory for such flows (which do not now have short correlation times) was developed by \citet{tobcatt:2008a,tobcatt:2008b}. This was achieved in two stages, the first step involved demonstrating that for such flows, knowledge of the form of the spectrum is not enough to determine the dynamo properties. In order to achieve this, they considered a flow with both long-lived vortices 
and a random component with a well-defined spectrum. They took advantage of the G.O.~Roberts trick for dynamos with $2.5$-dimensional velocities, by synthesising the flows from 
solutions to the active scalar equations. They  also generated a second (random) flow with the \textit{same spectrum} as the first
by randomising the phases of the spectral components. They found that the flow with coherent long-lived structures is a much more efficient dynamo than that which is purely random. Here, by efficient we mean that at the same $Rm$ the dynamo 
growth-rate is higher for the coherent flow. The systematic stretching from the long-lived coherent structures is pivotal in generating field efficiently. In particular helical vortices are able to generate magnetic field structures --- in this case the field generated takes the form of a helix reminiscent of that generated by a Ponomarenko dynamo.
 
It is now reasonable to assume that a multi-scale kinematic dynamo will be dominated by the coherent parts of the flow ---  the long-lived structures (rather than the temporally $\delta$-correlated random components) will determine the dynamo growth-rate and the form of the field. So, if a dynamo consists of a superposition of such coherent eddies with a range of spatial scales and turnover times (all shorter than their correlation time), what factors determine the small-scale dynamo growth-rate?

Progress can be made by assuming that each eddy acts as dynamo  largely independently of 
the eddies at very different scales. This assumption was validated in a model problem by \citet{ct2005}. A further step is 
to assume that 
each of the dynamo 
eddies are `quick dynamos'. As defined by \citet{tobcatt:2008a}, a `quick dynamo' is one that reaches a neighbourhood of its 
maximal growth-rate quickly as a function of $Rm$.
If this is the case then
the dynamo is driven by the coherent eddy 
which has the shortest turnover time $\tau$ and has $Rm \geq {\cal O}(1)$. Both the local $Rm$ and $\tau$ are functions of the spatial scale and so the location in the spectrum of the eddy 
responsible for dynamo action depends on the spectral slope as well as the magnetic Reynolds number on the integral scale.

The difference between dynamos that are dominated by random components and those that have long-lived coherent structures has recently been confirmed by
\citet{sesh:2017}. They considered the onset of dynamo action in a randomly forced flow subject to the effects of rotation. As the rotation is increased the flow develops more spatial and temporal coherence (as shown in Figure~\ref{sesha}), with the coherent vortices eventually winning out over the random element of the flow. This has the effect of reducing the critical $Rm$ at which dynamo action occurs, even at low $Pm$ --- the coherence engendered by the rotation turns a low $Pm$ dynamo into a high $Pm$ dynamo as predicted by \citet{tobcatt:2008a}.


\section{\label{sec:org}Organised magnetic field generation}
\medskip
As noted earlier, it is often the case that astrophysical  magnetic fields display some degree of organisation (or order)  either spatially (displaying order on spatial scales large compared with the typical eddies in the turbulent flows) or temporally (displaying temporal coherence on timescales much longer than typical timescales in the turbulence) or sometimes both (for example the spatiotemporal behaviour of the solar cycle).

It is therefore natural to consider theories that describe the evolution of the ``organised" components of the magnetic field (and potentially the velocity field). This could be for fields that are organised either spatially or temporally. In order to make progress it is necessary to give some mathematical precision as to the concept of  an organised field. This brings in the concept of averaging and forces the theorist to turn to statistical theories that are designed to provide information about these average quantities. Such theories are ubiquitous in fluid mechanics \citep[for example][]{frisch:1995,buhler:2014,kraichnan} though, owing to the historical path of research in dynamo theory, the statistical theories in the two disciplines have tended to develop along different tracks. I shall argue strongly that future progress in dynamo theory  can only be made by reconciling these approaches and returning to the paradigms preferred by fluid dynamicists. Nonetheless much can be learned from the dynamo approach, which I shall briefly review here.

\subsection{\label{averaging}The Nature of Averaging}

We shall be concerned with the derivation and solution of equations for the average properties of our state variables (for example {\bf B} and {\bf u}). We shall primarily be interested in decomposing state variables into mean (average) and fluctuating parts so that for example
\begin{equation} 
{\bf B} = {\overline {\bf B}} + {\bf b}'. 
\label{Rey:decomp}
\end{equation}
Here the overbar represents a linear averaging process that (in general) obeys the Reynolds averaging rules, i.e.,
\begin{equation}
\overline{{\bf B_1} + {\bf B_2}} = \overline{\bf B}_1 + \overline{\bf B}_2,
\end{equation}
and 
\begin{equation} 
{\overline{\overline{\bf B}}} = \overline{\bf B}.
\label{mean_mean}
\end{equation}
Hence averaging equation~(\ref{Rey:decomp}) gives
\begin{equation}
\overline{\bf b'} = 0.
\end{equation}
In terms of products it is also convenient, though not necessary,  if the averaging procedure satisfies
\begin{equation}
{\overline{\overline {\bf B}{\bf b'}}}=0,
\end{equation}
and
\begin{equation}
{\overline{{\overline{\bf B_1}}\,\,{\overline{\bf B_2}}}}={\overline{\bf B_1}}\,\,{\overline{\bf B_2}}.
\label{mean_prod}
\end{equation}
There are many different forms of averaging that satisfy the above Reynolds averaging rules, and some useful forms that do not. We briefly comment on a few here.

\subsubsection{\label{spatial}Spatial Averaging}
Perhaps the most utilised form of averaging assumes that the fluctuations are characterised by a length-scale $\ell_0$ (perhaps the scale of the energy containing eddies or fluctuating magnetic field). This is assumed small compared with the larger scale $L$ of the variation in the averaged quantities. For a system-scale dynamo $L$ will typically be ${\cal O}(R)$ where $R$ is the lengthscale of the domain. It is then possible to define an intermediate scale $a$ that satisfies $\ell_0 \ll a \ll L$. The spatial average can then be defined as \citep{Moffatt:1978}
\begin{equation}
{\overline {\bf B}} \equiv \langle {\bf B} \rangle_a \equiv \dfrac{3}{4 \pi a^3}\int_V {\bf B}({\bf x}+{\boldsymbol \xi},t)d^3 {\boldsymbol \xi}.
\end{equation}
\subsubsection{\label{temporal}Temporal Averaging}
A similar separation and averaging procedure is available if the quantity to be averaged varies on two time-scales. For example if the fluctuations have a characteristic short timescale $t_0$ (perhaps as defined by the correlation time) and the averaged quantities vary on a long timescale $T$ then one can average over an intermediate time-scale $\tau$ with $t_0 \ll \tau \ll T$ by defining
\begin{equation}
{\overline {\bf B}} \equiv \langle {\bf B} \rangle_\tau \equiv \dfrac{1}{2 \tau}\int_{-\tau}^{\tau} {\bf B}({\bf x},t+\tau')d \tau'.
\end{equation}
\subsubsection{\label{co-ordinate}Co-ordinate Averaging}
A fairly drastic, although perhaps also very natural, example of spatial averaging is to average over one spatial co-ordinate. For example the butterfly diagram of Figure~\ref{butterfly} is constructed by averaging the surface sunspot data over longitude and plotting the averaged field as a function of latitude and time. For this type of averaging there is a natural separation of spatial scales, with the axisymmetric component of field formally being separated in azimuthal wavenumber space from the non-zero wavenumbers. This type of `zonal' averaging is often utilised in planetary and stellar physics where zonal symmetry of the underlying turbulent statistics is a natural consequence \citep[see e.g.][]{bra:1975}. In local Cartesian numerical models, we shall see that it is sometimes natural to average over horizontal co-ordinates.
\subsubsection{\label{ensemble}Ensemble Averaging}
All of the averaging processes described above rely on some separation of scales (either spatial or temporal) between the average quantities and the fluctuations (this occurs naturally in the case of co-ordinate averaging). In most turbulent systems, this separation is difficult to achieve, and there is energy at all scales. Whether this really is a 
difficulty for the theory is open to debate \citep[see e.g.][]{bransub:2005}. A more natural averaging procedure, from a statistical viewpoint, is to take an average over realisations of the turbulence --- a so-called ensemble average. If the flow is ergodic then this method of averaging should (in combination with the underlying symmetries of the turbulence) yield the same answers as temporal or spatial averaging in the relevant asymptotic limits \citep{Moffatt:1978}.

\subsubsection{\label{filter}Averaging by Filtering and General Quasilinear (GQL) averaging}

A discussion of averaging procedures would not be complete without including filtering, either spectral or Gaussian \citep{germano_1992}. These filters are chosen to smooth small-scale spatial features and
are often representative of the action of observations at finite spatial resolution. It is worth noting that such procedures typically do not satisfy the Reynolds rules of averaging; for example Gaussian filtering does not satisfy equation~(\ref{mean_mean}) whilst spectral filtering typically does not satisfy equation~(\ref{mean_prod}). However, Gaussian filtering can be used in dynamo calculations \citep[see e.g.][]{Hollins:2017} whilst spectral filtering can be made to satisfy the Reynolds averaging rules by restricting the interactions to remove triad decimation in pairs \citep{kraichnan1985}. This filtering forms the basis of the Generalised Quasilinear Approximation introduced by \citet{mct:2016}.

\subsection{\label{kin:mfe}Kinematic Considerations --- Mean field Electrodynamics}
\medskip
\begin{center}
\textit{``Each success only buys an admission ticket to a more difficult problem."} \linebreak 
Henry Kissinger
\end{center}
\medskip

Armed with a suitable averaging procedure, it is possible to make significant progress in deriving equations for the evolution of averaged quantities. 
As before we shall proceed by solving for the evolution of the magnetic field via the averaged induction equation for a given velocity field. This formulation, which is now known as \textit{kinematic mean field electrodynamics} has  the potential for deep insight; however I shall argue that it is only via deriving the averaged equations for the full coupled Navier-Stokes/induction system that future progress will be made.

We proceed by splitting both the velocity and magnetic field into averaged and fluctuating parts so that 
\begin{equation} 
{\bf u} = {\overline {\bf U}} + {\bf u}', \quad {\bf B} = {\overline {\bf B}} + {\bf b}'. 
\label{UB_Rey:decomp}
\end{equation}
Now, taking the average of the induction equation yields
\begin{equation} 
\overline{\frac{\partial {\bf B}}{\partial t}} =
\overline{\nabla \times ({\bf u} \times {\bf B})} + \eta
\overline{\nabla^2 {\bf B}}, 
\end{equation}
which using the Reynolds averaging rules gives,
\begin{equation}
\frac{\partial \overline{\bf B}}{\partial t} =
\nabla \times \overline{({\bf u} \times {\bf B})}
+\eta \nabla^2 {\overline{\bf B}}.
\label{mean_ind}
\end{equation}
Here the non-trivial induction term is given by
$\displaystyle{\overline{({\bf u} \times {\bf B})}}$.
which can be expanded to give
\begin{eqnarray} 
\overline{{\bf u} \times {\bf B}}
&=&\overline{\overline{\bf U}\times \overline{\bf B} + \overline{\bf U}\times {\bf b'}
+ {\bf u'} \times \overline{\bf B}  + {\bf u'} \times {\bf b'}}, \\
&=&\overline{\bf U}\times \overline{\bf B} + \overline{{\bf u'} \times {\bf b'}}.
\end{eqnarray}
So there are two contributions to average induction; the first comes from induction of the average flow and average field (a so-called mean/mean interaction), whilst the second, 
\begin{equation}
\displaystyle{\bm{\mathcal{E}} = \overline{{\bf u'} \times {\bf b'}}},
\label{e_def}
\end{equation}arises from the \textit{average} interaction of the fluctuating velocity and magnetic field. This term is commonly known as the turbulent electromotive force (emf). 

Determination of this contribution to the evolution equation for the mean magnetic field forms a large part of the theoretical body of work in mean-field dynamo theory; in rather the same way that modelling the Navier-Stokes Reynolds stress tensor is crucial for many theories for mean flow/turbulence interactions in hydrodynamic turbulence.

Of course, in principle for a given $\overline{\bf U}$ and ${\bf u}'$ (remember we are in the kinematic approximation!), ${\bm{\mathcal{E}}}$ can be calculated exactly by solving the equation for the fluctuating magnetic field (which is derived by subtracting equation~(\ref{mean_ind}) from the full induction equation~(\ref{Induction})) to give
\begin{equation}
\dfrac{\partial {\bf b}'}{\partial t}=\nabla \times (\overline{{\bf U}} \times {\bf b}')+\nabla \times ({\bf u}' \times {\overline{\bf B}})+\nabla \times \bm{\mathcal G} + \eta \nabla^2 {\bf b}',
\label{fluc_eqn}
\end{equation}
where
\begin{equation}
\bm{\mathcal G} = {\bf u}' \times {\bf b}' - \overline{{\bf u}' \times {\bf b}'}.
\label{gdef}
\end{equation}
Notice that fluctuation equation~(\ref{fluc_eqn}) has four terms on the right hand side. The first two of these involve the interaction of means with fluctuations whilst the fourth is a linear term in the fluctuations. The third term given in (\ref{gdef}) involves the interaction of fluctuations with fluctuations (to yield fluctuations) and may prove to be problematic for theoretical progress. For this reason this has been termed the \textit{pain in the neck (PIN) term}. In the \ --- perhaps less expressive --- hydrodynamic literature fluctuations are often identified with eddies and the corresponding term in the fluctuating Navier-Stokes equation is often termed the eddy/eddy $\rightarrow$ eddy nonlinearity (or EENL for short.)

\subsection{\label{calc:emf}Calculation of the EMF}
At this point in the derivation no approximations have been made and there is little controversy. Indeed one could  solve equation~(\ref{mean_ind}) coupled with (\ref{fluc_eqn}) --- equivalent to solving the full induction equation\footnote{One could argue about whether it makes sense to separate the scales into averages and fluctuations, but if these are treated in the same manner then the point is moot as to whether this is a debate worth having.}.

However in order for the approach of mean-field electrodynamics to be useful, the solution of the full problem is to be avoided (and indeed is impossible with current computational resources). Progress can only be made by putting the averages and the fluctuations on a different theoretical footing --- and so treating them differently. Solving for the averages is not computationally difficult, and so this can be performed efficiently. However, in order to perform these calculations we need to be able to calculate the turbulent electromotive force, which arises from the average effect of the fluctuation/fluctuation interaction. In general it would be helpful to have a theory that relates the electromotive force to the average quantities, either via a differential equation or a simple prescription.

There are two possible ways to proceed. The first is to derive an evolution equation for the (generalised) electromotive force (which can be thought of as a low-order statistic of the flow). This approach has much in common with some techniques used in hydrodynamic turbulence theory.

The second approach is to milk the formal linearity of the kinematic induction equation for all it is worth, and see where it gets us. The answer is a surprising distance. As noted by \citet{Moffatt:1978} \textit{``\ldots there is now a satisfactory body of theory for the determination of $\bm{\mathcal E}$. The reason for this (comparative)  degree of success can be attributed to the linearity in ${\bf B}$ of the induction equation. There is no counterpart of this linearity in the dynamics of turbulence."}
It is important though to appreciate the limitations of this approach.

\subsubsection{\label{lin_ded} Deductions from linearity}

Clearly, calculation of the emf is possible from evaluation of ${\bf b}'$ (recalling that the velocity field is prescribed \textit{in the kinematic regime}), so the solution of equation~(\ref{fluc_eqn}) is desirable. However, as noted previously, the solution of this equation together with that of the mean equation is equivalent to solving the full problem. It is therefore useful to see what can be learned from the structure of equation~(\ref{fluc_eqn}). It is undoubtedly true that this equation is formally a linear equation for ${\bf b}'$, with a forcing term given by $\nabla \times ({\bf u}' \times {\overline{\bf B}})$. Solutions to this equation for ${\bf b}'$ are linearly (though not homogeneously) related to $\overline{\bf B}$; that is
\begin{equation}
{\bf b}' = {\bf b}'_{ss}+{\mathcal L}(\overline{\bf B}).
\end{equation}
Here ${\mathcal L}$ is a linear operator and ${\bf b}'_{ss}$ is that fluctuation field that exists in the absence of a mean field $\overline{\bf B}$. At high $Rm$ this ``small-scale dynamo field'' is inescapable as we have discussed at length above. However it is not clear that the presence of this field poses any difficulty for the calculation of the emf. Clearly ${\bf b}'_{ss}$ is, by definition, that field that exists in the absence of any large-scale field $\overline{\bf B}$. It seems unlikely therefore that, except in extremely contrived situations, it can contribute to an emf at high $Rm$; since its contribution to the emf would inevitably drive a large-scale field, which is supposed absent. In the nonlinear regime, this argument clearly does not hold (as the small-scale dynamo subspace may saturate and become unstable to large-scale perturbations as we shall discuss later). Indeed some mechanisms for large-scale field generation rely on interactions arising nonlinearly from a saturated small-scale dynamo \citep{sb2016}.

Hence it seems that, although the fluctuation dynamo field in the kinematic regime is not linearly and homogeneously related to $\overline{\bf B}$, the electromotive force is. This is not to say that the presence of small-scale dynamo action is irrelevant; indeed, as we shall see in section~\ref{getright}, in the kinematic regime at high $Rm$ the growth-rate of the dynamo (at both large and small scales) is completely controlled by the presence of a small-scale dynamo \citep{nigro:2017}.

The linear dependence of the electromotive force on the averaged field suggests an integral representation of the form
\begin{equation}
{\mathcal E}_i({\bf x},t) = 
\int \int K_{ij}({\bf x},t;\bm{\xi},\tau)
\overline{B}_j({\bf x}+\bm{\xi},t+\tau) 
\, d^3 \bm{\xi}\, d \tau,
\end{equation}
for some kernel $K_{ij}$. 

If a separation of spatial scales (for example) also pertains then it is possible to expand $\overline{\bf B}$ in a Taylor series, i.e.
\begin{equation}
\overline{B}_i ({\bf x+\bm{\xi}}) = \overline{B}_i ({\bf x}) + \xi_j \frac{\partial \overline{B}_i}{\partial x_j}({\bf x}) +
\frac{1}{2} \xi_j \xi_k \,\frac{\partial^2 \overline{B}_i }{\partial x_j \partial x_k}({\bf x}) + \cdots
\label{expansion}
\end{equation}
For scale separation, $|{\bm{\xi}}|$ is small compared with a typical scale 
of $\overline{\bf B}$, and higher order terms may be neglected. As noted by \citet{Moffatt:1978} terms in the Taylor expansion involving time derivatives of the mean field can be written in terms of the mean field and spatial derivatives through back-substitution into equation~(\ref{mean_ind}).

It is therefore natural, having made this approximation, to postulate a series expansion of the emf in terms of spatial gradients of the average field  of the form

\begin{equation}
\mathcal{E}_i = \alpha_{ij} \overline{B}_j + \beta_{ijk} \frac{\partial \overline{B}_j}{\partial x_k}+\cdots
\label{e_expansion}
\end{equation}
where the coefficients $\alpha_{ij}$ and $\beta_{ijk}$ are (pseudo-tensors) known as turbulent transport coefficients.

In general $\alpha_{ij}$ and $\beta_{ijk}$ need to be calculated from the turbulent flow statistics (usually this requires some approximations as we shall see). However some general statements about their form can be made immediately from equation~(\ref{e_expansion}).

We note that $\bm{\mathcal{E}}$ is a \textit{polar} vector, whereas $\overline{\bf B}$ is an \textit{axial} vector. From this we can immediately deduce that $\alpha_{ij}$ is a pseudo-tensor, which can be decomposed into symmetric and antisymmetric parts as
\begin{equation}
\alpha_{ij} = \alpha_{ij}^s - \epsilon_{ijk} \gamma_k.
\end{equation}
Hence the `$\alpha$-term' in equation~(\ref{e_expansion}) takes the form $\alpha_{ij}^s{}\overline{B}_j + (\bm{\gamma} \times \overline{{\bf B}})_i$. The anti-symmetric part of the $\alpha$-tensor therefore contributes an extra mean velocity to the mean field equations. The symmetric part of the $\alpha$-tensor obviously depends on the flow. In a system with preferred directions, given by say gravity ${\bf g}$ and rotation $\bm{\Omega}$, it is given by
\citep{krauraed:1980}
\begin{equation}
\alpha_{ij} = \alpha_1 \delta_{ij} \hat{\bf g} \cdot \hat{\bm{\Omega}} + \alpha_2 \hat{g}_i \hat{\Omega}_j + \alpha_3 \hat{g}_j \hat{\Omega}_i.
\label{general_alpha}
\end{equation}
This form is useful; however more enlightening is to consider the case where the turbulence is statistically isotropic (though still not reflectionally symmetric). In this case
\begin{equation}
\alpha_{ij} = \alpha \delta_{ij}
\label{iso_alpha}
\end{equation}
with $\bm{\gamma}=0$. The constant $\alpha$ is a \textit{pseudo-scalar}, which can only be non-zero if the turbulence lacks reflexional symmetry.

Similar considerations can lead to progress in understanding the second term in the expansion. For the simple case of isotropic turbulence, the $\beta_{ijk}$-tensor is also isotropic and so 
\begin{equation}
\beta_{ijk} = \beta \epsilon_{ijk}.
\label{iso_beta}
\end{equation}
Here $\beta$ is a pure scalar (i.e. it may be non-zero even for turbulence with no broken symmetry).

I stress here that, in general, these tensors (even if deemed to be useful) will not be homogeneous or isotropic. The presence of rotation, stratification and mean flows tend to give a preferred direction to the turbulence and the tensors will not take a particularly simple or enlightening form. There is now a significant body of work ascribing importance to the generation of an emf via particular parts of the tensors arising from different physical interactions. I shall argue later that what really matters are the low-order statistics of the MHD turbulence (in this case the electromotive force itself). Little progress can be made by ascribing and naming effects.

However it it still worth  following the theory through to its logical conclusion, it remains to determine the transport coefficients $\alpha_{ij}$ and $\beta_{ijk}$, ideally in terms of the prescribed flow. There have been many attempts at this that we shall summarise in section~\ref{trans_calc}. Before doing this, we shall briefly examine \textit{why} this theory has proven so popular by demonstrating solutions to the kinematic mean-field equations.

\subsubsection{Solution of the Mean-Field (filtered) equations \label{mf_sols}}

With the assumption that the transport coefficients are isotropic, so that $\alpha_{ij} = \alpha \delta_{ij}$ and $\beta_{ijk} = \beta \epsilon_{ijk}$ the mean field dynamo equations take the form
\begin{equation}
\frac{\partial \overline{\bf B}}{\partial t} =
\nabla \times (\overline{\bf u} \times \overline{\bf B})
+\nabla \times (\alpha \overline{\bf B}) 
- \nabla \times (\beta \nabla \times \overline{\bf B}) +\eta \nabla^2 {\overline{\bf B}}. 
\end{equation}
If $\beta$ is constant, $\nabla \times (\beta \nabla \times \overline{\bf B})
= - \beta \nabla^2 {\overline{\bf B}}$ so the 
equation becomes
\begin{equation}
\frac{\partial \overline{\bf B}}{\partial t} =
\nabla \times (\overline{\bf u} \times \overline{\bf B})
+\nabla \times (\alpha \overline{\bf B}) 
+(\eta+\beta) \nabla^2 {\overline{\bf B}}. 
\label{mf_eqns}
\end{equation}
We can now see the role of the transport coefficients clearly for this choice of isotropic turbulence. The $\nabla \times (\alpha \overline{\bf B})$ term is an inductive term leading to the generation of magnetic field (we shall spell this out in the next section) whilst for this simple configuration the $\beta$ term acts like
an enhanced diffusivity\footnote{Other components of the $\beta$ tensor can act so as to generate field for non isotropic flows. Given that there are 27 of these coefficients, it is fair to say that the role of them all has yet to be established. It is not even clear that it is desirable so to do, as argued above.}.

The adoption of the mean-field dynamo equations circumvents Cowling's Theorem! All the non-axisymmetric parts of the dynamo are encoded in the small scales, which are not solved for. Hence axisymmetric solutions for the \textit{mean} fields are allowed (and informative).
Combining equation~(\ref{mf_eqns}) with the axisymmetric formalism equations given in equations~(\ref{pol_eqn})-(\ref{tor_eqn}) one derives the axisymmetric mean field dynamo equations
\begin{eqnarray}
\frac{\partial A}{\partial t}
 + \frac{1}{s} ({\bf u}_P \cdot \nabla) (sA) &=& 
\alpha  B + \eta_T \left(\nabla^2 - \frac{1}{s^2}\right) A,\\
\frac{\partial B}{\partial t}
+ s ({\bf u}_P \cdot \nabla) \left(\frac{B}{s}\right) = 
\nabla \times \left(\alpha{\bf B}_P\right) \cdot \hat{\bm{\phi}} &+& \eta_T \left(\nabla^2 - \frac{1}{s^2}\right) B
+ s{{\bf B}_P}  \cdot \nabla \Omega,
\end{eqnarray}
where now it is understood that $A$ and $B$ represent the poloidal and toroidal parts of the mean magnetic field fields and $\eta_T = \eta + \beta$.

Cowling's antidynamo theorem therefore is defeated by the presence of a source term $\alpha B$ in the equation for the poloidal field. This source term, which is now called the $\alpha$-effect had been previously derived by \citet{parker:1955}. In the equation for the evolution of the azimuthal field both the shear (the $\omega$-effect) and the $\alpha$-effect can operate to generate this component of the field from the meridional field.
Depending on the relative importance of the $\alpha$ and $\omega$ effects, mean-field dynamos may be termed $\alpha^2$, $\alpha \omega$ or $\alpha^2 \omega$ dynamos.
There are two ways of generating azimuthal field $B$ from poloidal field ${\bf B}_P$: the
$\alpha$-effect or the $\omega$-effect. If the $\alpha$-effect dominates, 
the dynamo is called an
$\alpha^2$-dynamo. If the $\omega$-effect dominates its an $\alpha \omega$ dynamo.
We can also have $\alpha^2 \omega$ dynamos where both mechanisms operate equally.

The importance and utility of the mean-field ansatz can be seen by analysing the kinematic mean-field dynamo solutions in a two-dimensional local Cartesian geometry. We seek solutions independent of $y$, with no meridional flow and constant shear, $U'$,  in the $\alpha \omega$ limit (i.e. we set $\nabla \times \left(\alpha{\bf B}_P\right) \cdot \hat{\bm{\phi}} = 0$) , so that
\begin{equation}
{\bf B} = \left(- \dfrac{\partial A}{ \partial z}, B, \dfrac{\partial A}{ \partial x} \right) , \qquad
{\bf u} = \left(0, U' z, 0\right),
\end{equation}
\begin{equation}
\frac{\partial A}{\partial t}
  = 
\alpha  B + \eta_T \nabla^2  A,
\end{equation}
\begin{equation}
\frac{\partial B}{\partial t}
= 
J(A,U'z)  
+ \eta_T \nabla^2  B,
\end{equation}
where $J(f,g) = f_x g_z - g_x f_z$.
We seek wavelike solutions for $A$ and $B$ 
proportional $\exp (\sigma t + i {\bf k} \cdot {\bf x})$, with ${\bf k}= (k_x,0,k_z)$ to yield a dispersion relation of the form. 
\begin{equation}
(\sigma + \eta_T k^2)^2 = i k_x \alpha U', 
\end{equation}
with $k^2 = k_x^2+k_z^2$, and so
\begin{equation}
\sigma = \frac{1+i}{\sqrt{2}} (\alpha U' k_x)^{1/2} - \eta_T k^2
\end{equation}
In a finite Cartesian domain of size $d$ in the $z$-direction, one sets $k_z= \pi / d$ and large-scale dynamo action will occur if the dimensionless dynamo number $D = \alpha U' d^3/ \eta_T^2$ exceeds a threshold. The bifurcation to large-scale dynamo action occurs in a Hopf bifurcation to travelling waves. If $D > 0$ the waves travel in the negative $x$-direction, whilst if $D < 0$ they travel in the positive $x$-direction.

At this point it is immediately clear why mean-field theory has proved so popular. For an axisymmetric (indeed one-dimensional) model the theory can yield travelling wave solutions reminiscent of the waves of activity seen in the solar observations. The theory can easily be extended to bounded cartesian domains --- though care must be taken in carrying out the instability calculations owing to the difference between convective and absolute instabilities \citep{bassom:1997,tpk:1997}, and to spherical domains \citep{PHRoberts:1972}.

\subsubsection{Calculations of Kinematic Transport Coefficients: Theory and Computation \label{trans_calc}}

We described above how solution of the mean field equations  can lead to large-scale magnetic fields for simple choices of the transport coefficients. We also described in section~\ref{lin_ded} that basic properties of the transport coefficients ($\alpha_{ij}$ and $\beta_{ijk}$) can be determined via consideration of symmetries (and on making the assumptions of homogeneity).

However, it is important to calculate these transport coefficients for solvable and realistic models of turbulence (both in the kinematic and dynamic regime). We begin by describing the kinematic evaluation of these coefficients --- this is where the theory gets controversial, as all the current calculations of transport coefficients necessarily  involve some form of approximation.

\medskip
\noindent{\bf First Order Smoothing:} 
\medskip

The simplest ansatz in which to calculate the transport coefficients in the kinematic regime invokes the quasilinear approximation. This involves the discard of the fluctuation/fluctuation $\rightarrow$ fluctuation interactions in the equation for the fluctuating field; that is we set $\bm{\mathcal G}=0$ in equation~(\ref{fluc_eqn}) to yield
\begin{equation}
\dfrac{\partial {\bf b}'}{\partial t}=\nabla \times (\overline{{\bf U}} \times {\bf b}')+\nabla \times ({\bf u}' \times {\overline{\bf B}})+\eta \nabla^2 {\bf b}',
\label{fluc_eqn_with_time}
\end{equation}
The discard of the nonlinear term $\bm{\mathcal G}$ can be understood as assuming that the Kubo number of the MHD turbulence is small (in the sense that the fluctuation/fluctuation interaction is small compared with the mean/fluctuation interaction). This may occur in many circumstances of interest, though the conditions for which the approximation is valid is difficult to determine \textit{a priori}. Formally, however,  this is certainly the case when the fluctuating field is small compared with the mean field i.e. $|{\bf b}'| \ll |{\overline{\bf B}}|$. It is also applicable when $Rm \ll 1$ in which case one may also neglect the  $\partial {\bf b}'/\partial t$ term to leave
\begin{equation}
\eta \nabla^2 {\bf b}'=-\nabla \times (\overline{{\bf U}} \times {\bf b}')-\nabla \times ({\bf u}' \times {\overline{\bf B}}),
\label{fluc_eqn_fos}
\end{equation}
which can then easily be solved using standard linear transform techniques \citep[see][for a detailed description]{Moffatt:1978} to yield ${\bf b}'$ and hence (as ${\bf u}'$ is known) $\bm{\mathcal E}$.

The nonlinear term $\bm{\mathcal G}$ can also be formally neglected \textit{a priori} if the correlation time $\tau_c$ of the turbulence (which is given in the kinematic regime) is sufficiently short. In particular if the Kubo number $u_{rms} \tau_c/\ell \ll 1$ then equation~(\ref{fluc_eqn_with_time}) is justified. If in addition $Rm \gg 1$ then
\begin{equation}
\dfrac{\partial {\bf b}'}{\partial t}\approx\nabla \times (\overline{{\bf U}} \times {\bf b}')+\nabla \times ({\bf u}' \times {\overline{\bf B}}).
\label{fluc_eqn_with_time_high_Rm}
\end{equation}
Again this linear system may be solved by transform methods. In the simplified case with no local mean flow one may write
\begin{equation}
\overline{\bm{\mathcal{E}}}(t) = \overline{{\bf u}(t) \times \int_0^t \bnabla \times \left( {\bf u}(t') \times \overline{{\bf B}(t')} \right) dt'}.
\end{equation}
In a statistically steady state and for isotropic turbulence this simplifies to give
\begin{equation}
\overline{\bm{\mathcal{E}}}(t) =  \int_0^t \left(\hat{\alpha}(t-t') \overline{{\bf B}(t')} - \hat{\eta}_T(t-t') \overline{\bnabla \times {\bf B}(t')}\right) dt'.
\end{equation}
where $\hat{\alpha}(t-t') = -\frac{1}{3}\overline{\bm{u}'(t) \cdot \bm{\omega}'(t')}$ and $\hat{\eta}_T(t-t')= \frac{1}{3}\overline{\bm{u}'(t) \cdot \bm{u}'(t')}$.

For a mean field that is sufficiently slowly varying compared with the turbulence the integral kernels can be approximated by $\delta$-functions and so 
\begin{equation}
\overline{\bm{\mathcal{E}}}(t) =   \alpha \overline{{\bf B}} - \hat{\eta}_T \overline{\bnabla \times {\bf B}},
\end{equation}
with
\begin{eqnarray}
&\alpha& = - \dfrac{1}{3} \int_0^t \overline{\bm{u}'(t) \cdot \bm{\omega}'(t')} dt' \approx - \dfrac{1}{3} \tau_c \,\overline{\bm{u}' \cdot \bm{\omega}'}\\
&\eta_T& = \dfrac{1}{3} \int_0^t \overline{\bm{u}'(t) \cdot \bm{u}'(t')} dt' \approx - \dfrac{1}{3} \tau_c \,\overline{\bm{u}' \cdot \bm{u}'}
\label{transport_short}
\end{eqnarray}

First order smoothing, which is applied within the kinematic framework, is designed so that the electromotive force may be related to the mean magnetic field through properties of the prescribed velocity field. The smoothing is such that the fluctuating magnetic field is ``slaved" to the velocity field and the mean magnetic field and so the emf can be readily calculated. At high $Rm$, for correlation times that are not short, this slaving is not maintained and the magnetic fluctuations rapidly become decorrelated with the velocity field. This has two major effects. The first is that the emf can not be evaluated explicitly and one must resort to numerical procedures for its calculation (as described in the next section). The second, and more severe, effect is that the loss of correlation can lead to a diminution of the emf relative to the fluctuations. This may have serious consequences for large-scale dynamo theory in the kinematic regime at high $Rm$ as described below\footnote{though as we shall see in section there will be some effects that can mitigate the loss of correlation and to some extent rescue the theory}. 

\medskip
\noindent{\bf Numerical evaluation of the transport coefficients:}
\medskip

With the growth of computing power, it is now possible to evaluate the transport coefficients numerically, albeit at moderate $Rm$ for three-dimensional flows. This is of course possible both when the magnetic field is weak and does not act back on the flow and therefore the transport coefficients are in the kinematic regime, or in the fully nonlinear regime so that the effects of the Lorentz force on transport coefficients can be calculated --- an important topic discussed in section~\ref{systematic:sat}.

Constructing a reliable method for the calculation of the transport coefficients is not trivial (especially when $Rm$ is not small) since it involves the calculation of  averages and care must be taken to ensure that these averages are converged.
Calculation of the $\alpha$-coefficient is conceptually straightforward. Consider equations~(\ref{e_def}) and (\ref{e_expansion}). These are two separate expressions for the electromotive force. In the case where the averages are spatial, the flow and magnetic field can be decomposed into a spatial mean (in this case a uniform field) and spatial fluctuations. Now a uniform field has no spatial derivatives and so (\ref{e_expansion}) simplifies to
\begin{equation}
\mathcal{E}_i = \alpha_{ij} \overline{B}_j. 
\label{e_expansion2}
\end{equation}
Hence it is now possible to impose constant mean fields with different orientations and calculate $\mathcal{E}_i = \overline{\bm{u'} \times \bm{b'}}$ numerically to identify the different components of the $\alpha$-tensor. For a very turbulent flow at high(ish) $Rm$ it turns out that $ \overline{\bm{u'} \times \bm{b'}}(t)$ in the kinematic regime may be a rapidly varying timeseries which takes a broad distribution of values with a large variance and a relatively small mean \citep[see e.g.][who measured the alpha coefficient in a plane layer of rotating Boussineq convection]{catthughes:06}. This is an indication that the systematic behaviour in the kinematic regime may be in competition with (and perhaps dominated by) the disorganised or fluctuating dynamo behaviour (see section~\ref{competition} for a more complete discussion of this); however it is fair to say that the relative effectiveness of the fluctuating and organised magnetic fields may depend on both the form of the spectrum of the velocity and the value of $Rm$ as demonstrated by \citet{ct:2014} --- at very high $Rm$ the dynamo is controlled by smaller eddies from further down the spectrum with shorter correlation times, which means that averaging is a more effective procedure and the distribution of the emf is narrowed. Thus, in addition to spatial averaging, temporal (or ensemble) averaging should also be utilised to ensure well converged statistics.

Measuring the turbulent diffusivity $\beta_{ijk}$ is somewhat more problematic since that relates the emf to gradients of the organised magnetic field.
This causes problems as there is no simple method of imposing a  magnetic field with a spatial gradient that is not prone to grow exponentially via dynamo action at high $Rm$ in the kinematic regime. Three methods for evaluating the turbulent diffusivity have been suggested; the ``test field'' method, the ``turbulent  \AA ngstr\"om'' method and the ``Lagrangian Statistics'' method.

The test field method, introduced by \citet{schrinner2005} involves the solution of an auxiliary equation and is a simple extension of the procedure outlined above. Here the aim is to calculate the emf for a variety of applied test magnetic fields with different orientations and field gradients. If the test-fields are chosen to be constants then the $\alpha$-tensor can be recovered in a similar manner to that described above. In principle the $\beta$-coefficients can be backed out by imposing a series of spatially varying test fields and solving for the fluctuations, and hence the emf. The test field method has been criticised \citep{ch:2009} in that it requires a prohibitively large computational cost to evaluate the transport coefficients at high $Rm$. Nonetheless, at the moderate $Rm$ values that are currently achievable for current numerical computations convergence of results does appear achievable if ensemble averages are used in addition to spatial averages.

The turbulent  \AA ngstr\"om method \citep{tc2013} is based on that method used to measure experimentally the thermal conductivity of solids. The method consists of applying an oscillatory source of magnetic field (usually via an oscillating current) with a given frequency larger than the largest natural frequency of the turbulence. Here the turbulent diffusivity can be calculated from the response measured at the frequency of the imposed oscillating field. This method is attractive as the turbulent diffusivity can be calculated to any required precision, simply by analysing a long enough  timeseries for the magnetic field, though this may be computationally expensive.

The final method, which involves calculating Lagrangian statistics, is formally valid in the infinite $Rm$ limit and is outlined in \citet{Moffatt:1978}. Briefly both the $\alpha$ and $\beta$ tensors may be evaluated numerically by calculating averages over Lagrangian trajectories. This method can (and has) been extended to the finite $Rm$ case by adding stochastic fluctuations to the Lagrangian trajectories \citep{dh:1986}. This elegant method appears to merit further investigation.

\subsubsection{Organised Kazantsev Dynamos}
\label{sec:orgkaz}

We have described above how organised magnetic fields may be expected to emerge for flows that lack reflectional symmetry, and how, under certain assumptions, the evolution of the large-scale field may be determined in the kinematic regime by turbulent transport coefficients that encode the eddy/eddy $\rightarrow$ mean interactions. Even earlier, in section~\ref{kazantsev}, we also discussed how the Kazantsev model for random (short correlation time) flows could yield insight into the onset of small-scale dynamos. 

It turns out that the Kazantsev model can be extended to random flows that lack reflectional symmetry; in this case the velocity and magnetic  correlators are defined using two functions --- one, as before, related to the energy 
density (either kinetic or magnetic), the other  to the  helicity (either kinetic or magnetic) 
\citep{vainkit:1986}. The two parts of the magnetic correlator are then described by a pair of coupled Schr\"odinger-like 
equations \citep{vainkit:1986,bergerros:1995}. The 
analysis is now more complicated, yet it is possible to demonstrate that the system remains self-adjoint \citep{bolcattros:2005}. These authors demonstrated that, for sufficient kinetic helicity, extended states are found that  do not decay exponentially at infinity. These are the `large-scale dynamo solutions', analogous to the mean-field solutions described above. It can however be shown that at 
large $Rm$ the largest growth-rates remain associated with the localised bound states, so that the overall dynamo 
growth-rate remains controlled by small-scale dynamo action 
\citep{bolcattros:2005,malbol:2007,malbol:2008a,malbol:2008b}. This competition between large and small-scale dynamos in the kinematic regime will be a feature of our discussion in the next few sections.

\subsection{\label{competition}The competition between kinematic large and small-scale dynamos}

The discussion of organised \textit{random} dynamos in the last section described how the growth-rate of magnetic fields is determined by the small-scale dynamo, even if there is a large-scale component. In this simple case, the growth-rate for bound (small-scale) states can be shown to be larger than those for extended (large-scale) solutions.

In a non-random flow, the magnetic field must emerge as an eigenfunction to the kinematic dynamo problem with a well-defined (average) growth-rate $\sigma$; here large and small scales grow at this same rate. It is then appropriate to ask which processes control the growth-rate of a dynamo containing large and small scales? Moreover, if this is controlled by the small-scale dynamo action, then what is the role of large-scale (or mean-field) dynamo theory in determining the behaviour of the large scales.

The first of these questions has been addressed over a number of years by a variety of authors. The theoretical considerations of section~\ref{sec:small} show that in a flow with coherent structures over  spectrum of scales, kinematically the dynamo growth is dominated by dynamo action driven by the dynamo eddy which is just supercritical and has the fastest turnover time. As $Rm$ is increased, this dynamo scale moves further and further down the spectrum generating field at ever faster rates. Moreover the field that is generated is dominated by energy at scales smaller than the characteristic scale of the eddy. These considerations are all borne out by numerical calculations of kinematic dynamos at high $Rm$ \citep[see e.g.][]{tobcatt:2008a,tc:2013}.
The dynamo growth-rate \textit{is} therefore set by the small-scale dynamo; fundamentally, chaotic stretching on the advective timescale has a tendency to win out over an emf induced by correlations in the kinematic regime. In this kinematic regime there is nothing really to stop the generation of exceptionally small scales  
except for the action of diffusion on the resistive scales. Of course as all scales grow at the same rate this tendency for field to be generated on small scales manifests itself in the form of the eigenfunction, which is dominated by fields close to the resistive scale (all other things being equal).

The answer to the second question above, that of  the utility of kinematic mean field theory in describing the dynamics of the unfiltered equations, we shall defer to later (section~\ref{getright}). We shall first discuss how dynamo theorists have approached the problem of the competition between large and small-scale dynamos in the kinematic regime, and whether there are any strategies for promoting the generation of organised field over the small-scale, random fields in the kinematic regime.

\subsubsection{The role of shear: suppressing the small-scale dynamo \label{tc-dynamo}}

Most of the strategies to promote the large-scale dynamo over the small-scale dynamo involve the addition of a large-scale shear flow. As we have described above, in the absence of shear, the kinematic growth-rate is determined by the stretching of the small-scale dynamo. This effect, as discussed earlier, also manifests itself in the distribution of the emf, which in certain circumstances can be calculated to have a small mean and a large variance caused by the fluctuations. If one is to promote the generation of organised magnetic fields over that of small-scale fields then two strategies immediately suggest themselves; either utilise the shear to boost the mean electromotive force by introducing new effects (i.e. new terms in the transport coefficients) \textit{or} utilise the shear to suppress the small-scale dynamo.

Much attention has focussed on the first of these strategies \citep[see e.g.][among others]{Yetal:2008,kb:2009,srisin:2010}. Indeed it can be shown that the presence of a prescribed large-scale shear flow can lead to the presence of extra terms in the transport coefficients and hence the electromotive force. Moreover a shear flow can in principle interact with a weak and highly fluctuating emf to produce large-scale field \citep{proctor:2007,rp:2010}. All of these mechanisms can be shown to work, in the sense that they boost the generation of a mean emf, in certain circumstances. However, at high $Rm$, if this large-scale field is ever to be seen in the kinematic regime, there must be some mechanism present to suppress the small scales which have a tendency to dominate. It would also help to differentiate the two dynamos if the large scale dynamo were to exhibit different dynamics than that of the small scales.

\begin{figure}
\begin{center}
\centerline{(a)} 
\break
\quad \quad \includegraphics[height=4cm,width=8cm]{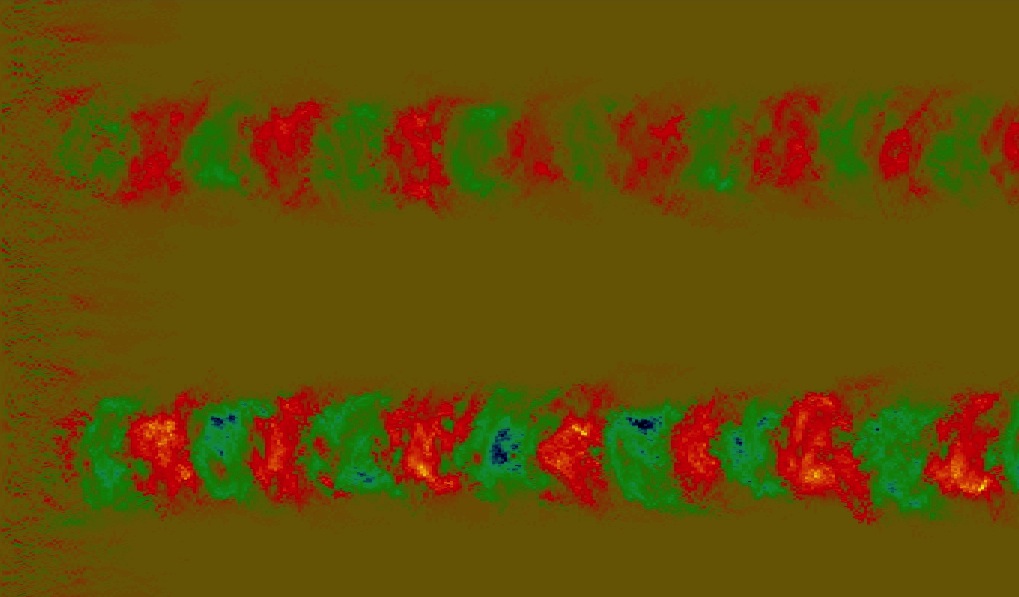}
\break
\centerline{(b)}
\includegraphics[height=5cm]{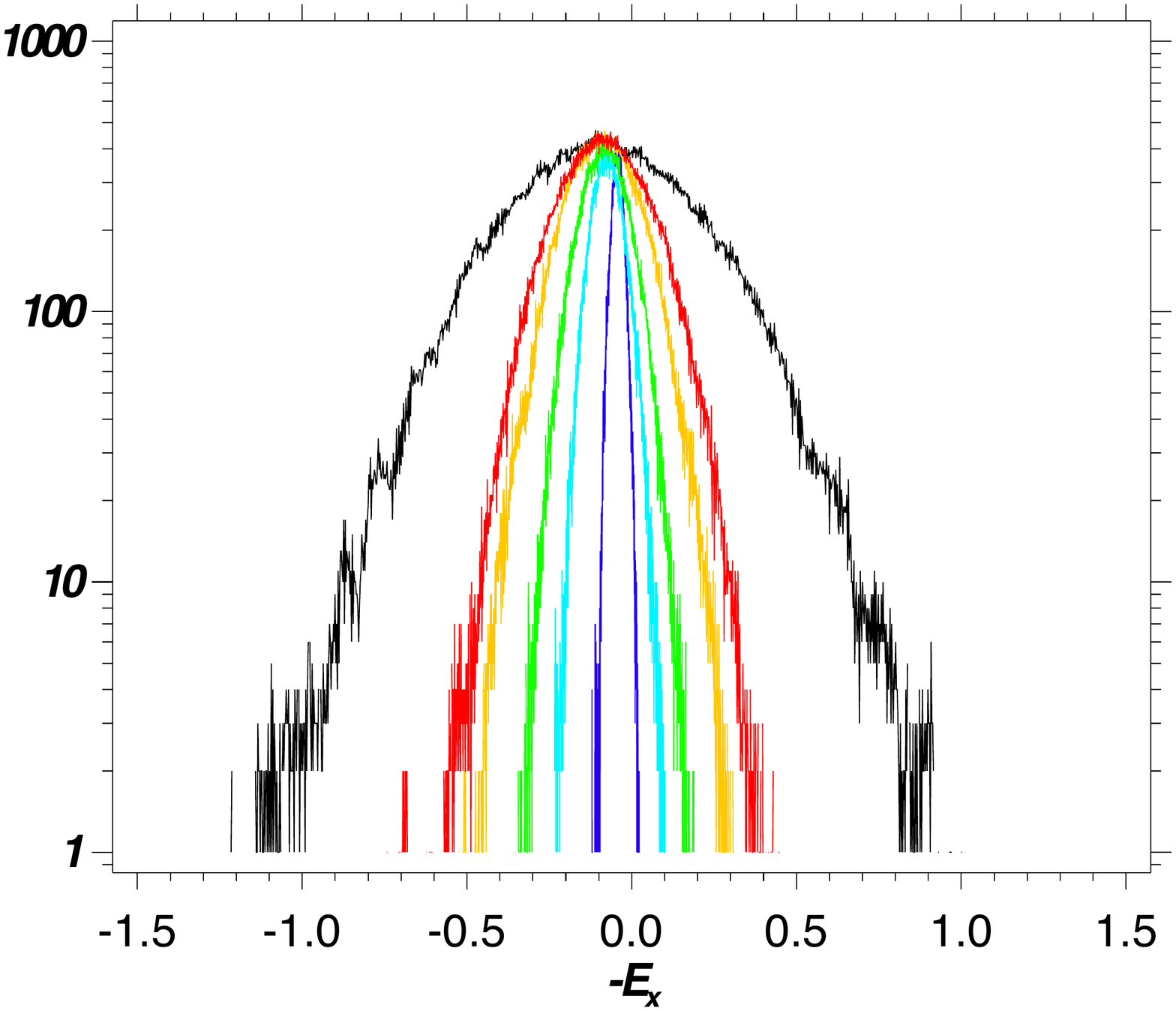}

\end{center}
\caption{(a) Space time $(t,y)$ plot of average magnetic field $\overline{B_x}(y,t)$ at fixed $z=0$ for dynamo with shear and small-scale flow with net helicity. \citep[after][]{ct:2014}. (b) Probability density function of electromotive force for a fixed small-scale helical flow and a range of strength of shear flow (defined in equation~(\ref{shear_define})) $V_0 = 0$ (black), $V_0 = 1$ (red), $V_0 = 2$ (yellow), $V_0 = 5$ (green), $V_0 = 10$ (cyan), and $V_0 = 20$ (blue). As the shear is increased the distribution narrows significantly with a small change in the mean emf \citep[after][]{ct:2014} \textcopyright AAS. Reproduced with permission.
}
\label{fig_ls}
\end{figure}

A rather different, and potentially more promising, way to proceed is to recognise that, as indicated by fast dynamo theory, the small-scale dynamo relies on the presence of chaotic stretching to survive at high $Rm$. In order to give the large-scale dynamo a chance \citep{ck:2009}, it is imperative to \textit{suppress the small-scale dynamo} by reducing the chaotic stretching in the flow. Such a strategy was proposed by \citet{tc:2013}. They argued that the addition of a shear flow to a chaotic (or turbulent) flow will increase the regions of integrability in the flow, thus reducing chaos and hence the effectiveness of the small-scale dynamo at high $Rm$. If such a dynamo also breaks reflectional symmetry and is capable of generating an electromotive force and hence large scale fields then it is possible that the suppression of the small scale dynamo will lead to the emergence of large-scale fields.

\citet{tc:2013} and \citet{ct:2014} investigated the properties of just such a dynamo. They considered a velocity field that was comprised of a combination of generalised Galloway-Proctor type flows imposed at a range of scales together with a shear flow. In particular they considered 
 Cartesian co-ordinates $(x,y,z)$ on a $2 \pi$-periodic domain, with a flow of the form
\begin{equation}
{\bf u}_k=A_k \left( \partial_y \psi_k, -\partial_x \psi_k, k \psi_k \right),
\end{equation}
where
\begin{equation}
\psi_k(x,y,t) = (\sin k ((x-\xi_k)+\cos \omega_k t)
+\cos k ((y-\eta_k)+ \sin \omega_k t)).
\end{equation}
This generalisation of the cellular GP-flow is also maximally helical and  takes the form of an infinite array of clockwise and anti-clockwise rotating helices
such that the origin of the pattern itself rotates in a circle with frequency $\omega_k$; $A_k$ is an amplitude and $\xi_k$ and $\eta_k$ are 
offsets that can be varied so as to decorrelate the pattern, with a decorrelation time $\tau_d$. 
In order to mimic the action of a turbulent spectrum, 
they considered a superposition of these flows on scales between $k_{min}$ and $k_{max}$, i.e.
\begin{equation}
{\bf u}_c= \sum_{k_{min}}^{k_{max}} {\bf u}_k,
\end{equation}
and selected $A(k)=k^{-\beta}$ (with $\beta=4/3$) so that $\omega_k=k^{2-\beta}$
and the decorrelation time is given by $\tau_d = \tau_0 k^{\beta-2}$. Hence  the decorrelation time scales in the same manner as the turnover time. With
these scalings  the turnover time $\tau_k$ and the
magnetic Reynolds number $Rm_k$ are themselves functions of $k$ 
given by \citep{ct2005}
\begin{eqnarray}
\tau_k &\sim& k^{\beta-2}, \\
R_m(k) &\sim& k^{-\beta}.
\label{more-scaling}
\end{eqnarray}
Thus for positive values of $\beta$, $R_m$ decreases with $k$, and for
$\beta <2$ so does the turnover time.  

To this time-dependent helical multi-scale flow  a steady unidirectional large-scale shear of the form
\begin{equation}
{\bf u}_s = \left(V_0 \sin y, 0, 0\right)
\label{shear_define}
\end{equation}
was added and dynamo solutions were sought (in a similar manner to the calculations for the Galloway-Proctor flow of section~\ref{fast_dynamo}) at high $Rm$.

In the absence of shear, the flow is an excellent small-scale dynamo.
For $\chi = Rm/Rm_c \gg 1$ all the magnetic energy is, as expected, concentrated at small scales  and generated on a typical timescale of a Galloway-Proctor eddy. There is no sign of a large scale organised component of the magnetic field. 
The situation changes significantly when shear is added. Although the small-scale dynamo is still operational its growth-rate is reduced \citep[see][]{tc:2013}. \textit{At high $Rm$ in the kinematic regime the role of an imposed shear is always to lower the growth-rate of a small-scale dynamo.} As explained above, this is due to the introduction of regions of integrability to the otherwise chaotic flow. This statement has caused some controversy in the literature, but is correct at high enough $Rm$ since the operation of such dynamos relies on the presence of chaos in the flow. Dynamos for which the shear increases the dynamo growth-rate are either poor small-scale dynamos to start with, or are simply not being investigated at high enough $\chi = Rm/Rm_c$.

Moreover, now that the small-scale dynamo is suppressed, it is reasonable to expect that the large-scale dynamo can be seen. Figure~\ref{fig_ls}(a), which shows $\overline{B_x}(y,t)$ the average (in $x$) of the toroidal (shear-aligned) field at a given $z$-level, demonstrates that this expectation is met. The average field can be seen to take the form of systematic dynamo waves propagating in the regions of strong shear (of the type envisaged by Parker). In the absence of helicity in the small-scale flow the shear still amplifies fields in bands but there is no systematic generation of spatio-temporally coherent field. That the shear reduces the fluctuations is clear on examination of the pdf for the electromotive force in  Figure~\ref{fig_ls}(b). This shows the distribution of emf as measured by imposing a mean field. In the absence of shear the distribution is dominated by the fluctuations with a large variance and a small mean (as shown by the black curve). As the strength of the shear is increased the variance of the distribution decreases markedly, though the mean shifts barely at all. This demonstrates (for this flow at least) that the crucial role of shear at high $Rm$ in the kinematic regime is to suppress the fluctuations, rather than modify the mean.

These results led \citet{ct:2014} to propose a general ``Suppression Principle'', namely 
\textit{``At high $Rm$ large-scale dynamo action can only be observed if there is a mechanism that suppresses the small-scale fluctuations."
}
As discussed earlier, the kinematic framework dynamo theory is linear and the solutions are superposable --- large-scale dynamo action can therefore only be observed if the small-scale dynamo is suppressed. In the cases discussed above the shear acted so as to suppress the fluctuations. It is possible however to  envisage other mechanisms that may act in this way. 
In most circumstances however dynamo action proceeds in the nonlinear regime (see section~\ref{sec:saturation}). There one can envisage a nonlinear mechanism that leads to either the suppression of the the small-scale dynamo or the saturation of the fluctuating field  
at a reasonable amplitude. In the saturated regime, whether the large-scale field can be observed depends on the ratio of the saturation amplitude for the large and small scales, which is itself a contentious issue in dynamo theory, as we shall see.

\subsubsection{\label{getright}What does mean-field theory get right in the kinematic regime?}

Mean-field theory describes the evolution of filtered or averaged quantities. In order for this theory to have utility in the kinematic regime, one would  expect that application of the filter to the eigenfunctions of  the full equations  should correspond to the eigenfunctions of the filtered equations. Moreover, the growth-rate of the large-scale structure of the solutions should coincide with the growth rate predicted by the filtered equations. As we have argued above, this second expectation is never met at high $Rm$ --- the growth-rate is determined by the stretching of the small scales \citep{tobcatt:2008a}. 

As noted above, it is possible to construct more complicated cases, for example with the addition of shear, where the filtered (mean-field) equations have a growth rate and a frequency \citep{parker:1955,tc:2013}. The shear breaks the isotropy of the filtered equations and the  isotropy of the statistics of the velocity in the unfiltered system and therefore one expects a breaking of symmetry of the statistics of the solutions. In that case one might expect the frequency, as manifested by the speed of propagation of the dynamo waves to be the same  both in the solutions of the filtered equations and those of the unfiltered equations in a statistical sense. Here symmetry breaking acts to perturb  the frequency  away from zero and \textit{is} controlled by a change in the symmetry of the large scales. 

\citet{nigro:2017} considered just such a case based on the Tobias-Cattaneo dynamo of section~\ref{tc-dynamo}. They determined a range of solutions with   a travelling helical (dynamo) wave, that remains coherent for long periods of time and whose frequency {\it is} determined by mean-field effects. The wave could only be identified from the rest of the structure by a persistent phase-coherent signal with the rest of the solution being incoherent in time. They argued from this that it is better to consider  a definition of large-scale dynamo action that considers the timescale of evolution of the pattern, rather than one that relies on  spatial scales alone, i.e.\ spatio-temporal averaging is the natural choice for determining organisation. Further, as  
kinematic large-scale dynamo action consists of a long-lived coherent pattern embedded in a sea of incoherent fluctuations; the filtered equations simply average over the fluctuations.

This gives hope for the construction of a sensible framework for deriving a filtered equation in the nonlinear regime. As argued by \citet{nigro:2017} as nonlinearity becomes important different scales may continue to grow at different rates and saturate at relative amplitudes completely different from those of the kinematic phase, controlling the fluctuations relative to the mean. If this were the case then they argue that the filtering technique used to identify spatio-temporally coherent large-scale fields in the kinematic regime may continue to  be utilised in the nonlinear regime.
It therefore behooves us to finally move into the dynamic regime, and examine solutions to the coupled Navier-Stokes/induction system.

\section{\label{sec:saturation}Turbulent Saturation of the Dynamo Instability}

\subsection{\label{sat:basic}Basic Considerations}

Consider first the``simpler" case where initially the magnetic field is weak, the Lorentz force (being quadratic in the magnetic field) can be neglected, and kinematic theory applies at least to the initial stages of the dynamo evolution. Given the arguments of the previous few sections at large enough $Rm$ one expects the magnetic field to grow exponentially with a well-defined growth-rate, with large and small scales growing at the same rate. The shape of the eigenfunction, and its coherence on timescales longer than typical timescales of the flow, may be controlled by the degree of breaking of reflectional symmetry of the flow. Eventually the kinematic phase must finish, as the Lorentz force becomes significant, and the dynamo instability saturates. Such a dynamo with a well-defined kinematic regime has been termed an \textit{essentially kinematic} dynamo \citep{tcb:2011}.

How the dynamo instability saturates depends on the form of the turbulence that underlies the initial instability. Presumably the dynamo equilibrates on a range of scales with a mean amplitude of the magnetic field that depends on scale (so there will exist a magnetic energy spectrum  alongside the kinetic energy spectrum). It may also be that different considerations are required, just as for the kinematic regime, to describe the saturation of dynamos at low and high $Pm$ and with or without broken reflectional symmetry. 

We shall first of all consider small-scale dynamo saturation, describing the current state of theory and numerical experiments. As might be expected, different considerations come into play for the saturation of organised magnetic fields and we introduce the contentious issue of the generation of large-scale field  in the nonlinear regime at high $Rm$ in section~\ref{systematic:sat}.
In order for the instability to saturate in a statistically steady state, on average the inductive terms in the induction equation must balance the diffusive terms. Given that, all other things being equal, the kinematic dynamo puts magnetic energy on the smallest scale possible, it is unlikely that an essentially kinematic dynamo will typically saturate by increasing the diffusion of the magnetic field --- since the magnetic field in the kinematic regime is already structured to maximise diffusion; clearly some modification of the inductive nature of the velocity field is needed. Therefore the field usually saturates by primarily modifying the $\nabla \times ({\bf u} \times {\bf B})$ term. The simplest way to achieve this is to modify the velocity field via the momentum equation.

The question then arises as to the nature of this modification. Is there an easily identifiable
property of the velocity field that in the kinematic phase leads to the exponential growth of the field whilst
in the saturated state yields a statistically stationary magnetic field? Discussions of dynamo saturation 
can be divided into three broad paradigms. The first is the equipartition argument, that the magnetic
energy will grow until it is comparable with the kinetic energy. This is a nice criterion as it can be applied 
with great generality, without knowing much about the details of the dynamo system; however it suffers from two main problems. 
It is easy to construct both examples where the saturation magnetic energy is substantially lower than the kinetic energy \citep{BrummCattTob:2001} and where it greatly exceeds it \citep{stellhans:04}. Even 
in systems where the energies are comparable, they may
not be comparable at all scales (Vainshtein \& Cattaneo 1992).  The second paradigm is a marginal stability argument that
the nonlinear effects of the magnetic field are to bring the system back to marginality. In some sense this must be the case as, on average, induction balances diffusion in the induction equation, as argued above. However this is not an equivalent criterion to reducing $Rm$ to its marginal value by modifying a typical amplitude of the velocity. This is linked with the limitation of the usefulness of the definition of the global $Rm$ based on the system scale discussed in section~\ref{Rmrole}.

The third paradigm is more sophisticated and invokes some subtle modification of the Lagrangian properties of the flow. Given the arguments of section~\ref{fast_dynamo} that chaos is required for a dynamo to operate at high $Rm$ it seems plausible that if the magnetic field is able to reduce the level of chaos then this can lead to saturation. Interestingly, it seems as though the general case is that saturation occurs without suppression of chaos at high $Rm$. Saturated dynamo flows have been shown to continue to stretch exponentially by \citet{ct2009} as evidenced by their ability to amplify exponentially a passive vector field. Indeed the set of vector fields tht do not grow exponentially when stretched by the turbulence appears to have measure zero.
Unfortunately this implies that there is no general theory for the saturation of dynamos and that, perhaps unsurprisingly, the physics of the momentum equation will play an important role.

\subsection{\label{sat:highPm}Saturation of high $Pm$ dynamos.}

\subsubsection{\label{highgc}High $Pm$ dynamos: General Considerations and Random Flows}

Just as for the kinematic regime,  there is a large difference between high and low $Pm$ regimes with regard to the saturation of dynamos. In the high $Pm$ regime the 
dynamo is operating at scales in the sub-inertial range of the velocity for which the kinetic Reynolds number is small and the inertial term in the
momentum equation can be neglected. The velocity can then formally be decomposed  into two parts ${\bf u} = {\bf u}_{F} + {\bf u}_M$ where ${\bf u}_{F}$ is the original velocity driven by the forcing (either mechanical or buoyancy) i.e.\ it is the velocity of the kinematic dynamo problem and has a characteristic scale $\gg l_{\eta}$, the other ${\bf u}_M$ is the magnetically driven velocity. These two components then evolve independently \citep{BrummCattTob:2001}.

To fix ideas,  consider the case where the velocity is kinematically driven on one scale, say the integral scale of the calculation. As noted in section~\ref{fast_dynamo}.
this will lead kinematically to a growing dynamo eigenfunction for the magnetic field that is concentrated at the resistive $Rm^{1/2}$ scale. If $Rm$ is large (i.e. in the kinematic fast dynamo regime) then this  is much smaller than the scale of the velocity. 
These small-scale filamentary fields 
now  drive flows on scales smaller than the velocity via the Lorentz force, eventually generating strong enough magnetic energy at the velocity scale to modify the original flow. Because of the simplicity of system at low $Re$, estimates for the strength of the generated magnetic field can be obtained where $B^2/U^2 \sim Rm^{1/2} Re^{-1}$, where $B$ is measured in units of the Alfv\'en velocity. This estimate is borne out at moderate $Rm$, but breaks down at higher $Rm$ \citep{BrummCattTob:2001}. At such low $Re$ and $Rm$ the saturation mechanism is simply to drive flows via the Lorentz force that can counteract the advective properties of the driven flow. 

For multi-scale velocity fields at moderate and large $Re$ the situation is more complicated and saturation can be investigated using semi-analytical models, phenomenological models and 
numerical experiments.  The semi-analytical models
are ultimately based on some closure of the MHD equations. For random flows described by the Kazantsev formalism, the magnetically driven velocity produces a change in the velocity 
correlator, which leads to the nonlinear saturation \citep{sub:1999,sub:2003}. Alternatively a 
Fokker-Planck equation for  the probability distribution function for magnetic fluctuations can be derived. It can be shown that the coefficients of this equation again are determined by 
the velocity correlation function which can be modified nonlinearly in a similar manner to above \citep{bol:2001}. However all of these results rely on some assumption of the role of the Lorentz force and that the flows remain $\delta$-correlated in time even in the presence of a saturating magnetic field.

At high $Pm$ dynamos an interesting phenomenological description has been proposed \citep{scheketal:2002-amodel} for multi-scale dynamos. 
The initial kinematic growth of field at the resistive scale (which is much smaller than the viscous 
scale) is modified  when ${\bf u} \cdot \nabla {\bf u} \approx {\bf B} \cdot \nabla {\bf B}$ at that scale. The 
left hand side is easily estimated to be $v^2/l_\nu$. To estimate the right hand side the authors use the foliated 
structure of the magnetic field to yield.
 $b^2/l_\nu$, 
where $b$ is a typical field strength at the scale 
$l_\eta$. The nonlinear saturation therefore  \textit{begins} when the magnetic energy comes into equipartition with the kinetic energy at the 
viscous scale and  dynamo growth associated with  eddies at the viscous scale is halted. This 
suppression is believed to the result of a  subtle 
modification of the eddy geometry; the Lorentz force causes the eddies to align with the local magnetic field, thus reducing induction. However, slightly larger eddies can still 
sustain growth and so growth will continue until the magnetic energy comes into 
equipartition with the energy of these eddies;  this process  continues until the magnetic 
energy reaches equipartition with the kinetic energy  at the integral scale. This nonlinear adjustment is 
characterised by a growth of the magnetic energy on an algebraic (rather than exponential) time with a shift to the generation of magnetic fields on larger scales.  It is possible to construct models where the final state only reaches a  fraction of global equipartition. 
This occurs when $Pm$ is large but $Pm \le Re^{1/2}$, and for this regime  $B^2/U^2 \approx 
Pm/Re^{1/2}$ \citep{scheketal:2002-amodel}.
In this scenario, however,  the characteristic scale of the saturated magnetic field is 
still smaller than the viscous scale.

\subsubsection{\label{highnum}High $Pm$ dynamos: Numerical Experiments}

There have been many simulations of dynamo saturation  at moderate to high $Pm$ that seem to confirm the phenomenological picture described above \citep{maronetal:2004}. In 
the kinematic phase the magnetic spectrum appears compatible with the $k^{3/2}$ prediction of the Kazantsev model, whilst the magnetic field does indeed appear foliated.  Dynamo saturation occurs when the magnetic and 
kinetic energies are comparable;  as the 
saturation progresses the magnetic spectrum grows and flattens with the creation of magnetic 
structures at larger scales.
The magnetic energy always exceeds the kinetic energy and the field is more intermittent than the velocity, with the pdf for the 
velocity field remaining close to that of a Gaussian whilst that for the magnetic field is better described by an 
exponential. The degree of intermittency is, however, reduced in the saturated state as compared with the kinematic 
state \citep{cattaneo:1999}. Interestingly, the  high $Pm$ formalism seems to persist for $Pm \sim {\cal O}(1)$ (see \citet{tcb11} for more details) though quite how well the phenomenological picture is replicated is unclear, owing to numerical limitations. It is clear though that at $Pm \sim {\cal O}(1)$ the magnetic fields are more intermittent that the velocity and the local magnetic field introduces statistical anisotropy in the flow in the nonlinear regime.

\subsection{\label{sat:lowPm}Saturation of low $Pm$ dynamos}


At present the low $Pm$ regime remains largely unexplored, owing to  the computational constraints described earlier for the kinematic problem. Because the dynamo operates in the inertial range 
of the turbulence inertial terms certainly remain important. Analytical progress can only be made by resorting to standard closure models such as EDQNM 
\citep{pouquetetal:1976} (see section~\ref{closures} for a discussion of EDQNM and other closures).  

However, dimensional arguments suggest that  for sufficiently low $Pm$ an asymptotic 
regime may be reached where the level of saturation becomes independent of $Pm$ \citep{petfauve:2008}. Unfortunately, even with current computational power, 
 simulations of  dynamos at low $Pm$ are difficuly, and it appears as though all simulations (whatever the quoted value of $Pm$)  are currently in the $Pm \sim {\cal O}(1)$  regime, 
with all the associated dynamics \citep[see e.g.][]{mininni:2006,iskakovetal:2007}. For reasons  discussed in \citet{tcb11}, any calculation that aims to test nonlinear saturation in the low $Pm$ regime (and that resolves the inertial range and the numerically growing eigenfunction) has a requirement in terms of gridpoints of several thousands (or tens of thousands). Using large-eddy simulations (LES) for the velocity should save a 
factor of ten in resolution \citep{pontyetal:2007}, but care must be taken in applying sub-grid modelling to MHD flows as emphasised by \citet{miesch:2015}.

\subsection{\label{closures}Analytic Closure Theories for nonlinear dynamos}

Analytical progress for nonlinear dynamos is  difficult and, just as for hydrodynamic turbulence, almost always relies on closure approximations. It is important to stress  that closure theories for MHD are on shakier ground than those for hydrodynamic flows;  whereas in 
hydrodynamic turbulence there is a vast body of experimental evidence and many systematic attempts to test closures and reduction hypotheses, there is no such thing in MHD.  Much more needs to be done in evaluating the accuracy and applicability of closure hypotheses for MHD; intuition for such problems is often misleading.
One other shortcoming of the procedure is that often expedient assumptions about the nature of the turbulence are made (for example homogeneity and isotropy) in order to make analytical progress in deriving the expressions for the electromotive force and hence the transport coefficients. For geophysical and astrophysical situations these assumptions are often poor and so it would be better to proceed without making such assumptions --- this is the philosophy behind Direct Statistical Simulation (DSS) as described in section~\ref{Future}.

Here we do not give details of the various possible closure schemes, as a discussion of these even for the hydrodynamic problem is well beyond the scope here. The interested reader is directed to the excellent review by \citet{yokoi19}. We only include here a flavour of the assumptions and shortcomings of the various closures.

The simplest closure scheme possible is to consider only quasilinear interactions in both the momentum equation \textit{and} the induction equations. That is to discard eddy/eddy $\rightarrow$ eddy interactions in both equations. This is expected to work well for the case where there are systematic mean flows and fields, but may not be appropriate when cascade and inverse cascade processes play a dominant role.

For cases where the eddy/eddy $\rightarrow$ eddy interactions are key,  more work is needed to capture the nature of the interactions. Following \citet{yokoi19} we explain the schemes using a model nonlinear system that schematically represents the hydrodynamic momentum equation\footnote{extensions of the formalism of the schemes to MHD (though not their applicability) is then straightforward. } in the form
\begin{equation}
    \dfrac{\partial u}{\partial t} = u u - p + \nu u.
\end{equation}
Here the $u u$ term is simply a prototypical nonlinear term. Hence 
\begin{equation}
    \dfrac{\partial \langle u \rangle}{\partial t} = \langle u u \rangle - \langle p \rangle + \nu \langle u \rangle.
\end{equation}
 The nonlinearity in this equation of course leads to the problem that the second-order correlation is required to solve for the mean and that the third-order correlation is also needed because
\begin{equation}
    \dfrac{\partial \langle u u \rangle}{\partial t} = \langle u u u \rangle + \nu \langle u u \rangle.
\end{equation}
The fourth-order correlation appears in the equation for the evolution of the third-order correlation because
\begin{equation}
    \dfrac{\partial \langle  u u u \rangle}{\partial t} = \left(\int \langle u u u u\rangle \right) + \langle u u u u\rangle + \nu \langle u u u \rangle,
\end{equation}
where the Poisson equation for the pressure has been used.

What is clear is that some closure is needed that can express the fourth-order correlations in terms of lower order correlations (either via a functional form or by deriving an evolution equation). For inhomogeneous systems this closure can be achieved by setting the values of higher order cumulants (which for the first three correspond to centred moments) to zero (see Section~\ref{Future}).

However if one also makes the assumption of homogeneity, then progress can be made by transforming to spectral (wavenumber) space and deriving the evolution equation for the spectral energy, $E(k)$, in terms of the spectral energy flux, $\Pi(k)$, and the spectral energy transfer, $T(k)$. Because the energy is a measure of the second-order correlations and the flux and energy transfer represent triple correlations, a closure can be achieved by expressing $T(k)$ or $\Pi(k)$ in terms of $E(k)$.

The most basic attempt to achieve this is to consider \textit{Quasi-normal models} \citep{moninyaglom75} in which the pdf of the velocity field is Gaussian and so the fourth-order correlation may be expressed as a sum of the products of the second-order correlations. However this harsh degree of approximation can lead to lack of realisability in the sense that negative energies may be achieved.

In order to alleviate this problem, which can be traced to the overestimation of triple correlations), \citet{orszag70} suggested placing a scale-dependent damping term $\tau(k)$ in the evolution equation for the triple correlations. Together with the assumption that the turbulence does not depend on its history (i.e. lacks memory via Markovianization) this leads to the eddy-damped quasi-normal Markovianized (EDQNM) approximation. The addition of the scale-dependent eddy-damping time can certainly resolve the issue of non-realisability; however the eddy-damping time depends critically on an assumed eddy damping rate $\mu({\bf k)}$, which is given as a parameter. Indeed simplifications (for example the so-called `minimal $\tau$ approximation', where the eddy-damping time is itself a parameter \textit{independent} of scale) have also been suggested and explored \citep{kleerogsok:2002}.

One of the central uncertainties of utilising closure models for MHD turbulence and dynamos arises because the  eddy-damping rate (and hence time) \textit{does} depend on the magnetic field strength and is not given as a parameter. Magnetic fields lend memory to the turbulence through the presence of restoring forces and waves, and this should be taken into account in any theory. For all the theories described above, some assumption must be made about this dependence, which does not arise self-consistently from the theory. It is certainly the case that, 
whereas in hydrodynamic turbulence there is overwhelming evidence that this quantity is of the order of the turnover 
time, there is no consensus on what this  quantity should be in  MHD theory. Indeed it may depend sensitively on the 
field strength and magnetic Reynolds number of the magnetised turbulence.

However one scheme that does enable some self-consistent derivation of the timescale for the turbulence is the \textit{Direct-interaction approximation (DIA)} introduced by \citet{kraichnan:1959}. Here, via Green's function techniques, a closed system of equations comprising an equation for the correlation function and one for the response function is derived. The method of derivation is long and involved (and beyond the scope of this paper - though see \citet{yokoi19}), taking in such mathematical joys as perturbation expansions, partial summations and renormalisation methods via Feynman diagrams. It goes without saying that, in order to close the system, some assumptions must be made even for this technically challenging procedure. There is much work to be done on understanding the basic response of turbulence in the presence of a magnetic field at different $Rm$ even for the case where the magnetic field has not been self-consistently generated by the turbulence.

Finally for this section, we stress that the closure models described above are all essentially formulated and solved in spectral space, and that, as noted, this approach relies on the turbulence being homogeneous (and for some simplifications isotropic). The astrophysical turbulence that leads to the generation of cosmical magnetic fields is almost never homogeneous (and certainly not isotropic) and so this approach may be limited in applicability. One approach to circumventing this issue is to derive statistical models that do not rely on homogeneity. Perforce, these models are less amenable to analytical descriptions and solution and numerical algorithms and techniques must be developed for their progression (see section~\ref{Future} for a discussion of Direct Statistical Simulation).
However some progress can be made in including inhomgeneity via the two-scale direct interaction approximation (TSDIA). This approach combines multiple scale analysis (of the type used to derive mean-field theory) with the DIA. In short, the method proceeds by introducing two timescales (one order one --- called fast for convenience --- and one slow) and spatial scales (one small and one order one), so that
\begin{equation}
    \bm{\xi} = \bm{x}, \quad \bm{X} = \epsilon \bm{x}; \quad \tau = t, \quad T = \epsilon t,
\end{equation}
with $\epsilon \ll 1$. All variable are then expanded into mean fields and perturbations so that 
\begin{equation}
    f = \overline{f}(\bm{X},T) + f'(\bm{\xi},\bm{X}; \tau,T).
\end{equation}
Progress is now made by assuming that the fluctuation field is homogeneous with respect to the fast space variable and expanding the fluctuating variables in a power series in the small parameter $\epsilon$, so that
$f' = f'_0+\epsilon f'_1+ \dots$, making assumptions about the statistical properties of $f_0$ and calculating the turbulent fluxes (for example Reynolds Stresses and electromotive forces) via DIA. This has the potential to be a formidable approach leading to breakthroughs for certain instability and dynamo problems \citep[see e.g.][]{yokoi_2018}.

\subsection{\label{systematic:sat} Saturation of systematic fields}

In this section we consider probably the most contentious issue in turbulent dynamo theory, the nonlinear saturation of the generation of systematic fields. It is fair to say that this issue is technically and computationally challenging and no consensus has emerged. It would be possible to construct a whole review of this topic without really doing it justice, so here I will only recount the salient arguments.

We have demonstrated how prescribed turbulent flows with broken reflection symmetry can lead to the generation of an electromotive force that leads to the direct generation of systematic magnetic fields (in a manner analogous to the driving of mean flows via a Reynolds stress in the momentum equation). There are two possible general mechanisms for how systematic fields may continue to be generated in the nonlinear regime. The first is that the kinematic mechanism of direct driving via an emf continues to proceed in the nonlinear regime. Here we stress that direct driving refers to eddy/eddy $\rightarrow$ mean interactions (in the language of fluid turbulence). If this were the case then it might be possible to develop a theory for nonlinear behaviour of this driving, including parameterising the role of the large-scale magnetic field in modifying the transport coefficients. A second possibility is that in the nonlinear regime a scale-by-scale inverse cascade proceeds whereby energy is successively transferred locally in scale to larger spatial scales, eventually reaching the system scale of the object. It is fair to say that currently both of these mechanisms are believed to contribute.

As noted above, even in the nonlinear regime, the induction equation is still formally linear in the magnetic field and so it can be argued that the machinery of kinematic mean-field electrodynamics can, and should, still be of utility. Indeed, the arguments there were based on solution of the induction equation alone, which is an important component of the nonlinear dynamo system. 

It should even be possible to continue to couch the direct generation of systematic magnetic fields in the nonlinear regime in terms of the calculation of transport coefficients similar to those outlined in section~\ref{trans_calc}. What would be needed here is to utilise the {\it saturated velocity field} in the equations for the evolution of the fluctuating magnetic field (equation~\ref{fluc_eqn_with_time}) to calculate the evolution of the electromotive force. Formally this of course would yield the correct prescription, as this is simply a proxy for solving the induction equation. There are two major problems with this approach, however. The first is that there is no general theory for the turbulent response of the fluctuating velocity field to the presence of a magnetic field (either large or small-scale) as should have become clear in our discussions earlier. The second is that even were such a theory possible, we have seen that determination of the transport coefficients can only proceed if certain approximations (for example low $Rm$ or short correlation time, $\tau_c$, for the turbulence) are utilised. So for example, if the turbulence is assumed to maintain a short correlation time then the transport coefficients can be described by equations such as equation~(\ref{transport_short}), where now $\bm{u}$ and $\bm{\omega}$ must be interpreted as the saturated velocity and vorticity.
Astrophysical dynamos are certainly not at low $Rm$. Moreover it has often been argued that a significant consequence of the presence of magnetic field in turbulence is to add memory to the flow, thereby increasing the correlation time ($\tau_c$) of the turbulence and making the approximations of first order smoothing less valid in the nonlinear regime (even if they were assumed valid in the kinematic regime). It seems as though the desire to couch everything in terms of a mean-field theory based on kinematic considerations is causing considerable complications. 

\subsubsection{\label{catastrophic}Catastrophic (Vainshtein-Cattaneo) Quenching}
\medskip

Perhaps the greatest challenge to mean-field theory in the dynamic regime arises from arguments in the early nineties proposed by Vainshtein and Cattaneo \citep{cv1991,VainCatt:1992} and extended by \citet{grdi1994}. The arguments concern the level of saturation of the organised field that can be generated by an electromotive force, before the electromotive force is itself suppressed by the nonlinear effects of Lorentz force in the momentum equation.

One way to proceed would be to assume on energetic grounds that the Lorentz force can only begin to suppress a kinematic effect once the energy in the magnetic field is comparable with that of the turbulence. Traditionally (and naively) it has been assumed that this occurs when the magnetic energy of the \textit{mean field} is in equipartition with the
kinetic energy of the flow, i.e.\ when
\begin{equation}
 \langle \rho u^2 \rangle \sim B_0^2 / \mu,
\label{eq:equi_no_rm}
\end{equation}
where $B_0^2 = \overline{\bm{B}} \cdot \overline{\bm{B}}$ is the energy in the mean field.
Hence a simple phenomenological model would then be to take the nonlinear dependence of the transport coefficients as 
\begin{equation}
\alpha = \dfrac{\alpha_0}{1+ B_0^2/  {\cal B}^2 }, \qquad
\beta = \dfrac{\beta_0}{1+ B_0^2/   {\cal B}^2 },
\label{eq:no_rm_quench}
\end{equation}
where $\alpha_0$ and $\beta_0$ are the values of the transport coefficients derived in the kinematic approximation and ${\cal B}^2$ is the non-dimensional equipartition energy.  Note that this prescription implicitly assumes that the mean field dominates (or at least is the same order of magnitude as) the fluctuating field in the saturated regime, and so it is the mean field that leads to saturation of the transport. The expressions in equation~(\ref{eq:no_rm_quench}) are attractive as they lead to a saturation of the mean field for an energy comparable with that of the turbulence. Moreover these expressions rely only on the calculation of the mean or organised fields, which requires little in the way of computational expense.

However the assumption that the magnetic energy is dominated by the organised component is questionable at high $Rm$ as pointed out clearly by \citet{cv1991,VainCatt:1992}. The distribution of magnetic energy with spatial scale at high $Rm$ is a key issue for understanding nonlinear dynamo saturation. One can proceed on either physical or mathematical grounds to understand this distribution --- computational models, although suggestive and informative, are not at the point of unambiguously settling this issue. 

On physical grounds, it is argued that if there is a large-scale component of the field generated kinematically in a turbulent dynamo then  turbulence will amplify magnetic fields on the small-scale via the action of eddies with short turnover times. Hence one might expect  the small-scale fields (kinematically) to be more energetic than those on
large scales. If diffusion plays a role in setting the ratio of these energies then one might expect a relationship of the form
\begin{equation}
\overline{b^2}  \sim Rm^p B_0^2 ,
\label{eq:ratio_of_energies}
\end{equation}
where $b^2 = \bm{b'}\cdot \bm{b'}$ and $0 \le p \le 2$ is a flow- and geometry-dependent coefficient.

If this is the case, and assuming that magnetic fields with this energy do distort the eddies with correlations that lead to the generation of organised field (i.e.\ those with significant broken reflectional symmetry), then this may indicate that  the transport coefficients are altered significantly once
this strong \textit{small-scale} field reaches equipartition with the
kinetic energy of the turbulence, i.e.\ when 
\begin{equation}
\langle \rho u^2 \rangle \sim Rm^p B_0^2 / \mu. 
\label{eq:equi_rm}
\end{equation}
 One might then expect that  the expressions in equation~(\ref{eq:no_rm_quench}) to be replaced by
\begin{equation}
\alpha = \frac{\alpha_0}{1+Rm^p B_0^2/ {\cal B}^2 } ,
\quad  \beta = 
\frac{\beta_0}{1+Rm^p B_0^2/ {\cal B}^2 } .
\label{eq:rm_quench}
\end{equation}
At high $Rm$ one can then argue that the efficiency of the transport coefficients is strongly suppressed even when the organised field $B_0$ is small; a phenomenon known as  `catastrophic' quenching \citep{cv1991,VainCatt:1992}. It certainly seems plausible that the transport coefficients begin to be suppressed when the mean field is weak.
However, the generation of a mean electromotive force relies on correlations in the turbulence between the fluctuating magnetic field and eddies, which is a subtle interplay, so arguments on physical grounds may be misleading. Hence other approaches to this issue have been employed. 

The first of these is to derive
analytical expressions for the electromotive force (and hence the transport coefficients) utilising closure approximations (such as those discussed in section~\ref{closures}) and applying conservation laws that are valid in the non-dissipative system.  This approach can yield significant insight (in fact even understanding the role played by conservation laws is important) but is only as good as the approximations that are utilised. For a more extensive discussion see \cite{dhk2005,bransub:2005}. We give a brief explanation here.

Let us start by describing a closed domain, with no fluxes in or out of the region. We noted in equation~(\ref{hel_evolution}) that the total magnetic helicity changes only via dissipation or through surface fluxes from the domain. Hence for an ideal, closed system magnetic helicity is a quadratic (though not sign-definite) invariant quantity. For a closed dissipative system with no surface fluxes, global magnetic helicity can only be created or destroyed via irreversability introduced by diffusive processes, i.e.\
\begin{equation}
\dfrac{d}{dt} \langle \bm{A}\cdot \bm{B}\rangle = - 2 \eta \langle \bm{j}\cdot \bm{B} \rangle.
\label{total_hel_cons}
\end{equation}
We recall here that angular brackets refer to volume averages. Note at this point that, for this closed system, this immediately allows some inference about $\langle \bm{j}\cdot \bm{B}\rangle$ in the steady state, namely that $\langle \bm{j}\cdot \bm{B}\rangle = 0$.

Furthermore if one defines an intermediate average (such as co-ordinate averaging in the mean-field sense as discussed in section~\ref{averaging}) and denotes it by an overbar then
in the steady state
\begin{equation}
    \langle \bm{j}' \cdot \bm{b}' \rangle + \langle \overline{\bm{j}} \cdot \overline{\bm{B}} \rangle = 0,
\end{equation}
so that the average current helicity from the fluctuations is of the opposite sign to that from the means in the steady state, and if there is no mean current then $\langle \bm{j}' \cdot \bm{b}' \rangle = 0$.

Indeed, if there is no mean current another exact result emerges from consideration of Ohm's law, giving
\begin{equation}
{\bm{\mathcal{E}}} \cdot \bm{B}_0 = -\dfrac{1}{\sigma} \langle \bm{j'} \cdot \bm{b'} \rangle
+ \langle \bm{e'} \cdot \bm{b'} \rangle = \alpha B_0^2,   
\label{alpha_from_ohm}
\end{equation}
where $\bm{B}_0$ is the uniform mean field, with no associated current. This exact result relates the transport coefficient $\alpha$ to the fluctuating current density and the average correlation between the fluctuating magnetic field and emf.

Finally, for a closed domain one may obtain evolution equations for the magnetic helicity contributions from large and small scales, namely
\begin{equation}
   \dfrac{d}{dt} \langle \overline{\bm{A}}\cdot \overline{\bm{B}}\rangle = 2 \langle \overline{\bm{{\cal E}}}\cdot \overline{\bm{B}} \rangle - 2 \eta \langle \overline{\bm{j}}\cdot \overline{\bm{B}} \rangle,
   \label{ls_helicity}
\end{equation}
and
\begin{equation}
   \dfrac{d}{dt} \langle {\bm{a}'}\cdot {\bm{b}'}\rangle = -2 \langle \overline{\bm{{\cal E}}}\cdot \overline{\bm{B}} \rangle - 2 \eta \langle {\bm{j}'}\cdot {\bm{b}'} \rangle.
   \label{ss_helicity}
\end{equation}
Hence the $\langle \overline{\bm{{\cal E}}}\cdot \overline{\bm{B}} \rangle$ term acts to transfer magnetic helicity between fluctuating and mean fields. It is clear then for this closed system that if the large-scale magnetic helicity is to grow it must be at the expense of
the small-scale magnetic helicity; recall though that magnetic helicity is not sign definite so it is certainly possible for these to grow together if they are of oppostite signs.

We note here that the expressions derived above allow evaluation of an expression for the 
projection of the average emf onto the mean field, i.e.\ $\displaystyle{\bm{\mathcal{E}} \cdot \bm{B}_0}$, in terms of correlations between $\bm{j}'$ and $\bm{b}'$, instead of correlations between $\bm{u}'$ and $\bm{b}'$. However in the nonlinear regime we know none of $\bm{u}'$, $\bm{b}'$ or $\bm{j}'$, so what has this gained us?

The answer arises from utilising results from calculations employing the closure approximations described in section~\ref{closures} --- specifically calculations employing the EDQNM approximation, and generalisations thereof. These calculations \citep[see e.g.][]{pouquetetal:1976,kleerogsok:2002} yield an approximate relationship between the $\alpha$-coefficient for homogeneous isotropic turbulence and the kinetic and current helicity, namely that
\begin{equation}
\alpha \sim - \dfrac{\tau_c}{3} \langle \bm{u}' \cdot \bm{\omega}' - \bm{j}' \cdot \bm{b}' \rangle.
\label{eq:pouquet}
\end{equation}
At this point it is worth commenting again on the status and interpretation of this expression. Unlike the earlier global balance relationships, which are exact, it is to be stressed that this expression can only be derived via approximate closure relationships or via linearisation about a pre-existing turbulent MHD state at low $Rm$ and short $\tau_c$ \citep{proctor2003}. Equation~(\ref{eq:pouquet}) has clearly replaced the  calculation of correlations between $\bm{u}'$ and $\bm{b}'$ with those between $\bm{j}'$ and $\bm{b}'$. This is only possible if some dynamical relationship between  $\bm{j}'$ and $\bm{u}'$ has been  approximated.

One key limitation in this expression is the assumption that the correlation time of the turbulence, $\tau_c$ remains unaffected by the presence of large-scale magnetic field, which is not necessarily true at high $Rm$, as discussed earlier.
Nonetheless, what is clear is that the expression~(\ref{eq:pouquet}) does reduce to to that given in equation~(\ref{transport_short}) in the kinematic regime; the correction to the kinematic result often being termed a magnetically-driven $\alpha$-effect,
$\alpha_M$.

Because, in this approximation, the current helicity from the small scales plays a key role determining the correction to the transport coefficient, it is now possible to combine this with the exact results of equations~(\ref{ss_helicity}) to derive an evolution equation for the transport coefficient $\alpha$. For details of this see \cite{grdi1994,krr1995,bb2002}; which include details of the further assumptions/simplifications (such as relating $\langle \bm{a}' \cdot \bm{b}'\rangle$ to $\langle \bm{j}' \cdot \bm{b}'\rangle$ via a spectral relationship) required to derive the evolution equation.

The evolution equation is given in full in \cite{bb2002} who also show that, in the saturated steady state for large scale fields with no associated currents (i.e. uniform fields),
\begin{equation}
    \alpha = \frac{\alpha_0}{1+Rm B_0^2/ {\cal B}^2 },
    \label{cvgd}
\end{equation}
a result first derived by \citet{grdi1994}, which is in agreement with the catastrophic quenching result of \citet{cv1991} in equation~(\ref{eq:rm_quench}). There is therefore now little debate as to the nature of the  quenching of the $\alpha$-effect in closed systems with no large-scale currents; particularly as we shall see that the quenching formula is consistent with the results of numerical experiments (albeit at low and moderate $Rm$.)

Physically this result can be interpreted as the conservation of magnetic helicity placing a topological constraint on the nature of the magnetic field (as discussed in section~\ref{consmh}). In order to grow the large scale magnetic field, the knotted small-scale field that is naturally generated kinematically must be untangled, which is impossible if the topological constraint is maintained. If the presence of small magnetic diffusion is the only source of irreversability allowing the untangling of magnetic field then this places a severe constraint on the level of the mean field that can be generated --- or at least the timescale on which it can be generated.

One interpretation of a formula such as that of equation~(\ref{cvgd}) is that large-scale field generation is completely suppressed for mean fields an order of magnitude smaller than equipartition ($B_0 \sim \mathcal{O}(Rm^{-1/2} \cal B)$). However consideration of the full evolution equation given in say \citet{bb2002} demonstrates that even with this `catastrophic' quenching formula the energy in the large-scale magnetic field continues to grow on a timescale controlled by the magnetic diffusion (an ohmic time). In the long run (for example in the case of the Sun a timescale comparable with the age of the star) reaching a statistically steady state with significant large-scale field may be reached, see also \citet{MoffattDormy:2019}. However,  \textit{in the long run we are all dead} \citep{keynes:1923}.

\subsubsection{\label{fluxes}Large-scale currents and Helicity Fluxes: Can they help?}
\begin{center} \textit{``E pur si muove"} Attributed (perhaps incorrectly) to Galileo. \end{center}
\medskip

Catastrophic quenching as envisaged by \citet{VainCatt:1992} and \citet{grdi1994} places a severe constraint on the applicability of mean-field theory at high $Rm$. Large-scale organised fields can only be generated from a weak seed field on an ohmic timescale, this being the time taken for the action of ohmic dissipation to untangle the field. However it may be that other processes can be identified that break reversibility and hence the topological constraints placed on the field, which we shall discuss below. Another option is that the magnetic field may not have been generated all the way from kinematic values by a turbulent emf, but may simply be sustained by an emf that arises from an instability of the field itself. These more laminar ``essentially nonlinear" dynamos will be  described in section~\ref{ess:non}.

When a mean current $\overline{\bm{j}}$ is allowed, this naturally leads to the presence of large-scale current helicity  $\langle \overline{\bm{j}} \cdot \overline{\bm{B}}\rangle$\footnote{Care does need to be taken here as a the presence of a large-scale current leads to both questions of the nature of the separation of scales and the presence of a turbulent diffusion because of the presence of large-scale field gradients.}. The inclusion of a mean current may seems natural since it it difficult to see how a dynamo can generate a magnetic field with no mean current. In that case equation~(\ref{cvgd}) (subject to the same assumptions and approximations) is modified to 
\begin{equation}
    \alpha = \frac{\alpha_0}{1+Rm B_0^2/ {\cal B}^2 } + \frac{Rm \, \beta  \langle \overline{\bm{j}} \cdot \overline{\bm{B}}\rangle}{{{\cal B}^2 }+Rm \langle \overline{\bm{B}}^2 \rangle},
    \label{gdk}
\end{equation}
as first derived by \citet{grdi1994}\footnote{Here $\beta$ is the magnitude of the turbulent diffusivity.}. Thus, although the kinematic $\alpha$-effect is catastrophically quenched, if one can (somehow) generate large-scale fields with a non-trivial current helicity then they can be maintained via a magnetically driven emf, the second term in equation~(\ref{gdk}) (see also section~\ref{ess:non}).

However, it is not at all clear how one generates these significant large-scale fields from a weak seed field (e.g. in a galaxy). It may be that shear flows play a major role here in producing a large-scale field strong enough so that nonlinear effects can kick in. However if one is relying on turbulence (i.e.\ fluctuation-fluctuation interactions) to do the job then there must be another process breaking reversability for this to proceed quickly.

\citet{bf2000} proposed that catastrophic quenching, or the associated generation of large-scale fields on an ohmic timescale, may be circumvented if irreversability is introduced to the system by allowing magnetic helicity fluxes into and out of the domain. For example if magnetic helicity associated with small scales is preferentially lost on a timescale $\tau_{flux}$ then the rate of generation of large-scale field may occur on this timescale. Mathematically this can be seen by reinstating the helicity flux term into equation~(\ref{total_hel_cons}) and hence also into equations~(\ref{ls_helicity}-\ref{ss_helicity}). The presence of this term breaks the reliance on ohmic processes to reconfigure (untangle) the magnetic field, and hence may alleviate the catastrophic quenching or allow the more rapid generation of large-scale field.
There are two, physically distinct, contributions to the helicity flux; one diffusive and one ideal. Clearly the diffusive flux should become analogously weak to the ohmic dissipation of magnetic helicity in the limit of high $Rm$, so we are relying on the ideal flux to do the job!

There are a number of issues with this picture that require further investigation. The first is that the ideal flux term takes the form of a surface integral; it is not obvious whether a term that relies on losses through a surface can lead to a significant impact on the generation of large-scale fields throughout a large volume. The second is whether the ideal fluxes manage to avoid scaling with the magnetic Reynolds number (not directly, since the diffusivity does not appear explicitly in these terms, but through the form of the magnetic field). This may be checked using numerical calculations as described in the next subsection, at least at moderate $Rm$; this potential lifebelt for the generation of large-scale fields at high $Rm$ should be investigated more closely in the near future. Finally, it may be that it is enough to separate spatially the region of large-scale field generation from that of large-scale field storage. Faced with the proposal of catastrophic quenching, \citet{Parker:1993} introduced the concept of interface dynamos, which made use of precisely this idea. More investigation of dynamos with inhomogeneous turbulence and shear flows may reveal whether this model can function at high $Rm$. Shear-enhanced diffusion and other methods of introducing irreversability on a fast timescale may also play an important role here.

\subsubsection{Numerical investigations of catastrophic quenching.}

The theoretical arguments proposed above can be somewhat tested by devising numerical experiments. These, in general, take two complementary forms: dynamo experiments examine the nature of the generated magnetic field when it is allowed to evolve freely in a turbulent flow, whereas turbulent transport experiments typically impose a large-scale field (perhaps a uniform field or a large-scale field with a large-scale current) and measure the response of the turbulence in creating an emf. Of course, the first of these approaches is more natural, with the magnetic field  self-consistently evolving with the turbulence. The second approach, though, offers more control and the potential ability to determine the functional dependence of the emf on parameters such as the imposed field strength, rotation rate, and magnetic Prandtl number.

However it must be stressed that such numerical simulations are limited in the range of $Rm$ that are accessible and it is difficult to extrapolate the results to the astrophysically relevant high $Rm$ regime. For example, a spectral calculation performed with $10^9$ degrees of freedom with a reasonable separation of scales between large and small scales is able to reach magnetic Reynolds numbers of $\mathcal{O}(10^3)$ \citep{ht2010}.

Simulations are usually configured in one of two ways. In the first a flow is driven via a prescribed body force, typically chosen to drive a relatively small-scale  flow with significant kinetic helicity (to maximise the chances of driving an emf and hence a large-scale dynamo) and potentially a shear flow. In the second driving is via buoyancy in thermal convection, where helicity emerges naturally via the interaction with rotation; clearly the second of these is of more relevance astrophysically though there is less control over the form of the flow.

The list of numerical experiments performed over the past twenty or so years is long, and a complete review is well beyond the scope here. The interested reader is directed to \citet{bransub:2005} for a thorough review of the early attempts and \citet{brandenburg:2018} for a briefer  review of later attempts.

In a closed domain all simulations (albeit at low and moderate $Rm$) are consistent with the picture of catastrophic quenching of the emf (or diffusive growth of the mean field). For example \citet{CattHughes:1996} utilised a forcing that, in the absence of magnetic field, was designed to drive a small-scale Galloway-Proctor flow (here small-scale is defined as $k=1$) and measured the response of the turbulent emf to an imposed uniform ($k=0$) magnetic field in a triply periodic box. They found results consistent with equation~(\ref{eq:rm_quench}). Analogous results were found for dynamo calculations driven by helical turbulence by \citet{bb2002} (and see references therein), where for closed systems the mean field grew ohmically after an exponential kinematic phase. It seems as though there is now reasonable agreement for homogeneous, confined systems; the magnetic field can only emerge after an ohmic timescale.

For open systems, where boundary conditions that allow magnetic helicity to escape or enter the domain, the situation is more controversial, and more detailed calculations at higher $Rm$ are needed. Briefly, it does appear that the dynamos in these open domains do seem to have a different character to those in closed domains --- at least at moderate $Rm$. For example, \citet{hb:2010} found that, as $Rm$ is increased both the diffusive volume term and helicity flux contribution to irreversability decrease, though intriguingly the diffusive volume term decreases more rapidly as a function of $Rm$ (at least for moderate $Rm$). Linear extrapolation would seem to indicate that the boundary term would dominate at sufficiently high $Rm$. Similar results have also been found by \citet{dsgb:2013}. However it does seem in all current cases as though the ratio of mean field to fluctuating field continues to decrease with $Rm$ \citep{brandenburg:2018}. Clearly this issue is difficult to address computationally, though the status of mean-field theory in the nonlinear regime may become clearer in the next few years.

\section{\label{ess:non}Essentially Nonlinear Dynamos}

Up to this point, we have focussed on dynamos (whether large or small scale) that have a well-defined kinematic regime; the role of the Lorentz force is simply to  saturate the dynamo. These types of dynamo have been termed \textit{essentially kinematic} (whether in the linear \textit{or} nonlinear regime), and much attention has focussed on whether such dynamos may act to generate organised magnetic fields, as we have seen.

However, another type of dynamo is possible; one where the  components of the flow that lead to the generation of the field are themselves driven by the instabilities of, or the intervention of, a finite amplitude magnetic field. Such a dynamo would not exist in the limit of vanishingly small magnetic field (it is necessary to have a finite amplitude magnetic perturbation to sustain such a dynamo). These dynamos, which have been termed \textit{essentially nonlinear}, often use the presence of a finite amplitude magnetic field to facilitate the extraction of energy from an energy reservoir (typically a shear flow that can not act as a dynamo in its own right). 

It has been argued \citep{tcb:2011} that such \textit{essentially nonlinear} dynamos are better candidates for generating organised magnetic fields at high $Rm$ than \textit{essentially kinematic} dynamos. The line of reasoning here is that organised magnetic fields are generated through correlations between the small-scale velocity and small-scale magnetic fields. If the small-scale velocity field is itself driven by say the instability of a magnetic field then it is more likely to be correlated with the field perturbations that arise in the same instability than if driven by some external force. This is particularly true at high $Rm$.

Some examples of \textit{essentially nonlinear} magnetically driven dynamos are have been examined in detail. One of particular interest is that dynamo that exists through the interaction of a shear flow with the magnetic buoyancy instability \citep{cbc2003}. In this scenario large-scale toroidal magnetic field is generated from large-scale poloidal field by a shear flow.  The regeneration of poloidal field occurs owing to the combined action of magnetic buoyancy and Kelvin-Helmholtz instabilities; this can only occur if  the initial magnetic fields exceed a critical threshold. Furthermore the instabilities lead to fluctuating velocities that are strongly correlated with the unstable magnetic fields --- unsurprising as these velocities are themselves driven by the field. This leads to the generation of a large-scale magnetic field and the cycle can begin again. To a fluid dynamicist the description should be reminiscent of a self-sustaining process, particularly that described by \citet{waleffe:1997} for nonlinear transition in wall-bounded shear flows.

\begin{figure}
\begin{center}
\includegraphics[height=4cm,width=6.5cm]{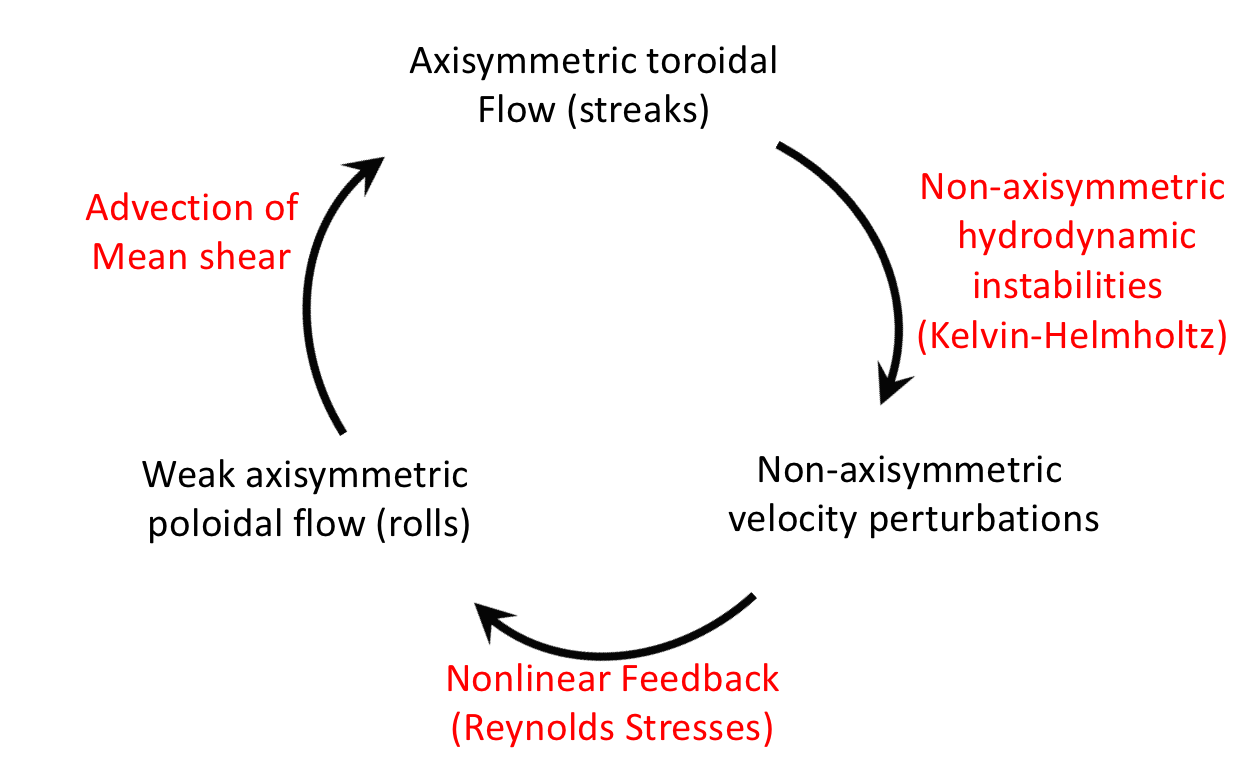}
\includegraphics[height=3.9cm,width=6.5cm]{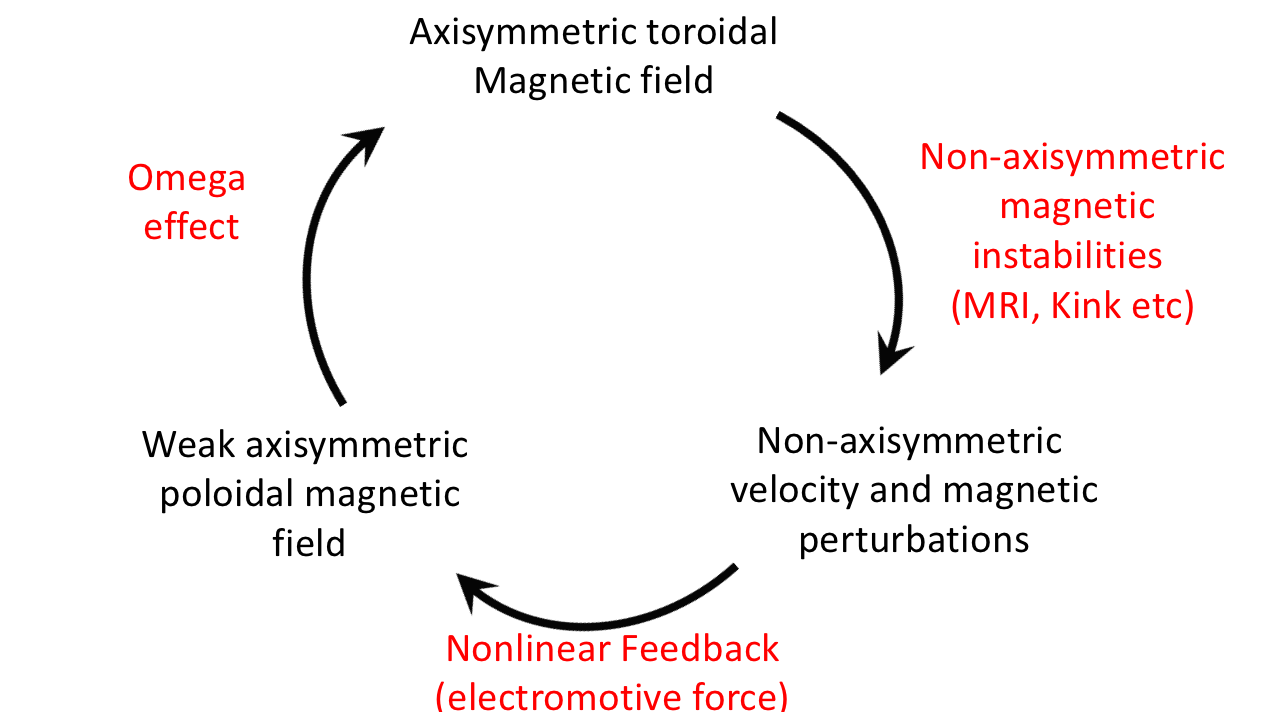}

\end{center}
\caption{(a) Hydrodynamic self-sustaining process \citep{waleffe:1997} and (b) self-sustaining
dynamo processes in shear flows prone to MHD instabilities \citep{rincon2007}. After \citet{rincon2013}
}
\label{fig_rincon}
\end{figure}

This analogy has been carried much further in the description of another \textit{essentially nonlinear} dynamo in an excellent series of papers by Rincon and collaborators \citep[see e.g][and others]{rincon2007,rincon2013}. Here the dynamo is driven by the magnetorotational instability, which is a joint instability of differential rotation and magnetic field. Here the Keplerian shear flow is hydrodynamically stable via Rayleigh's criterion. However, in a finite domain and in the presence of dissipation, a finite amplitude magnetic field can lead to the destabilisation of the shear and the generation of turbulence. This turbulence may itself lead to the sustainment of the magnetic field sufficient to maintain the instability. Such a dynamo bears all the hallmarks of a self-sustaining process, and the strong analogies with nonlinear transition in shear flows have been emphasised, as shown in Figure~\ref{fig_rincon}, which is adapted from \citet{rincon2013}. 

It remains to be seen whether these mechanisms continue to work at high $Rm$, but this is an extremely promising avenue for research in astrophysical dynamos. Other such dynamos that may benefit from such an analysis include dynamos driven by magnetic buoyancy, current-driven instabilities, joint instabilities of differential rotation and magnetic fields and possibly the geodynamo. Methods from transition could be useful in aiding progress in all of these cases.

\section{\label{balances}Balances in Rapidly Rotating Magnetised Convection and the Geodynamo}

This perspective has focussed mainly on the key problem of generating magnetic field in the astrophysically relevant limit of high Magnetic Reynolds number. However, for the construction of models of planetary magnetic fields, the conceptual difficulty lies not in the solution of the induction equation at high $Rm$, but rather that of the momentum equation in the rapidly rotating limit. This difficulty takes the form of the solution of equations with variability on a vast range of temporal scales. Here we shall focus on descriptions of the geodynamo, though of course the theoretical difficulties will propagate through to models of other rapidly rotating solar-system planets and exoplanets.

\subsection{Timescales in the Geodynamo. \label{geo_time}}

Earth's magnetic field is believed to be generated in the outer core by the motion of the iron-rich fluid there. The driving mechanism for this motion is still the subject of debate; most models utilise driving by thermal or compositional convection, though the role of instabilities driven by precession or tides are still of interest.
The Earth, of course, rotates with a period $(T_\Omega)$ of one day, which sets the shortest timescale of interest for the geodynamo. Once the magnetic field is generated by the dynamo it allows magnetically mediated waves (so called MAC and torsional waves) with periods of years and decades. It is believed that a typical convective time $T_c$ for the Earth is on the order of $10^2$~years, whilst the magnetic field diffuses on a timescale $T_\eta \sim 10^4$ years. Finally the viscous and thermal timescales are extremely long $T_\kappa \sim T_\nu \sim 10^9$ years. The ratio of these timescales is encoded in the relevant non-dimensional parameters, i.e.\
\begin{eqnarray*}
Ro &=& \frac{T_\Omega}{T_c} \sim 10^{-4}, \quad \quad  Re = \frac{T_\nu}{T_c} \sim 10^7, \quad \quad Rm = \frac{T_\eta}{T_c} \sim 10^2-10^3,\\
Pm &=& \frac{T_\eta}{T_{\nu}} \sim 10^{-5}, \quad \quad  Pr = \frac{T_\kappa}{T_\nu} \sim 1, \quad \quad \quad E = \frac{T_\Omega}{T_\nu}\sim 10^{-15},
\end{eqnarray*}
where $Ro$ is the Rossby number and $E = Ro/Re$ is the Ekman number.
Note that most of these numbers are either very small or very large. However the magnetic Reynolds number, the source of all the problems for stellar and galactic dynamos, is not prohibitively large. Therefore we do not anticipate too much difficulty in the solution of the induction equation and our attention should be focussed on the momentum equation.

The vast range of temporal scales provides the most significant obstacle to the construction of a theoretical framework for the geodynamo. However extra difficulties arise from a lack of knowledge of the strength and the form of the magnetic field in the core of the Earth, which leads to some strong debate about the nature of the important terms in the balance of the momentum equation.

\subsection{\label{gmbalances}Geostrophic and Magnetostrophic Balance}

Consider the the momentum equation~(\ref{NS}), but now in a frame rotating with angular velocity $\bm{\Omega}$ and where the body force $\bm{F}$ arises from buoyancy in a gravity ${\mathbf g}$. In this case the equation takes the form
\begin{eqnarray}
\rho \left(
\dfrac{\partial \mbu}{\partial t}
+ \mbu \cdot \bm{\nabla} \mbu \quad  + 2 \bm{\Omega} \times \mbu \right) = &\quad&- \nabla p \quad \quad +\, \mbj \times \mbB \quad + \rho \nu \nabla^2 \mbu \quad + \quad \rho {\mathbf g}, \\
Inertia\quad \quad \quad Coriolis\quad \quad &&Pressure \quad Lorentz \quad  Viscous \quad Buoyancy. \nonumber
\label{rotNS}
\end{eqnarray}
In the Earth's interior it is well-established that, on timescales of relevance to the dynamo, the inertial and viscous terms do not contribute to the leading order balance of terms in the momentum equation owing to the smallness of the Rossby and Ekman numbers respectively. The Coriolis and Pressure terms certainly do enter into the leading order balance, with the buoyancy force acting to drive the dynamics. The debate surrounds the role of the magnetic field in the balance, since this arises from the dynamo generated field and can not be theoretically determined \textit{a priori}.

The discussion below follows similar lines of argument to those given in \citet{ns2013,ak2017} and see the recent paper of \citet{sga19,aubert:2019}. The importance of the Lorentz force  relative to the Coriolis force can be determined by calculating the ratio of the two, given by
\begin{equation}
\Lambda_{C} = \dfrac{|\bm{j} \times \bm{B}|}{| 2 \rho \Omega \times \bm{u} |} \sim \dfrac{B^2}{2 \mu \rho \ell_b \Omega u},
\label{ratio_geo}
\end{equation}
where $B$ is a typical magnetic field strength and the magnitude of the current density has been estimated as $|\bm{j}|\sim B/ (\mu \ell_B)$\footnote{Here the subscript $C$ is used for `correct' as this measure gives the correct ratio of the relative importance of the Lorentz and Coriolis forces.}. Of course the typical amplitude of $B$ and $u$ is a scale-dependent quantity that emerges from the dynamo calculations, but it is of interest to estimate the typical size of this ratio for the measured field ($\sim 40 \mu T$) and inferred flow ($\ 10^{-3} m s^{-1}$) in the Earth's core. For these values $\Lambda_{C} \sim 0.05 - 0.1$. This would seem to indicate that the Lorentz force, at least at the largest scale, is subdominant to the pressure and the Coriolis terms, and that a so-called leading order \textit{geostrophic balance occurs} where
\begin{eqnarray}
 2 \rho\, \bm{\Omega} \times \mbu \quad \sim &\quad&- \nabla p. \quad \quad \\
\quad \quad \quad Coriolis\quad \quad &&Pressure \nonumber
\label{geostrophic}
\end{eqnarray}

Given the difficulty of estimating the typical amplitudes as a function of spatial scale  of the magnetic fields and flows (and indeed the currents) progress is often made by making one more assumption in the induction equation to link the three. If the fluid is assumed to be at asymptotically low $Rm$ then one may approximate the Lorentz force as
\begin{equation}
\bm{F}_L \sim \frac{1}{\mu}\overline{\bm{B}}\cdot \bm{\nabla}   \bm{b'},
\end{equation}
where $\overline{\bm{B}}$ is the large-scale field and $\bm{b'}$ is the small induced field given by
\begin{equation}
0 \sim \overline{\bm{B}}\cdot \bm{\nabla}   \bm{u} + \eta \nabla^2 \bm{b'}, 
\end{equation}
so that $B \sim \overline{B} u \ell_b/ \eta$.
The attractiveness of this approach becomes clear on realising that, on making the approximation, all the unknown terms disappear in the ratio~(\ref{ratio_geo}) and for this limit the ratio of the Lorentz term to the Coriolis term is given by the Elsasser number
\begin{equation}
\Lambda =  \dfrac{\overline{B}^2}{2 \mu \rho \eta \Omega },
\label{ratio_elsasser}
\end{equation}
Here $\overline{B}$ is the typical size of the large-scale magnetic field and can be measured, whilst all the other quantities are known (or semi-known) \textit{a priori}. $\Lambda$ is known as the \textit{Elsasser number} and is nearly always quoted as \textit{the} measure of the relative importance of the Lorentz to Coriolis force. In the Earth's core $\Lambda \sim 1$. Taken at face value this would place the Earth's core in \textit{magnetostrophic balance} at largest scales, i.e.
\begin{eqnarray}
2 \rho \, \bm{\Omega} \times \mbu \quad \sim &\quad&- \nabla p \quad \quad +\, \mbj \times \mbB \quad, \\
Coriolis\quad \quad &&Pressure \quad Lorentz. \nonumber
\label{magnetostroph}
\end{eqnarray}

Even though estimates based on the modified Elsasser number give the leading order balance as geostrophic at the largest scale, it is clear from equation~(\ref{ratio_geo}) that the importance of the Lorentz force in the momentum equation is scale-dependent. At intermediate or small scales, the precise definition of which depends on the assumed form of the spectrum for the velocity and magnetic field, it is likely that magnetostrophic balance is re-established \citep{ns2013,ak2017}. Of course, because $Pm$ is small, the lengthscale for which $Rm \sim 1$ ($\sim 10^4 - 10^5$m) is much larger than the viscous scale  (and indeed the Rossby scale). For scales smaller than this one enters a diffusive regime with a strong applied field. For scales larger than the Rossby scale the flow is heavily influenced by rotation, whilst for smaller scales the dynamics is prototypical diffusive MHD \citep{ns2013}; the kinetic energy spectrum $E_K(k) \sim k^{-3}$, whilst the magnetic energy spectrum is even steeper $E_M(k) \sim k^{-5}$.


It is worth mentioning here some theoretical arguments in favour of magnetostrophic balance in the geodynamo. Much of these arise from considerations of the linear theory of magnetoconvection, both in a plane layer and spherical shell. For magnetonvection studies the magnetic field is imposed and the critical Rayleigh number $Ra_c$ is calculated as a function of rotation rate and magnetic field strength. These are measured by the non-dimensional input parameters $E = \nu/2 \Omega L^2$, 
$Q =  B_0^2/\mu \rho \nu \eta$ or $\Lambda = Q\,E = B_0^2/(2 \mu \rho \eta \Omega)$. Briefly, as noted in Chandrasekhar's monograph \citep{chan:61}, both rotation and magnetic fields separately act so as to suppress convection, with (in a plane layer) $Ra_c \sim E^{-4/3}$ as $E \rightarrow 0$ so that $Ra_c$ gets large; the associated critical wavenumber $k_c \sim E^{-1/6}$ so that the preferred mode is at small scales. In the non-rotating, magnetised case $Ra_c \sim Q$ as $Q \rightarrow \infty$ so that $Ra_c$ gets large with large field. Here also the associated mode is at small scales so that $k_c \sim Q^{1/6}$. Remarkably when both strong rotation and magnetic fields are present, the preferred mode has $\Lambda \sim 1$ and neither $Ra_c$ nor $k_c$ are asymptotically large. The magnetic field breaks the constraints imposed by rotation to allow efficient convection, there is a sweet spot in parameter space for linear convection to onset.

The existence of this sweet spot has formed the basis of the argument for the adoption of magnetostrophy in dynamo calculations. It is argued that, all other things being equal, the dynamo will try to locate itself in a state in which convection is optimised. This is certainly a plausible argument, though it should be stressed that it is not necessarily the case that convection is optimised in the nonlinear regime for the same parameters that optimise it in the linear regime; recall that investigating the linear regime for magnetoconvection is implicitly a low $Rm$ calculation. Moreover the breaking of the rotational constraint does not need to take place at all scales to allow efficient convection, allowing the possibility that geostrophic balance may proceed at large scales whilst magnetostrophy reigns at smaller scales. 

At this point we note that it now sometimes traditional to designate a system as being in   MAC balance when the primary balance is geostrophic (i.e. Coriolis force balanced by pressure) but the component of the Coriolis force that is not balanced by pressure is balanced by buoyancy and the Lorentz force at the next order, this then takes the form of balancing
\begin{eqnarray}
 2 \rho \,\bm{\Omega} \times \mbu \quad \sim &\quad&- \nabla p \quad \quad \quad + \mbj \times \mbB \quad + \quad \quad \quad \rho {\mathbf g}, \\
Coriolis\quad \quad &&Pressure \quad \,\quad Lorentz  \quad \quad Buoyancy. \nonumber
\label{rot_MAC}
\end{eqnarray}
It is interesting to note that if equation~(\ref{rot_MAC}) holds then the magnetic field generated should be a function of the modified Rayleigh number ($\tilde{Ra} = (\alpha \Delta T L)/\eta \Omega$) only \citep{dop:2018}.

Recently \citet{sga19} have argued that this balance should be termed QG-MAC balance to indicate that the primary balance is geostrophic whilst the Lorentz force enters in at the next order, though here we shall use the more usual terminology.
However, it is fair to say that there is no overwhelming evidence or argument for assuming a particular force balance in the Earth's core.  The investigation of these balances, in the relevant parameter regime of low $Pm$, is a very worthwhile topic of future investigations. Though this is difficult with current computational resources for the dynamo problem, much can be learned by studying balances in the nonlinear rotating magnetoconvection problem \citep[see e.g][]{stellhans:04}.

\subsubsection{\label{Taylor}Consequences of MAC Balance: Taylor's constraint}

Amazingly enforcing MAC balance imposes a serious constraint on the nature of the possible dynamo solutions. Here we derive the simplest constraint on the form of the integrated Lorentz force known as Taylor's constraint.

Taking the $\phi$-component of equation~(\ref{rot_MAC}) and integrating over cylinders aligned with the rotation axis one obtains
\begin{equation}
 \int 2 \rho \,(\bm{\Omega} \times \mbu)_\phi \,dS \sim - \int (\nabla p)_\phi \,dS  + \int (\mbj \times \mbB)_\phi \, dS +  \int (\rho {\mathbf g})_\phi \, dS,
\label{Tay1}
\end{equation}
where $dS = s d \phi dz$. As buoyancy is purely radial the final term is zero and the pressure term integrates to zero. Finally the Coriolis term can also be shown to be zero by using the divergence theorem over the cylinder. Hence the only term to contribute to equation~(\ref{Tay1}) is the Lorentz force and so
\begin{equation}
\int (\mbj \times \mbB)_\phi \, dS = \dfrac{1}{\mu} \int ((\bm{\nabla}   \times \mbB) \times \mbB)_\phi \, dS = 0.
\label{Tay1_worked}
\end{equation}
This Taylor constraint ensures that the Lorentz torque on any cylindrical surface parallel to the rotation axis is zero. 

It is possible to construct dynamos that solve the induction equation in addition to maintaining Taylor's constraint both in two \citep{rw2018} and three \citep{ljl2018} dimensions. This elegant formalism certainly provides a way forward for finding exact solutions that satisfy MAC balance. Numerical computations usually do solve the full momentum equation (albeit in the wrong parameter regime) and utilise the size of the departures from the solution to equation~(\ref{Tay1_worked}) (the degree of `Taylorization') as a measure of the degree to which they have been successful in obtaining solutions that are in MAC balance.


\subsection{Computational Models of the Geodynamo \label{comp_geo}}


For all the reasons described above, computationally modelling the geodynamo remains a formidable venture. The modern era of massively parallel computational geodynamo modelling was started by the seminal work of \citet{gr1995}. They simulated the interaction of rotating, anelastic convection with magnetic fields to produce dynamo models that bore striking resemblance to the observed magnetic field; they even managed to observe reversing magnetic fields. That this tour de force calculation was so successful is surprising given the large discrepancy between the parameters that could be reached computationally and those that pertained to the Earth. With the increase in computational power, a large number of subsequent papers have attempted to refine and improve on the Glatzmaier \& Roberts models. The culmination of these efforts so far are  two recent high resolution models \citep{agf2017,sjnf2017}.

The first of these models \citep{agf2017} uses large-eddy simulations for the momentum equation (though the induction equation is fully resolved). The parameters are selected to get as close as computationally possible to the values in the Earth whilst maintaining a MAC balance and keeping a constant $Rm$. This is achieved by scaling the input parameters with a parameter $\epsilon$ (in this case the convective power) in a consistent manner --- this is similar in spirit to the distinguished limit calculations pioneered by \citet{Dormy2016} discussed below. Here, in the simplest set-up, $Pm \propto \epsilon^{1/2}$, $E\propto \epsilon$, $Pr \propto 1$, $Rm \propto 1$. Here $\epsilon \sim 1$ corresponds to the conditions found successfully to produce MAC balance in a moderately rotating model, whilst $\epsilon$ small is appropriate for the Earth. The impressive numerical calculations are performed for a range of parameters on the path, $\epsilon=1$ corresponds to $E \sim 3 \times 10^{-5}$ and  $\epsilon = 3 \times 10^{-4}$, gives $E = 10^{-8}$. As expected, the solutions take the form of a MAC balanced dynamo field following diffusivity-free power based scaling laws.

\citet{sjnf2017} also consider numerical simulations of dynamo action is a similar parameter regime ($E \sim 10^{-7}$, $Pm \sim 0.1$, $Rm > 500$). They find magnetic fields in MAC balance, with the magnetic
energy  one order of magnitude larger than the kinetic energy. Interestingly they find inhomogeneity of the solution with a dynamical contrast between the
interior and the exterior of the the cylinder parallel to the axis of rotation that
circumscribes the inner core, the so-called tangent cylinder. Inside the tangent cylinder, the strong magnetic field is generated by a   polar vortex.  Outside the tangent cylinder, however, the kinetic energy is mostly non-zonal, with zonal winds being suppressed by the Lorentz force. At different spatial and temporal scales the flow may be either geostrophic (for large-scale and low frequency flows and for small-scale convection in weak field regions) or in MAC balance (for high frequency large-scale modes or convective modes at intermediate scales).

\subsubsection{\label{distinguished}The Search For A Distinguished Limit} 

As noted above, geodynamo models are usually integrated in completely the wrong parameter range, but sometimes yield solutions that resemble the geomagnetic field. When and why is this the case, and how can one continue to achieve solutions in MAC balance as computers get faster? Furthermore, it is tempting to try to compute numerical models with all the parameters set to be as close as possible to their correct geophysical values (in particular low $E$ and low $Pm$). Is this the most sensible approach, given current computational limitations?

An alternative, and more considered approach, was pioneered by \citet{Dormy2016} who studied the properties of a large set of dynamo models at moderate Ekman number (with $E \approx \mathcal{O}(10^{-4}$)). He argued that, at such moderate values amenable to rapid computation, it was necessary to \textit{increase} $Pm$ from its order one value to achieve the correct balance for the generated magnetic field; indeed  it is also of interest to examine rapidly rotating dynamos where inertia is completely suppressed \citep{hc2016} --- in this regime $Pm$ is infinite. Of course, when $E$ is reduced, as computational resources become more lavish, one should move away from these models and $Pm$ should be reduced in tandem with the Ekman number. 
Dormy argued that a distinguished limit should be found where all parameters, including $Pm$ scale with the Ekman number as $E \rightarrow 0$ to leave one on the balanced dynamo branch. This is also the approach taken by \citet{agf2017}, as described above, though the scalings selected for the variation of the parameters are slightly different.

\subsection{Asymptotic Models of the Geodynamo
\label{red_geo}}

An alternative approach to performing DNS is to construct asymptotic models valid in the regime of  strong rotation (i.e. asymptotically small Rossby number $Ro$). This has much in common with the approach advocated by \citet{Dormy2016,agf2017}. The difference here is that the models are predicated on the fact that the separation of timescales discussed earlier leads naturally to  reduced dynamics. There are two reasons why one might seek an asymptotic solution to a complicated problem. The first is that  performing an asymptotic expansion often leads to the development of a model where analytic progress is possible and an elegant solution emerges; there are many examples of this use of such models. The second reason is that it is \textit{the right thing to do} for a given problem; performing the expansion does not lead to a reduction in the complexity of the problem (for example a three-dimensional PDE may stay a three-dimensional PDE) but performing the expansion does, by design, lead to a set of reduced equations that no longer contain the small parameters, which is therefore more amenable to solution\footnote{Such models have sometimes been viewed with suspicion, ironically sometimes even by researchers happy to utilise the Boussinesq approximation as an asymptotic framework for modelling}. 

The models for the geodynamo summarised here are of the second kind. The first multi-scale asymptotic models were constructed by \citet{chilsow:1972}, who performed a weakly nonlinear analysis of Boussinesq convective dynamos in a plane layer; their analysis showed how weak large-scale magnetic fields were readily generated. The Childress-Soward model has been developed, first by \citet{fauchil1982} who extended the analysis to intermediate field strengths, and more  recently to include the effects of multiple length scales perpendicular to the rotation axis to enable the description of strongly forced convection. In a series of papers \citep[see][and the references therein]{cjta2015,cjt:2017,calkins2018,pcjt2018} the asymptotic theory for asymptotic modelling of Quasigeostrophic dynamo models (QGDM) was developed. 

\begin{figure}
 \centering
 \includegraphics[width=0.9\textwidth]{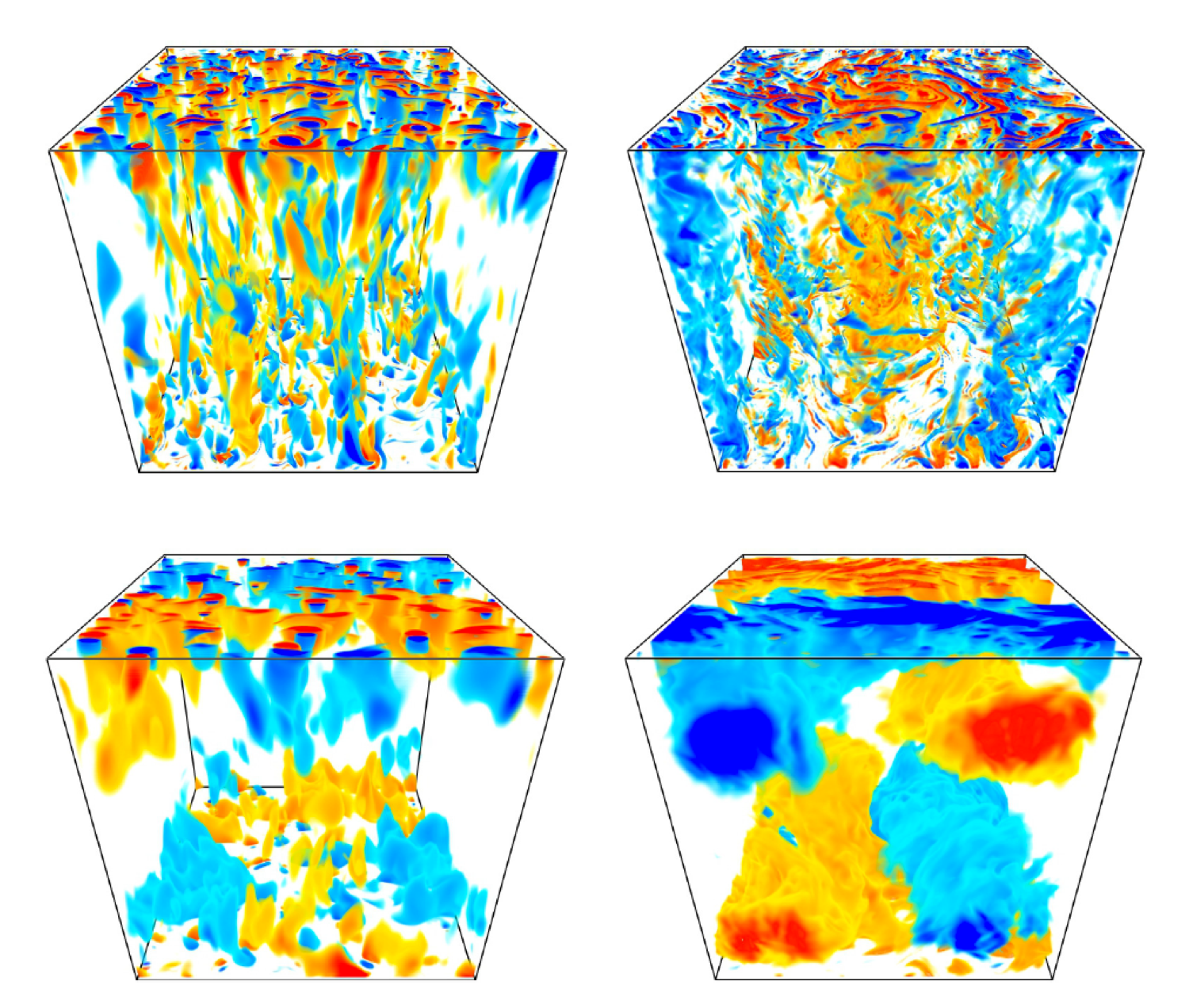}
  \caption{(Color online) Volumetric renderings of the small-scale vertical vorticity (top panels) and current density (bottom panels)  illustrating two flow regimes observed in the reduced convection simulations, namely the plume regime (left panels) and the geostrophic turbulence regime (right panels).}
  \label{fig:reduced}
 \end{figure} 

Briefly, these models  utilise the separation of timescales to construct an asymptotic representation of the dynamics. The asymptotic parameter is  related to the Rossby number by setting $\epsilon = Ro = E^{1/3}$. The QGDM is then allowed to  evolve on three timescales, a fast convective timescale $t$, an intermediate magnetic timescale $\tau = \epsilon^{3/2} t$ and a slow thermal timescale $T = \epsilon^2 t$. Note that these timescales are in the correct ordering as described in section~\ref{geo_time}. In the most realistic of these models both $\tilde{Ra} = \epsilon^{4}Ra$ and $\tilde{Pm} = \epsilon^{1/2} Pm$ are also scaled. Importantly, the magnetic field is also scaled so that magnetic energy is much larger than the kinetic energy, which is physically realistic. In addition to the separation in timescales, multiple spatial scales both perpendicular and parallel to the rotation axis are  identified. The method of multiple scales is used to derive three-dimensional reduced PDEs for large-scale and small-scale quantities. 

The dynamics is such that the QGDM is geostrophic to leading order, with the strong magnetic field entering into a prognostic equation that determines the dynamics.

This dynamics in plane-layer models is beginning to be investigated in both the kinematic and nonlinear dynamo regimes. 
Hydrodynamically, QG convection in a plane layer passes through  four typical configurations, (a) cellular convection, (b) columnar convection, (c) plume convection and (d) geostrophic turbulence as the Rayleigh number increases.
Figure~\ref{fig:reduced} shows the vorticity for  the final two such asymptotic solutions in the kinematic regime, and the small-scale current generated by the corresponding kinematic dynamo. Interestingly, all the convection regimes act as large-scale dynamos, with the properties of the large-scale dynamo remaining  relatively insensitive to the precise form of the convection. These models have recently been extended into the nonlinear regime by \citet{pcjt2018}.

It is to be hoped that these models will be able to give some insight as to the behaviour of rapidly rotating planetary dynamos at low Rossby and magnetic Prandtl number. Of crucial importance is to extend these models to more realistic geometries such as spherical shells and thermal annuli, and to investigate the dynamics of reduced models with other dynamic balances.

\section{\label{expts}A word on dynamo experiments}

Although this perspective is concerned with the \textit{theory} of turbulent dynamos, it would be wrong to conclude without some discussion of the amazing dynamo experiments that have been performed over the past thirty or so years. Primarily these have been constructed so as to address the question of the nature of dynamo action near to onset, i.e.\ for $Rm$ close to $Rm_c$. 

A typical laboratory dynamo is usually designed  to drive a large-scale flow with desirable laminar 
dynamo properties; i.e. they are usually optimised so that $Rm_c$ is as small as possible. For example, the Riga dynamo experiment \citep[see e.g.][]{getal2002}, is designed to mimic the Ponomarenko dynamo described in section~\ref{kinematic}, whilst the Karlsruhe experiment \citep[see e.g.][]{stmu:2001} was largely based on the Roberts dynamo of the same section. These constrained designs were optimised and dynamo action, of the type predicted by computation, was found.

All of the experimental dynamos utilise liquid metals for which, $Pm \sim {\mathcal O} (10^{-6} - 
10^{-5}) \ll 1$, as discussed earlier. Hence even though
 $Rm_{c}$ is moderate.
 the corresponding Reynolds numbers are vast. 
 This has a significant impact on the experimental search for dynamo action in less constrained flows such as the VKS sodium experiment \citep{monchauxetal:2007}. There the flow of liquid sodium was driven in a cylindrical container by rotating propellors. The desired large-scale ``French washing machine'' flow is known from computation to be a decent dynamo. However, because $Re$ is large, this flow is unstable and turbulent fluctuations are found that are comparable in magnitude with the mean flow. As discussed in section~\ref{lowpmkin}, this has a negative effect on the dynamo efficiency of the experiment, increasing $Rm_c$ and rendering dynamo action impossible unless magnetic material is used for the driving disks. Nonetheless, once this material is used the system undergoes a bewildering array of nonlinear dynamics, including the generation and reversals of a large-scale field. Whether or not the system can be classed as a dynamo in the strictest sense, much can be learned of the nonlinear behaviour of large scale magnetic fields in turbulent systems.
 
 It is important to also mention the ``whirling dervish" experiments of the Grenoble group \citep[see e.g.][]{whirling2011}. This liquid sodium experiment investigates magnetised spherical couette flow, with a strongly magnetised inner core. Although the field is imposed, so that this is not strictly a dynamo experiment, much has been learned about the possible balances in rapidly rotating MHD systems from this experiment. 
 
 Finally, and extremely impressively, turbulent transport coefficients have been calculated in liquid metal experiments in a sphere by the Madison group \citep{cary:2012}. Although of course these experiments are perforce at low $Rm$, they give a crucial insight into the behaviour of driven low $Pm$ fluids.

\section{\label{Future}Discussion: What of the future?}
\begin{center} \textit{``I may not have gone where I intended to go, but I think I have ended up where I intended to be."} \linebreak
Douglas Adams
\end{center}

The previous discussion has emphasised that the current status of the theory of turbulent dynamos is uncertain, both for the geodynamo and for its astrophysical counterparts.
Despite the heroic efforts to solve the relevant equations numerically, computational efforts are limited to parameter regimes that are far from those that pertain to naturally occuring dynamos. Perhaps this is best emphasised by a recent calculation by Petri K\"apyl\"a (Private Communication). He demonstrates that, even if the computational resources were available to undertake numerical simulation of turbulent dynamos, the power required to simulate a star such as the Sun is  $10^{22}$W. This is equivalent to the luminosity of a M9V main sequence red dwarf. Although this is a fairly cool star, the point is well made!

The lack of (computational) power for numerical simulations manifests itself as a different type of deficiency for geodynamo models from astrophysical dynamo models. For astrophysical models, the key point is that the flows that can be simulated are too laminar and too diffusive and there is not enough separation of scales between the resistive and viscous scales. Hence, for dynamo theorists even solving the induction equation is a formidable task, which is beyond the computational power of the near future.
For planetary dynamos, such as the geodynamo, the numerical solutions are too viscous and too slowly rotating (although the low $Pm$ problem is also manifest here). It is very difficult to achieve solutions that are in the relevant asymptotic regime of geostrophy or magnetostrophy.

So what is the correct way to proceed? For planetary dynamos, since the viscosity is known to play a minor role, it is imperative to examine models that do not rely on viscosity, such as those that are driven by waves \citep[see e.g.][]{dr2018} or those that saturate via the enforcement of Taylor's Constraint --- or a modified Taylor's Constraint that includes the temporal variation of the zonal flow --- \citep[see e.g.][]{wurob2015,ljl2018}. Another promising avenue to pursue is to determine a distinguished limit (or a path through parameter space) that maintains the correct balances, even though the computational resources are not adequate currently to simulate the vast range of temporal scales; this is the approach take by \citet{Dormy2016,agf2017}. Furthermore, given the vast separation in spatial and temporal scales for the Earth's interior, a programme based on the derivation and solution of asymptotic models is warranted \citep{calkins2018}. This programme will require the development of a new class of timestepping methods that are capable of efficiently integrating reduced equations that evolve on very different timescales (i.e.\ slow-fast dynamics). 
One issue that remains puzzling has to be the mechanism that leads to the reversals of the Earth's field. Presumably such an abrupt transition relies on nonlinear effects \citep[see e.g.][]{JONES:2008}. One cannot rely on inertia, which is small, to be the source of nonlinearity, so that leaves potentially the Lorentz force or the temperature advection and there are constraints on the action of both of these. 

For the Sun and other astrophysical bodies, the problem remains the lack of techniques to deal with flows on a vast range of spatial scales. Clearly any numerical code is capable of only resolving an extremely limited range of these scales, and some procedure must be developed for the parameterisation of the low-order statistics of the scales below the smallest that can be resolved. As noted above, such sub-grid models are extremely difficult to construct for magnetohydrodynamic turbulence \citep{miesch:2015}. Moreover, we have stressed here that it is key that  such parameterisations (even if couched in terms of the construction of turbulent transport coefficients) respect the inherent conservation laws of the original system in the relevant limit. As we have shown, for example, conservation of magnetic helicity plays a vital part in constraining the dynamo solutions. One possible avenue is to utilise \textit{Direct Statistical Simulation (DSS)}  via cumulant expansions \citep[see e.g.][]{mqt2019}. 

This approach involves the direct solution of differential equations derived for the low-order statistics of fluid flows (such as mean flows and two-point correlation functions).  In this methodology the equations take the form of a truncated (equal-time) cumulant expansion that respects the inhomogeneity and anisotropy of the underlying dynamo system.  This approach is therefore a generalisation of the methods discussed in section~\ref{closures} to anisotropic and inhomogeneous systems. Because of the increased complexity, a computational approach is required to facilitate this approach \citep[see e.g.][]{tdm2011}.

Briefly, the method proceeds by considering the dynamo system as a dynamical system and via deriving evolution equations for the cumulants of the probability distribution function of the turbulent state. This is much less computationally expensive than solving the Fokker-Planck equation for the whole pdf though that approach may also be worth pursuing \citep[see e.g.][]{venturi2018}. Briefly, consider a dynamical system where the state variable $\boldsymbol{q}(\boldsymbol{x},t)$ (in the case of dynamos this may include information about the velocity, pressure, magnetic field, temperature etc of the fluid) evolves via
\begin{equation}
\boldsymbol{q}_t = \mathcal{L}(\boldsymbol{q}) + \mathcal{N}(\boldsymbol{q,q}) + \boldsymbol{f}(\boldsymbol{x},t).
\end{equation}
Here $\mathcal{L}$ is a linear operator, whilst $\mathcal{N}$ is a nonlinear (in the simplest case quadratic) operator and $\boldsymbol{f}(\boldsymbol{x},t)$ is an (often stochastic) driving term.

The method proceeds by defining the low order cumulants which, up to third order, are given by
\begin{eqnarray}
    c_1(\boldsymbol{r}) &=& \langle \boldsymbol{q} \rangle, \\
    c_2(\boldsymbol{r}_1,\boldsymbol{r}_2) &=& \langle \boldsymbol{q}'(\boldsymbol{r}_1)\, \boldsymbol{q}'(\boldsymbol{r}_2) \rangle, \\
    c_3(\boldsymbol{r}_1,\boldsymbol{r}_2,\boldsymbol{r}_3) &=& \langle \boldsymbol{q}'(\boldsymbol{r}_1)\, \boldsymbol{q}'(\boldsymbol{r}_2)\, \boldsymbol{q}'(\boldsymbol{r}_3)\rangle,
\end{eqnarray}
so that to this order the first cumulant is given by the mean. Here, as before, the averaging process may be taken over a co-ordinate or be defined as an ensemble mean. The second and third cumulants are the centred moments (or two and three point correlation functions).

Equations for the evolution of the cumulants may be derived either via brute force (potentially symbolic manipulators) or via the Hopf functional technique. Schematically these take the form of a \textit{cumulant hierarchy}
\begin{eqnarray}
\dfrac{\partial {c_1}}{\partial t} &=& \mathcal{L}(c_1) + \mathcal{N}(c_1^2 +c_2), \\
\dfrac{\partial {c_2}}{\partial t} &=& \mathcal{L}(c_2) + \mathcal{N}(c_1 c_2 +c_3) + \Gamma, \\
\dfrac{\partial {c_3}}{\partial t} &=& \mathcal{L}(c_3) + \mathcal{N}(c_1 c_3 + c_2 c_2 + c_4).
\end{eqnarray}
Here $\Gamma$ encodes the net action of the stochastic driving on the second cumulant \citep[see e.g.][for more details]{tdm2011}. As for the models discussed in section~\ref{closures}, progress can only be made by truncating the hierarchy. Usually this is done by setting $c_3 = 0$ (in which case this approach is known as CE2\footnote{CE2 is formally equivalent to the stochastic structural stability theory (SSST) of \citet{fi2003}}) or by discarding $c_4$ and either including an eddy-damping term or by ensuring realisability by projecting out the eigenvectors associated with the unphysical negative eigenvalues of the second cumulant \citep{mqt2019}. In the dynamo (and magnetorotational instability) context CE2 has been utilised in the landmark paper of \citet{sb2016}.

Various classes of DSS, relying on different statistical formulations for the interactions, have been derived \citep{mqt2019}. DSS provides a self-consistent evolution of the large and small-scales, respecting the conservation laws of the original system. As we have noted, conservation of magnetic helicity plays a key role in the nonlinear saturation of dynamos and so deriving statistical models that respect this (and other) quadratic conservation laws is important.
Current research into DSS includes the use of reduced bases to increase computational efficiency and of machine learning methods into basis selection. The efficacy of the various truncations of DSS may also be evaluated with relation to the resolvent analysis of \citet{ms2010}.

Finally for statistical models and methods, it is important to derive a firm theoretical foundation on which to build algorithms and computational models. As for pure turbulence and climate research, where statistical models are more widely utilised, there needs to be a program of research that attempts to answer the questions as to whether  it possible to construct a non-equilibrium theory that describes the interaction of mean flows and magnetic fields with turbulence in inhomogeneous, anisotropic dissipative systems (such as dynamo sytems). For example, is there a well-defined procedure for connecting fluctuations to dissipation in turbulent magnetised flows that are not homogeneous and isotropic (and may be stratified and rotating)? These will probably take the form of generalised fluctuation-dissipation theorems; progress in this area could lead to the development of meaningful statistical theories for the turbulent generation of magnetic field.

Whatever the successful approach turns out to be, I believe it will not emerge from a blind application of computing power. However that is not to say that there is no role for direct numerical solution of dynamo systems. Such an approach should be regarded as the construction of silicon-based thought experiments, rather than the direct modelling of the astrophysical objects. Much can be learned from the investigation of such models; though, given the restrictions and the sensitivity of the dynamo system, extreme care must be taken in extrapolating conclusions to the relevant regimes for geophysics or asrophysics.

\section*{Acknowledgements}
I would like to thank Emmanuel Dormy and Jeff Oishi for a critical reading of the manuscript and extremely helpful comments.
I would like to acknowledge support of  funding from the European Research Council (ERC) under the European Union’s Horizon 2020 research and innovation programme (grant agreement no. D5S-DLV-786780).

%
\bibliographystyle{jfm}
\bibliography{References_all}

\begin{thebibliography}{191}
\expandafter\ifx\csname natexlab\endcsname\relax\def\natexlab#1{#1}\fi

\bibitem[{Arlt} \& {Weiss}(2015)]{aw:2015}
{\sc {Arlt}, R. \& {Weiss}, N.} 2015 {\em {Solar Activity in the Past and the
  Chaotic Behaviour of the Dynamo}\/}, p. 525.

\bibitem[{Arnold} \& {Korkina}(1983)]{arnkork:1983}
{\sc {Arnold}, V.~I. \& {Korkina}, E.~I.} 1983 {The growth of a magnetic field
  in the three-dimensional steady flow of an incompressible fluid}. {\em
  Moskovskii Universitet Vestnik Seriia Matematika Mekhanika\/} pp. 43--46.

\bibitem[{Arponen} \& {Horvai}(2007)]{arphorvai:2007}
{\sc {Arponen}, H. \& {Horvai}, P.} 2007 {Dynamo Effect in the Kraichnan
  Magnetohydrodynamic Turbulence}. {\em Journal of Statistical Physics\/} {\bf
  129}, 205--239.

\bibitem[{Aubert}(2019)]{aubert:2019}
{\sc {Aubert}, J.} 2019 {Approaching Earth's core conditions in high-resolution
  geodynamo simulations}. {\em arXiv e-prints\/} p. arXiv:1905.06049.

\bibitem[{Aubert} {\em et~al.\/}(2017){Aubert}, {Gastine} \&
  {Fournier}]{agf2017}
{\sc {Aubert}, Julien, {Gastine}, Thomas \& {Fournier}, Alexandre} 2017
  {Spherical convective dynamos in the rapidly rotating asymptotic regime}.
  {\em Journal of Fluid Mechanics\/} {\bf 813}, 558--593.

\bibitem[{Aubert} {\em et~al.\/}(2010){Aubert}, {Tarduno} \&
  {Johnson}]{aubertetal:2010}
{\sc {Aubert}, J., {Tarduno}, J.~A. \& {Johnson}, C.~L.} 2010 {Observations and
  Models of the Long-Term Evolution of Earth's Magnetic Field}. {\em Space
  Science Reviews\/} {\bf 155}, 337--370.

\bibitem[{Aurnou} \& {King}(2017)]{ak2017}
{\sc {Aurnou}, J.~M. \& {King}, E.~M.} 2017 {The cross-over to magnetostrophic
  convection in planetary dynamo systems}. {\em Proceedings of the Royal
  Society of London Series A\/} {\bf 473}, 20160731.

\bibitem[Backus(1958)]{backus:1958}
{\sc Backus, George} 1958 A class of self-sustaining dissipative spherical
  dynamos. {\em Annals of Physics\/} {\bf 4}~(4), 372 -- 447.

\bibitem[{Baliunas} {\em et~al.\/}(1995){Baliunas}, {Donahue}, {Soon}, {Horne},
  {Frazer}, {Woodard-Eklund}, {Bradford}, {Rao}, {Wilson}, {Zhang}, {Bennett},
  {Briggs}, {Carroll}, {Duncan}, {Figueroa}, {Lanning}, {Misch}, {Mueller},
  {Noyes}, {Poppe}, {Porter}, {Robinson}, {Russell}, {Shelton}, {Soyumer},
  {Vaughan} \& {Whitney}]{baliunasetal:1995}
{\sc {Baliunas}, S.~L., {Donahue}, R.~A., {Soon}, W.~H., {Horne}, J.~H.,
  {Frazer}, J., {Woodard-Eklund}, L., {Bradford}, M., {Rao}, L.~M., {Wilson},
  O.~C., {Zhang}, Q., {Bennett}, W., {Briggs}, J., {Carroll}, S.~M., {Duncan},
  D.~K., {Figueroa}, D., {Lanning}, H.~H., {Misch}, T., {Mueller}, J., {Noyes},
  R.~W., {Poppe}, D., {Porter}, A.~C., {Robinson}, C.~R., {Russell}, J.,
  {Shelton}, J.~C., {Soyumer}, T., {Vaughan}, A.~H. \& {Whitney}, J.~H.} 1995
  {Chromospheric variations in main-sequence stars}. {\em Astrophysical
  Journal\/} {\bf 438}, 269--287.

\bibitem[{Barkley}(2016)]{barkley:2016}
{\sc {Barkley}, D.} 2016 {Theoretical perspective on the route to turbulence in
  a pipe}. {\em Journal of Fluid Mechanics\/} {\bf 803}, P1.

\bibitem[{Bassom} \& {Gilbert}(1997)]{bassom:1997}
{\sc {Bassom}, A.~P. \& {Gilbert}, A.~D.} 1997 {Nonlinear equilibration of a
  dynamo in a smooth helical flow}. {\em J. Fluid Mech.\/} {\bf 343}, 375--406.

\bibitem[{Batchelor}(1959)]{batchelor:1959}
{\sc {Batchelor}, G.~K.} 1959 {Small-scale variation of convected quantities
  like temperature in turbulent fluid. Part 1. General discussion and the case
  of small conductivity}. {\em Journal of Fluid Mechanics\/} {\bf 5}, 113--133.

\bibitem[{Beck}(2015)]{beck:2015}
{\sc {Beck}, R.} 2015 {Magnetic Fields in Galaxies}. In {\em Magnetic Fields in
  Diffuse Media\/} (ed. A.~{Lazarian}, E.~M. {de Gouveia Dal Pino} \&
  C.~{Melioli}), {\em Astrophysics and Space Science Library\/}, vol. 407, p.
  507.

\bibitem[{Beer} {\em et~al.\/}(1998){Beer}, {Tobias} \& {Weiss}]{btw:1998}
{\sc {Beer}, J., {Tobias}, S. \& {Weiss}, N.} 1998 {An Active Sun Throughout
  the Maunder Minimum}. {\em Solar Physics\/} {\bf 181}, 237--249.

\bibitem[{Beer} {\em et~al.\/}(2018){Beer}, {Tobias} \& {Weiss}]{btw:2018}
{\sc {Beer}, J., {Tobias}, S.~M. \& {Weiss}, N.~O.} 2018 {On long-term
  modulation of the Sun's magnetic cycle}. {\em Monthly Notices of the Royal
  Astronomical Society\/} {\bf 473}, 1596--1602.

\bibitem[{Berger} \& {Rosner}(1995)]{bergerros:1995}
{\sc {Berger}, M. \& {Rosner}, R.} 1995 {The evolution of helicity in the
  presence of turbulence}. {\em Geophysical and Astrophysical Fluid Dynamics\/}
  {\bf 81}, 73--99.

\bibitem[{Biskamp}(2003)]{biskamp:2003}
{\sc {Biskamp}, D.} 2003 {\em {Magnetohydrodynamic Turbulence}\/}.

\bibitem[{Blackman} \& {Brandenburg}(2002)]{bb2002}
{\sc {Blackman}, E.~G. \& {Brandenburg}, A.} 2002 {Dynamic Nonlinearity in
  Large-Scale Dynamos with Shear}. {\em \apj\/} {\bf 579}, 359--373.

\bibitem[{Blackman} \& {Field}(2000)]{bf2000}
{\sc {Blackman}, E.~G. \& {Field}, G.~B.} 2000 {Constraints on the Magnitude of
  {$\alpha$} in Dynamo Theory}. {\em \apj\/} {\bf 534}, 984--988.

\bibitem[{Boldyrev}(2001)]{bol:2001}
{\sc {Boldyrev}, S.} 2001 {A Solvable Model for Nonlinear Mean Field Dynamo}.
  {\em \apj\/} {\bf 562}, 1081--1085.

\bibitem[{Boldyrev} \& {Cattaneo}(2004)]{bolcatt:2004}
{\sc {Boldyrev}, S. \& {Cattaneo}, F.} 2004 {Magnetic-Field Generation in
  Kolmogorov Turbulence}. {\em Physical Review Letters\/} {\bf 92}~(14),
  144501--+.

\bibitem[{Boldyrev} {\em et~al.\/}(2005){Boldyrev}, {Cattaneo} \&
  {Rosner}]{bolcattros:2005}
{\sc {Boldyrev}, S., {Cattaneo}, F. \& {Rosner}, R.} 2005 {Magnetic-Field
  Generation in Helical Turbulence}. {\em Physical Review Letters\/} {\bf
  95}~(25), 255001--+.

\bibitem[Bouya \& Dormy(2015)]{bouyadormy}
{\sc Bouya, Isma\"el \& Dormy, Emmanuel} 2015 Toward an asymptotic behaviour of
  the abc dynamo. {\em EPL (Europhysics Letters)\/} {\bf 110}~(1), 14003.

\bibitem[{Braginskii}(1975)]{bra:1975}
{\sc {Braginskii}, S.~I.} 1975 {Nearly axially symmetric model of the
  hydromagnetic dynamo of the earth. I}. {\em Geomagnetism and Aeronomy\/} {\bf
  15}, 149--156.

\bibitem[{Brandenburg}(2018)]{brandenburg:2018}
{\sc {Brandenburg}, Axel} 2018 {Advances in mean-field dynamo theory and
  applications to astrophysical turbulence}. {\em Journal of Plasma Physics\/}
  {\bf 84}, 735840404.

\bibitem[{Brandenburg} \& {Subramanian}(2005)]{bransub:2005}
{\sc {Brandenburg}, A. \& {Subramanian}, K.} 2005 {Astrophysical magnetic
  fields and nonlinear dynamo theory}. {\em Physics Reports\/} {\bf 417},
  1--209.

\bibitem[Brito {\em et~al.\/}(2011)Brito, Alboussi\`ere, Cardin, Gagni\`ere,
  Jault, La~Rizza, Masson, Nataf \& Schmitt]{whirling2011}
{\sc Brito, D., Alboussi\`ere, T., Cardin, P., Gagni\`ere, N., Jault, D.,
  La~Rizza, P., Masson, J.-P., Nataf, H.-C. \& Schmitt, D.} 2011 Zonal shear
  and super-rotation in a magnetized spherical couette-flow experiment. {\em
  Phys. Rev. E\/} {\bf 83}, 066310.

\bibitem[{Brummell} {\em et~al.\/}(2001){Brummell}, {Cattaneo} \&
  {Tobias}]{BrummCattTob:2001}
{\sc {Brummell}, N.~H., {Cattaneo}, F. \& {Tobias}, S.~M.} 2001 {Linear and
  nonlinear dynamo properties of time-dependent ABC flows}. {\em Fluid Dyn.
  Res.\/} {\bf 28}, 237--265.

\bibitem[{Brun} \& {Browning}(2017)]{brunbrowning:2017}
{\sc {Brun}, A.~S. \& {Browning}, M.~K.} 2017 {Magnetism, dynamo action and the
  solar-stellar connection}. {\em Living Reviews in Solar Physics\/} {\bf 14},
  4.

\bibitem[B{\"u}hler(2014)]{buhler:2014}
{\sc B{\"u}hler, O.} 2014 {\em Waves and Mean Flows\/}. Cambridge University
  Press.

\bibitem[{Bullard} \& {Gellman}(1954)]{bg:1954}
{\sc {Bullard}, E. \& {Gellman}, H.} 1954 Homogeneous dynamos and terrestrial
  magnetism. {\em Philosophical Transactions of the Royal Society of London A:
  Mathematical, Physical and Engineering Sciences\/} {\bf 247}~(928), 213--278.

\bibitem[{Calkins}(2018)]{calkins2018}
{\sc {Calkins}, Michael~A.} 2018 {Quasi-geostrophic dynamo theory}. {\em
  Physics of the Earth and Planetary Interiors\/} {\bf 276}, 182--189.

\bibitem[{Calkins} {\em et~al.\/}(2017){Calkins}, {Julien} \&
  {Tobias}]{cjt:2017}
{\sc {Calkins}, Michael~A., {Julien}, Keith \& {Tobias}, Steven~M.} 2017
  {Inertia-less convectively-driven dynamo models in the limit of low Rossby
  number and large Prandtl number}. {\em Physics of the Earth and Planetary
  Interiors\/} {\bf 266}, 54--59.

\bibitem[{Calkins} {\em et~al.\/}(2015){Calkins}, {Julien}, {Tobias} \&
  {Aurnou}]{cjta2015}
{\sc {Calkins}, Michael~A., {Julien}, Keith, {Tobias}, Steven~M. \& {Aurnou},
  Jonathan~M.} 2015 {A multiscale dynamo model driven by quasi-geostrophic
  convection}. {\em Journal of Fluid Mechanics\/} {\bf 780}, 143--166.

\bibitem[{Cattaneo}(1999)]{cattaneo:1999}
{\sc {Cattaneo}, F.} 1999 {On the origin of magnetic fields in the quiet
  Photosphere}. {\em Astrophys. J. lett.\/} {\bf 515}, L39--L42.

\bibitem[{Cattaneo} \& {Hughes}(1996)]{CattHughes:1996}
{\sc {Cattaneo}, F. \& {Hughes}, D.~W.} 1996 {Nonlinear saturation of the
  turbulent \&alpha; effect}. {\em Physical Review E\/} {\bf 54}, 4532--.

\bibitem[{Cattaneo} \& {Hughes}(2006)]{catthughes:06}
{\sc {Cattaneo}, F. \& {Hughes}, D.~W.} 2006 {Dynamo action in a rotating
  convective layer}. {\em Journal of Fluid Mechanics\/} {\bf 553}, 401--418.

\bibitem[Cattaneo \& Hughes(2009)]{ch:2009}
{\sc Cattaneo, F. \& Hughes, D.~W.} 2009 Problems with kinematic mean field
  electrodynamics at high magnetic reynolds numbers. {\em Monthly Notices of
  the Royal Astronomical Society: Letters\/} {\bf 395}~(1), L48--L51.

\bibitem[{Cattaneo} \& {Tobias}(2005)]{ct2005}
{\sc {Cattaneo}, F. \& {Tobias}, S.~M.} 2005 {Interaction between dynamos at
  different scales}. {\em Physics of Fluids\/} {\bf 17}~(12),
  127105--127105--6.

\bibitem[Cattaneo \& Tobias(2009)]{ct2009}
{\sc Cattaneo, Fausto \& Tobias, Steven~M.} 2009 Dynamo properties of the
  turbulent velocity field of a saturated dynamo. {\em Journal of Fluid
  Mechanics\/} {\bf 621}, 205–214.

\bibitem[{Cattaneo} \& {Tobias}(2014)]{ct:2014}
{\sc {Cattaneo}, F. \& {Tobias}, S.~M.} 2014 {On Large-scale Dynamo Action at
  High Magnetic Reynolds Number}. {\em \apj\/} {\bf 789}, 70.

\bibitem[{Cattaneo} \& {Vainshtein}(1991)]{cv1991}
{\sc {Cattaneo}, Fausto \& {Vainshtein}, Samuel~I.} 1991 {Suppression of
  Turbulent Transport by a Weak Magnetic Field}. {\em Astrophysical Journal\/}
  {\bf 376}, L21.

\bibitem[{Chandrasekhar}(1961)]{chan:61}
{\sc {Chandrasekhar}, S.} 1961 {\em {Hydrodynamic and hydromagnetic
  stability}\/}. International Series of Monographs on Physics, Oxford:
  Clarendon, 1961.

\bibitem[{Chertkov} {\em et~al.\/}(1999){Chertkov}, {Falkovich}, {Kolokolov} \&
  {Vergassola}]{chertetal:1999}
{\sc {Chertkov}, M., {Falkovich}, G., {Kolokolov}, I. \& {Vergassola}, M.} 1999
  {Small-Scale Turbulent Dynamo}. {\em Physical Review Letters\/} {\bf 83},
  4065--4068.

\bibitem[Childress(1969)]{childress:1969}
{\sc Childress, S.} 1969 {\em Théorie Magnétohydrodynamique de l'Effet
  Dynamo\/}. Institut Henri Poincar\'e.

\bibitem[Childress \& Gilbert(1995)]{chilgil:1995}
{\sc Childress, S. \& Gilbert, A.~D.} 1995 {\em Stretch, Twist, Fold: The Fast
  Dynamo\/}. Berlin: Springer.

\bibitem[{Childress} \& {Soward}(1972)]{chilsow:1972}
{\sc {Childress}, S. \& {Soward}, A.~M.} 1972 {Convection-Driven Hydromagnetic
  Dynamo}. {\em Physical Review Letters\/} {\bf 29}, 837--839.

\bibitem[{Christensen} {\em et~al.\/}(1999){Christensen}, {Olson} \&
  {Glatzmaier}]{christensenetal:1999}
{\sc {Christensen}, U., {Olson}, P. \& {Glatzmaier}, G.~A.} 1999 {Numerical
  modelling of the geodynamo: a systematic parameter study}. {\em Geophysical
  Journal International\/} {\bf 138}, 393--409.

\bibitem[{Cline} {\em et~al.\/}(2003){Cline}, {Brummell} \&
  {Cattaneo}]{cbc2003}
{\sc {Cline}, Kelly~S., {Brummell}, Nicholas~H. \& {Cattaneo}, Fausto} 2003
  {Dynamo Action Driven by Shear and Magnetic Buoyancy}. {\em \apj\/} {\bf
  599}, 1449--1468.

\bibitem[Connerney {\em et~al.\/}(2018)Connerney, Kotsiaros, Oliversen, Espley,
  Joergensen, Joergensen, Merayo, Herceg, Bloxham, Moore, Bolton \&
  Levin]{Cetal:2018}
{\sc Connerney, J. E.~P., Kotsiaros, S., Oliversen, R.~J., Espley, J.~R.,
  Joergensen, J.~L., Joergensen, P.~S., Merayo, J. M.~G., Herceg, M., Bloxham,
  J., Moore, K.~M., Bolton, S.~J. \& Levin, S.~M.} 2018 A new model of
  jupiter's magnetic field from juno's first nine orbits. {\em Geophysical
  Research Letters\/} {\bf 45}~(6), 2590--2596.

\bibitem[{Courvoisier} \& {Kim}(2009)]{ck:2009}
{\sc {Courvoisier}, A. \& {Kim}, E.-J.} 2009 {Kinematic {$\alpha$} effect in
  the presence of a large-scale motion}. {\em Phys. Rev. E.\/} {\bf 80}~(4),
  046308.

\bibitem[{Cowling}(1933)]{cowling:1933}
{\sc {Cowling}, T.~G.} 1933 {The magnetic field of sunspots}. {\em Mon. Not.
  Roy. Ast. Soc.\/} {\bf 94}, 39--48.

\bibitem[{Cowling}(1978)]{cowling:1978}
{\sc {Cowling}, T.~G.} 1978 {\em {Magnetohydrodynamics.}\/}.

\bibitem[{Davidson} \& {Ranjan}(2018)]{dr2018}
{\sc {Davidson}, P.~A. \& {Ranjan}, A.} 2018 {Are planetary dynamos driven by
  helical waves?} {\em Journal of Plasma Physics\/} {\bf 84}, 735840304.

\bibitem[{Del Sordo} {\em et~al.\/}(2013){Del Sordo}, {Guerrero} \&
  {Brandenburg}]{dsgb:2013}
{\sc {Del Sordo}, F., {Guerrero}, G. \& {Brandenburg}, A.} 2013 {Turbulent
  dynamos with advective magnetic helicity flux}. {\em \mnras\/} {\bf 429},
  1686--1694.

\bibitem[{Diamond} {\em et~al.\/}(2005){Diamond}, {Hughes} \& {Kim}]{dhk2005}
{\sc {Diamond}, P.~H., {Hughes}, D.~W. \& {Kim}, Eun-Jin} 2005 {Self-consistent
  mean field electrodynamics in two and three dimensions}. In {\em Fluid
  Dynamics and Dynamos in Astrophysics and Geophysics\/} (ed. Andrew~M.
  {Soward}, Christopher~A. {Jones}, David~W. {Hughes} \& Nigel~O. {Weiss}), p.
  145.

\bibitem[{Dombre} {\em et~al.\/}(1986){Dombre}, {Frisch}, {Henon}, {Greene} \&
  {Soward}]{dom:1986}
{\sc {Dombre}, T., {Frisch}, U., {Henon}, M., {Greene}, J.~M. \& {Soward},
  A.~M.} 1986 {Chaotic streamlines in the ABC flows}. {\em Journal of Fluid
  Mechanics\/} {\bf 167}, 353--391.

\bibitem[{Donati} \& {Landstreet}(2009)]{donland:2009}
{\sc {Donati}, J.-F. \& {Landstreet}, J.~D.} 2009 {Magnetic Fields of
  Nondegenerate Stars}. {\em Annual Reviews of Astronomy and Astrophysics\/}
  {\bf 47}, 333--370.

\bibitem[{Donati} {\em et~al.\/}(2005){Donati}, {Paletou}, {Bouvier} \&
  {Ferreira}]{donatietal:2005}
{\sc {Donati}, J.-F., {Paletou}, F., {Bouvier}, J. \& {Ferreira}, J.} 2005
  {Direct detection of a magnetic field in the innermost regions of an
  accretion disk}. {\em Nature\/} {\bf 438}, 466--469.

\bibitem[{Dormy}(2016)]{Dormy2016}
{\sc {Dormy}, E.} 2016 {Strong-field spherical dynamos}. {\em Journal of Fluid
  Mechanics\/} {\bf 789}, 500--513.

\bibitem[{Dormy} {\em et~al.\/}(2018){Dormy}, {Oruba} \&
  {Petitdemange}]{dop:2018}
{\sc {Dormy}, Emmanuel, {Oruba}, Ludivine \& {Petitdemange}, Ludovic} 2018
  {Three branches of dynamo action}. {\em Fluid Dynamics Research\/} {\bf
  50}~(1), 011415.

\bibitem[Dormy \& Soward(2007)]{dormysoward:2007}
{\sc Dormy, E. \& Soward, A.M.} 2007 {\em {Mathematical aspects of natural
  dynamos}\/}. CRC Press/Taylor \& Francis.

\bibitem[{Dormy} {\em et~al.\/}(2000){Dormy}, {Valet} \&
  {Courtillot}]{dvc:2000}
{\sc {Dormy}, Emmanuel, {Valet}, Jean-Pierre \& {Courtillot}, Vincent} 2000
  {Numerical models of the geodynamo and observational constraints}. {\em
  Geochemistry, Geophysics, Geosystems\/} {\bf 1}~(10), 1037--42.

\bibitem[{Drummond} \& {Horgan}(1986)]{dh:1986}
{\sc {Drummond}, I.~T. \& {Horgan}, R.~R.} 1986 {Numerical simulation of the
  alpha-effect and turbulent magnetic diffusion with molecular diffusivity}.
  {\em Journal of Fluid Mechanics\/} {\bf 163}, 425--438.

\bibitem[{Dudley} \& {James}(1989)]{dud89}
{\sc {Dudley}, M.~L. \& {James}, R.~W.} 1989 {Time-dependent kinematic dynamos
  with stationary flows}. {\em Proceedings of the Royal Society of London
  Series A\/} {\bf 425}, 407--429.

\bibitem[{Farrell} \& {Ioannou}(2003)]{fi2003}
{\sc {Farrell}, Brian~F. \& {Ioannou}, Petros~J.} 2003 {Structural Stability of
  Turbulent Jets.} {\em Journal of Atmospheric Sciences\/} {\bf 60}~(17),
  2101--2118.

\bibitem[{Fautrelle} \& {Childress}(1982)]{fauchil1982}
{\sc {Fautrelle}, Y. \& {Childress}, S.} 1982 {Convective dynamos with
  intermediate and strong fields}. {\em Geophysical and Astrophysical Fluid
  Dynamics\/} {\bf 22}, 235--279.

\bibitem[{Finn} \& {Ott}(1988)]{finnott:1988}
{\sc {Finn}, J.~M. \& {Ott}, E.} 1988 {Chaotic flows and fast magnetic
  dynamos}. {\em Physics of Fluids\/} {\bf 31}, 2992--3011.

\bibitem[Frisch(1995)]{frisch:1995}
{\sc Frisch, U.} 1995 {\em Turbulence: the legacy of A.~N.~Kolmogorov\/}.
  Cambridge: Cambridge University Press.

\bibitem[{Gailitis} {\em et~al.\/}(2002){Gailitis}, {Lielausis}, {Platacis},
  {Gerbeth} \& {Stefani}]{getal2002}
{\sc {Gailitis}, Agris, {Lielausis}, Olgerts, {Platacis}, Ernests, {Gerbeth},
  Gunter \& {Stefani}, Frank} 2002 {Colloquium: Laboratory experiments on
  hydromagnetic dynamos}. {\em Reviews of Modern Physics\/} {\bf 74}, 973--990.

\bibitem[{Galloway} \& {Frisch}(1984)]{gallfrisch:1984}
{\sc {Galloway}, D.~J. \& {Frisch}, U.} 1984 {A numerical investigation of
  magnetic field generation in a flow with chaotic streamlines}. {\em
  Geophysical and Astrophysical Fluid Dynamics\/} {\bf 29}, 13--18.

\bibitem[Galloway \& Proctor(1992)]{Galloway:1992}
{\sc Galloway, D.~J. \& Proctor, M. R.~E.} 1992 Numerical calculations of fast
  dynamos in smooth velocity fields with realistic diffusion. {\em Nature\/}
  {\bf 356}, 691--693.

\bibitem[Germano(1992)]{germano_1992}
{\sc Germano, M.} 1992 Turbulence: the filtering approach. {\em Journal of
  Fluid Mechanics\/} {\bf 238}, 325–336.

\bibitem[Gilbert(1988)]{Gilbert:1988}
{\sc Gilbert, A.~D.} 1988 Fast dynamo action in the {Ponomarenko} dynamo. {\em
  Geophys. Astrophys. Fluid Dyn.\/} {\bf 44}, 241--258.

\bibitem[{Glatzmaier} \& {Roberts}(1995)]{gr1995}
{\sc {Glatzmaier}, Gary~A. \& {Roberts}, Paul~H.} 1995 {A three-dimensional
  convective dynamo solution with rotating and finitely conducting inner core
  and mantle}. {\em Physics of the Earth and Planetary Interiors\/} {\bf 91},
  63--75.

\bibitem[{Gruzinov} \& {Diamond}(1994)]{grdi1994}
{\sc {Gruzinov}, A.~V. \& {Diamond}, P.~H.} 1994 {Self-consistent theory of
  mean-field electrodynamics}. {\em Physical Review Letters\/} {\bf 72},
  1651--1653.

\bibitem[{Gubbins} {\em et~al.\/}(2000){Gubbins}, {Barber}, {Gibbons} \&
  {Love}]{gub2000}
{\sc {Gubbins}, D., {Barber}, C.~N., {Gibbons}, S. \& {Love}, J.~J.} 2000
  {Kinematic dynamo action in a sphere. I. Effects of differential rotation and
  meridional circulation on solutions with axial dipole symmetry}. {\em
  Proceedings of the Royal Society of London Series A\/} {\bf 456}, 1333.

\bibitem[{Hathaway}(2015)]{hathaway:2015}
{\sc {Hathaway}, D.~H.} 2015 {The Solar Cycle}. {\em Living Reviews in Solar
  Physics\/} {\bf 12}, 4.

\bibitem[Hollins {\em et~al.\/}(2017)Hollins, Sarson, Shukurov, Fletcher \&
  Gent]{Hollins:2017}
{\sc Hollins, J.~F., Sarson, G.~R., Shukurov, A., Fletcher, A. \& Gent, F.~A.}
  2017 Supernova-regulated {ISM}. v. space and time correlations. {\em The
  Astrophysical Journal\/} {\bf 850}~(1), 4.

\bibitem[{Hubbard} \& {Brandenburg}(2010)]{hb:2010}
{\sc {Hubbard}, Alexander \& {Brandenburg}, Axel} 2010 {Magnetic helicity
  fluxes in an {\ensuremath{\alpha}}2 dynamo embedded in a halo}. {\em
  Geophysical and Astrophysical Fluid Dynamics\/} {\bf 104}, 577--590.

\bibitem[{Hughes} \& {Cattaneo}(2016)]{hc2016}
{\sc {Hughes}, David~W. \& {Cattaneo}, Fausto} 2016 {Strong-field dynamo action
  in rapidly rotating convection with no inertia}. {\em \pre\/} {\bf 93}~(6),
  061101.

\bibitem[{Hughes} \& {Tobias}(2010)]{ht2010}
{\sc {Hughes}, D.~W. \& {Tobias}, S.~M.} 2010 {\em {An Introduction to Mean
  Field Dynamo Theory}\/}, pp. 15--48. World Scientific Publishing Co.

\bibitem[{Iskakov} {\em et~al.\/}(2007){Iskakov}, {Schekochihin}, {Cowley},
  {McWilliams} \& {Proctor}]{iskakovetal:2007}
{\sc {Iskakov}, A.~B., {Schekochihin}, A.~A., {Cowley}, S.~C., {McWilliams},
  J.~C. \& {Proctor}, M.~R.~E.} 2007 {Numerical Demonstration of Fluctuation
  Dynamo at Low Magnetic Prandtl Numbers}. {\em Physical Review Letters\/} {\bf
  98}~(20), 208501--+.

\bibitem[Jackson {\em et~al.\/}(2000)Jackson, Jonkers \& Walker]{Jackson:2000}
{\sc Jackson, Andrew, Jonkers, Art R.~T. \& Walker, Matthew~R.} 2000 Four
  centuries of geomagnetic secular variation from historical records. {\em
  Philosophical Transactions: Mathematical, Physical and Engineering
  Sciences\/} {\bf 358}~(1768), 957--990.

\bibitem[Jones(2008)]{JONES:2008}
{\sc Jones, Chris~A.} 2008 Dynamo theory. In {\em Dynamos\/} (ed. Ph. Cardin \&
  L.F. Cugliandolo), {\em Les Houches\/}, vol.~88, pp. 45 -- 135. Elsevier.

\bibitem[{Jones} \& {Holme}(2017)]{jonesholme:2017}
{\sc {Jones}, C.~A. \& {Holme}, R.} 2017 {A close-up view of Jupiter's magnetic
  field from Juno: New insights into the planet's deep interior}. {\em
  Geophysical Research Letters\/} {\bf 44}, 5355--5359.

\bibitem[{K{\"a}pyl{\"a}} \& {Brandenburg}(2009)]{kb:2009}
{\sc {K{\"a}pyl{\"a}}, P.~J. \& {Brandenburg}, A.} 2009 {Turbulent Dynamos with
  Shear and Fractional Helicity}. {\em Astrophysical Journal\/} {\bf 699},
  1059--1066.

\bibitem[{Kazantsev}(1968)]{kaz:1968}
{\sc {Kazantsev}, A.~P.} 1968 {Enhancement of a Magnetic Field by a Conducting
  Fluid}. {\em Soviet Journal of Experimental and Theoretical Physics\/} {\bf
  26}, 1031.

\bibitem[Keynes(1923)]{keynes:1923}
{\sc Keynes, J.M.} 1923 {\em A tract on monetary reform\/}. Macmillan and Co.,
  limited.

\bibitem[Klapper \& Young(1995)]{Klapper:1995}
{\sc Klapper, I. \& Young, L.~S.} 1995 Rigorous bounds on the fast dynamo
  growth-rate involving topological entropy. {\em Comm. Math. Phys.\/} {\bf
  175}, 623--646.

\bibitem[{Kleeorin} {\em et~al.\/}(1995){Kleeorin}, {Rogachevskii} \&
  {Ruzmaikin}]{krr1995}
{\sc {Kleeorin}, N., {Rogachevskii}, I. \& {Ruzmaikin}, A.} 1995 {Magnitude of
  the dynamo-generated magnetic field in solar-type convective zones}. {\em
  \aap\/} {\bf 297}, 159--167.

\bibitem[{Kleeorin} {\em et~al.\/}(2002){Kleeorin}, {Rogachevskii} \&
  {Sokoloff}]{kleerogsok:2002}
{\sc {Kleeorin}, N., {Rogachevskii}, I. \& {Sokoloff}, D.} 2002 {Magnetic
  fluctuations with a zero mean field in a random fluid flow with a finite
  correlation time and a small magnetic diffusion}. {\em Physical Review E.\/}
  {\bf 65}~(3), 036303--+.

\bibitem[{Kraichnan}(1959)]{kraichnan:1959}
{\sc {Kraichnan}, R.~H.} 1959 {The structure of isotropic turbulence at very
  high Reynolds numbers}. {\em Journal of Fluid Mechanics\/} {\bf 5}, 497--543.

\bibitem[{Kraichnan}(1965)]{kraichnan}
{\sc {Kraichnan}, R.~H.} 1965 {Inertial-Range Spectrum of Hydromagnetic
  Turbulence}. {\em Physics of Fluids\/} {\bf 8}, 1385--1387.

\bibitem[Kraichnan(1985)]{kraichnan1985}
{\sc Kraichnan, R.~H.} 1985 {Decimated amplitude equations in turbulence
  dynamics}. In {\em Theoretical Approaches to Turbulence\/} (ed. D.~L. Dwoyer,
  M.~Y. Hussaini \& R.~G. Voight), pp. 91--135. New York: Springer.

\bibitem[Krause(1993)]{krause:1993}
{\sc Krause, F.} 1993 The cosmic dynamo: From t = -infinity to cowling's
  theorem a review on history. In {\em The Cosmic Dynamo\/} (ed. F.~Krause,
  K.-H. R{\"a}dler \& G.~R{\"u}diger), pp. 487--499. Dordrecht: Springer
  Netherlands.

\bibitem[{Krause} \& {Raedler}(1980)]{krauraed:1980}
{\sc {Krause}, F. \& {Raedler}, {K.-H.}} 1980 {\em {Mean-field
  magnetohydrodynamics and dynamo theory}\/}. Oxford, Pergamon Press, Ltd.,
  1980.~271 p.

\bibitem[{Kulsrud} \& {Anderson}(1992)]{kuland:1992}
{\sc {Kulsrud}, R.~M. \& {Anderson}, S.~W.} 1992 {The spectrum of random
  magnetic fields in the mean field dynamo theory of the Galactic magnetic
  field}. {\em \apj\/} {\bf 396}, 606--630.

\bibitem[{Kulsrud} \& {Zweibel}(2008)]{kulsrudzweibel:2008}
{\sc {Kulsrud}, R.~M. \& {Zweibel}, E.~G.} 2008 {On the origin of cosmic
  magnetic fields}. {\em Reports on Progress in Physics\/} {\bf 71}~(4),
  046901--+.

\bibitem[{Kumar} \& {Roberts}(1975)]{kumrob75}
{\sc {Kumar}, S. \& {Roberts}, P.~H.} 1975 {A three-dimensional kinematic
  dynamo}. {\em Proceedings of the Royal Society of London Series A\/} {\bf
  344}, 235--258.

\bibitem[{Li} {\em et~al.\/}(2018){Li}, {Jackson} \& {Livermore}]{ljl2018}
{\sc {Li}, K., {Jackson}, A. \& {Livermore}, P.~W.} 2018 {Taylor state dynamos
  found by optimal control: axisymmetric examples}. {\em Journal of Fluid
  Mechanics\/} {\bf 853}, 647--697.

\bibitem[{Malyshkin} \& {Boldyrev}(2007)]{malbol:2007}
{\sc {Malyshkin}, L. \& {Boldyrev}, S.} 2007 {Magnetic Dynamo Action in Helical
  Turbulence}. {\em \apjl\/} {\bf 671}, L185--L188.

\bibitem[{Malyshkin} \& {Boldyrev}(2008{\natexlab{{\em a\/}}})]{malbol:2008a}
{\sc {Malyshkin}, L. \& {Boldyrev}, S.} 2008{\natexlab{{\em a\/}}}
  {Amplification of magnetic fields by dynamo action in Gaussian-correlated
  helical turbulence}. {\em ArXiv e-prints\/} .

\bibitem[{Malyshkin} \& {Boldyrev}(2008{\natexlab{{\em b\/}}})]{malbol:2008b}
{\sc {Malyshkin}, L. \& {Boldyrev}, S.} 2008{\natexlab{{\em b\/}}} {On magnetic
  dynamo action in astrophysical turbulence}. {\em ArXiv e-prints\/} .

\bibitem[{Maron} {\em et~al.\/}(2004){Maron}, {Cowley} \&
  {McWilliams}]{maronetal:2004}
{\sc {Maron}, J., {Cowley}, S. \& {McWilliams}, J.} 2004 {The Nonlinear
  Magnetic Cascade}. {\em \apj\/} {\bf 603}, 569--583.

\bibitem[{Marston} {\em et~al.\/}(2016){Marston}, {Chini} \&
  {Tobias}]{mct:2016}
{\sc {Marston}, J.~B., {Chini}, G.~P. \& {Tobias}, S.~M.} 2016 {Generalized
  Quasilinear Approximation: Application to Zonal Jets}. {\em Physical Review
  Letters\/} {\bf 116}~(21), 214501.

\bibitem[{Marston} {\em et~al.\/}(2019){Marston}, {Qi} \& {Tobias}]{mqt2019}
{\sc {Marston}, J.~B., {Qi}, Wanming \& {Tobias}, S.~M.} 2019 {Direct
  Statistical Simulation of a Jet}. In {\em Zonal Jets: Phenomenology, Genesis,
  and Physics\/} (ed. B.~{Galperin} \& P.~{Read}).

\bibitem[{McKeon} \& {Sharma}(2010)]{ms2010}
{\sc {McKeon}, B.~J. \& {Sharma}, A.~S.} 2010 {A critical-layer framework for
  turbulent pipe flow}. {\em Journal of Fluid Mechanics\/} {\bf 658}, 336--382.

\bibitem[{Miesch} {\em et~al.\/}(2015){Miesch}, {Matthaeus}, {Brandenburg},
  {Petrosyan}, {Pouquet}, {Cambon}, {Jenko}, {Uzdensky}, {Stone}, {Tobias},
  {Toomre} \& {Velli}]{miesch:2015}
{\sc {Miesch}, M., {Matthaeus}, W., {Brandenburg}, A., {Petrosyan}, A.,
  {Pouquet}, A., {Cambon}, C., {Jenko}, F., {Uzdensky}, D., {Stone}, J.,
  {Tobias}, S., {Toomre}, J. \& {Velli}, M.} 2015 {Large-Eddy Simulations of
  Magnetohydrodynamic Turbulence in Heliophysics and Astrophysics}. {\em Space
  Science Reviews\/} {\bf 194}, 97--137.

\bibitem[{Mininni}(2006)]{mininni:2006}
{\sc {Mininni}, P.~D.} 2006 {Turbulent magnetic dynamo excitation at low
  magnetic Prandtl number}. {\em Physics of Plasmas\/} {\bf 13}~(5), 056502--+.

\bibitem[Moffatt \& Dormy(2019)]{MoffattDormy:2019}
{\sc Moffatt, H.K. \& Dormy, E.} 2019 {\em Self-Exciting Fluid Dynamos\/}.
  Cambridge University Press, Cambridge.

\bibitem[{Moffatt}(1969)]{moffatt1969}
{\sc {Moffatt}, H.~K.} 1969 {The degree of knottedness of tangled vortex
  lines}. {\em Journal of Fluid Mechanics\/} {\bf 35}, 117--129.

\bibitem[{Moffatt}(1978)]{Moffatt:1978}
{\sc {Moffatt}, H.~K.} 1978 {\em {Magnetic field generation in electrically
  conducting fluids}\/}. Cambridge, England, Cambridge University Press,
  1978.~353 p.

\bibitem[{Moffatt} \& {Proctor}(1985)]{moffatproctor:1985}
{\sc {Moffatt}, H.~K. \& {Proctor}, M.~R.~E.} 1985 {Topological constraints
  associated with fast dynamo action}. {\em Journal of Fluid Mechanics\/} {\bf
  154}, 493--507.

\bibitem[{Monchaux} {\em et~al.\/}(2007){Monchaux}, {Berhanu}, {Bourgoin},
  {Moulin}, {Odier}, {Pinton}, {Volk}, {Fauve}, {Mordant}, {P{\'e}tr{\'e}lis},
  {Chiffaudel}, {Daviaud}, {Dubrulle}, {Gasquet}, {Mari{\'e}} \&
  {Ravelet}]{monchauxetal:2007}
{\sc {Monchaux}, R., {Berhanu}, M., {Bourgoin}, M., {Moulin}, M., {Odier}, P.,
  {Pinton}, J.-F., {Volk}, R., {Fauve}, S., {Mordant}, N., {P{\'e}tr{\'e}lis},
  F., {Chiffaudel}, A., {Daviaud}, F., {Dubrulle}, B., {Gasquet}, C.,
  {Mari{\'e}}, L. \& {Ravelet}, F.} 2007 {Generation of a Magnetic Field by
  Dynamo Action in a Turbulent Flow of Liquid Sodium}. {\em Physical Review
  Letters\/} {\bf 98}~(4), 044502--+.

\bibitem[Monin {\em et~al.\/}(1975)Monin, Yaglom \& Lumley]{moninyaglom75}
{\sc Monin, A.S., Yaglom, A.M. \& Lumley, J.L.} 1975 {\em Statistical Fluid
  Mechanics: Mechanics of Turbulence\/}. MIT Press.

\bibitem[{Moore} {\em et~al.\/}(2017){Moore}, {Bloxham}, {Connerney},
  {J{\o}rgensen} \& {Merayo}]{mooreetal:2017}
{\sc {Moore}, K.~M., {Bloxham}, J., {Connerney}, J.~E.~P., {J{\o}rgensen},
  J.~L. \& {Merayo}, J.~M.~G.} 2017 {The analysis of initial Juno magnetometer
  data using a sparse magnetic field representation}. {\em {Geophysical
  Research Letters}\/} {\bf 44}, 4687--4693.

\bibitem[{Moreau}(1961)]{moreau1961}
{\sc {Moreau}, J.~J.} 1961 {Constantes d’un lot tourbillonnaire en fluid
  parfait barotrope}. {\em C.R. Acad. Sci. Paris\/} {\bf 252}, 2810--2812.

\bibitem[{Nataf} \& {Schaeffer}(2015)]{ns2013}
{\sc {Nataf}, H. \& {Schaeffer}, N.} 2015 {Turbulence in the core}. In {\em
  Treatise on Geophysics\/} (ed. P.~{Olson} \& G.~{Schubert}).

\bibitem[{Nigro} {\em et~al.\/}(2017){Nigro}, {Pongkitiwanichakul}, {Cattaneo}
  \& {Tobias}]{nigro:2017}
{\sc {Nigro}, G., {Pongkitiwanichakul}, P., {Cattaneo}, F. \& {Tobias}, S.~M.}
  2017 {What is a large-scale dynamo?} {\em \mnras\/} {\bf 464}, L119--L123.

\bibitem[{Orszag}(1970)]{orszag70}
{\sc {Orszag}, S.~A.} 1970 {Analytical theories of turbulence}. {\em Journal of
  Fluid Mechanics\/} {\bf 41}, 363--386.

\bibitem[Otani(1988)]{otani:1988}
{\sc Otani, NlELS~F} 1988 Computer simulation of fast kinematic dynamos. {\em
  EOS Transactions, American Geophysical Union\/} {\bf 64}~(44), 1366.

\bibitem[Otani(1993)]{otani:1993}
{\sc Otani, Niels~F.} 1993 A fast kinematic dynamo in two-dimensional
  time-dependent flows. {\em Journal of Fluid Mechanics\/} {\bf 253},
  327–340.

\bibitem[{Parker}(1955)]{parker:1955}
{\sc {Parker}, E.~N.} 1955 {Hydromagnetic Dynamo Models.} {\em \apj\/} {\bf
  122}, 293.

\bibitem[{Parker}(1993)]{Parker:1993}
{\sc {Parker}, E.~N.} 1993 {A Solar Dynamo Surface Wave at the Interface
  between Convection and Nonuniform Rotation}. {\em \apj\/} {\bf 408}, 707.

\bibitem[{P{\'e}tr{\'e}lis} \& {Fauve}(2008)]{petfauve:2008}
{\sc {P{\'e}tr{\'e}lis}, F. \& {Fauve}, S.} 2008 {Chaotic dynamics of the
  magnetic field generated by dynamo action in a turbulent flow}. {\em Journal
  of Physics Condensed Matter\/} {\bf 20}, 4203--+.

\bibitem[{Plumley} {\em et~al.\/}(2018){Plumley}, {Calkins}, {Julien} \&
  {Tobias}]{pcjt2018}
{\sc {Plumley}, Meredith, {Calkins}, Michael~A., {Julien}, Keith \& {Tobias},
  Steven~M.} 2018 {Self-consistent single mode investigations of the
  quasi-geostrophic convection-driven dynamo model}. {\em Journal of Plasma
  Physics\/} {\bf 84}, 735840406.

\bibitem[{Ponomarenko}(1973)]{Ponomarenko:1973}
{\sc {Ponomarenko}, Yu.B.} 1973 {On the theory of hydromagnetic dynamos}. {\em
  English translation: J. Appl. Mech. Tech. phys.\/} {\bf 14}, 775--778.

\bibitem[{Ponty} {\em et~al.\/}(2007){Ponty}, {Mininni}, {Pinton}, {Politano}
  \& {Pouquet}]{pontyetal:2007}
{\sc {Ponty}, Y., {Mininni}, P.~D., {Pinton}, J.-F., {Politano}, H. \&
  {Pouquet}, A.} 2007 {Dynamo action at low magnetic Prandtl numbers: mean flow
  versus fully turbulent motions}. {\em New Journal of Physics\/} {\bf 9},
  296--+.

\bibitem[{Pouquet} {\em et~al.\/}(1976){Pouquet}, {Frisch} \&
  {Leorat}]{pouquetetal:1976}
{\sc {Pouquet}, A., {Frisch}, U. \& {Leorat}, J.} 1976 {Strong MHD helical
  turbulence and the nonlinear dynamo effect}. {\em Journal of Fluid
  Mechanics\/} {\bf 77}, 321--354.

\bibitem[Prior \& Yeates(2014)]{PriorYeates:2014}
{\sc Prior, C. \& Yeates, A.~R.} 2014 On the helicity of open magnetic fields.
  {\em The Astrophysical Journal\/} {\bf 787}~(2), 100.

\bibitem[{Proctor}(2003)]{proctor2003}
{\sc {Proctor}, M.R.E.} 2003 {\em {Dynamo processes: the interaction of
  turbulence and magnetic fields}\/}, pp. 143--158.

\bibitem[Proctor(2015)]{proctor:2015}
{\sc Proctor, M.R.E.} 2015 Energy requirement for a working dynamo. {\em
  Geophysical \& Astrophysical Fluid Dynamics\/} {\bf 109}~(6), 611--614.

\bibitem[{Proctor}(2007)]{proctor:2007}
{\sc {Proctor}, M.~R.~E.} 2007 {Effects of fluctuation on {$\alpha$}{$\Omega$}
  dynamo models}. {\em \mnras\/} {\bf 382}, L39--L42.

\bibitem[{Rahbarnia} {\em et~al.\/}(2012){Rahbarnia}, {Brown}, {Clark},
  {Kaplan}, {Nornberg}, {Rasmus}, {Zane Taylor}, {Forest}, {Jenko} \&
  {Limone}]{cary:2012}
{\sc {Rahbarnia}, Kian, {Brown}, Benjamin~P., {Clark}, Mike~M., {Kaplan},
  Elliot~J., {Nornberg}, Mark~D., {Rasmus}, Alex~M., {Zane Taylor}, Nicholas,
  {Forest}, Cary~B., {Jenko}, Frank \& {Limone}, Angelo} 2012 {Direct
  Observation of the Turbulent emf and Transport of Magnetic Field in a Liquid
  Sodium Experiment}. {\em \apj\/} {\bf 759}~(2), 80.

\bibitem[{Ribes} \& {Nesme-Ribes}(1993)]{ribesnesmeribes:1993}
{\sc {Ribes}, J.~C. \& {Nesme-Ribes}, E.} 1993 {The solar sunspot cycle in the
  Maunder minimum AD1645 to AD1715}. {\em Astronomy and Astrophysics\/} {\bf
  276}, 549.

\bibitem[{Richardson} \& {Proctor}(2010)]{rp:2010}
{\sc {Richardson}, K.~J. \& {Proctor}, M.~R.~E.} 2010 {Effects of
  {$\alpha$}-effect fluctuations on simple nonlinear dynamo models}. {\em
  Geophysical and Astrophysical Fluid Dynamics\/} {\bf 104}, 601--618.

\bibitem[{Rincon}(2019)]{rincon:2019}
{\sc {Rincon}, Francois} 2019 {Dynamo theories}. {\em arXiv e-prints\/} p.
  arXiv:1903.07829.

\bibitem[{Rincon} {\em et~al.\/}(2007){Rincon}, {Ogilvie} \&
  {Cossu}]{rincon2007}
{\sc {Rincon}, F., {Ogilvie}, G.~I. \& {Cossu}, C.} 2007 {On self-sustaining
  processes in Rayleigh-stable rotating plane Couette flows and subcritical
  transition to turbulence in accretion disks}. {\em \aap\/} {\bf 463},
  817--832.

\bibitem[{Riols} {\em et~al.\/}(2013){Riols}, {Rincon}, {Cossu}, {Lesur},
  {Longaretti}, {Ogilvie} \& {Herault}]{rincon2013}
{\sc {Riols}, A., {Rincon}, F., {Cossu}, C., {Lesur}, G., {Longaretti}, P.~Y.,
  {Ogilvie}, G.~I. \& {Herault}, J.} 2013 {Global bifurcations to subcritical
  magnetorotational dynamo action in Keplerian shear flow}. {\em Journal of
  Fluid Mechanics\/} {\bf 731}, 1--45.

\bibitem[{Roberts}(1972{\natexlab{{\em a\/}}})]{Roberts:1972}
{\sc {Roberts}, G.~O.} 1972{\natexlab{{\em a\/}}} {Dynamo action of fluid
  motions with two-dimensional periodicity}. {\em Phil. Trans. Roy. Soc. Lond.
  A.\/} {\bf 271}, 411--454.

\bibitem[{Roberts}(1972{\natexlab{{\em b\/}}})]{PHRoberts:1972}
{\sc {Roberts}, P.~H.} 1972{\natexlab{{\em b\/}}} {Kinematic Dynamo Models}.
  {\em Philosophical Transactions of the Royal Society of London Series A\/}
  {\bf 272}, 663--698.

\bibitem[{Roberts}(1994)]{roberts:1994}
{\sc {Roberts}, P.~H.} 1994 {Fundamentals of dynamo theory}. In {\em Lectures
  on Solar and planetary Dynamos\/} (ed. M.~R.~E. {Proctor} \& A.~D.
  {Gilbert}), pp. 1--58. Cambridge University Press.

\bibitem[{Roberts} \& {Soward}(1992)]{robsow:1992}
{\sc {Roberts}, P.~H. \& {Soward}, A.~M.} 1992 {Dynamo theory}. {\em Annual
  Reviews Fluid Mech.\/} {\bf 24}, 459--512.

\bibitem[{Roberts} \& {Wu}(2018)]{rw2018}
{\sc {Roberts}, P.~H. \& {Wu}, C.-C.} 2018 {On magnetostrophic mean-field
  solutions of the geodynamo equations. Part 2}. {\em Journal of Plasma
  Physics\/} {\bf 84}~(4), 735840402.

\bibitem[{Schaeffer} {\em et~al.\/}(2017){Schaeffer}, {Jault}, {Nataf} \&
  {Fournier}]{sjnf2017}
{\sc {Schaeffer}, N., {Jault}, D., {Nataf}, H.~C. \& {Fournier}, A.} 2017
  {Turbulent geodynamo simulations: a leap towards Earth's core}. {\em
  Geophysical Journal International\/} {\bf 211}, 1--29.

\bibitem[{Schekochihin} {\em et~al.\/}(2002{\natexlab{{\em
  a\/}}}){Schekochihin}, {Cowley}, {Maron} \& {Malyshkin}]{schek:2002}
{\sc {Schekochihin}, A., {Cowley}, S., {Maron}, J. \& {Malyshkin}, L.}
  2002{\natexlab{{\em a\/}}} {Structure of small-scale magnetic fields in the
  kinematic dynamo theory}. {\em \pre\/} {\bf 65}~(1), 016305--+.

\bibitem[{Schekochihin} {\em et~al.\/}(2002{\natexlab{{\em
  b\/}}}){Schekochihin}, {Cowley}, {Hammett}, {Maron} \&
  {McWilliams}]{scheketal:2002-amodel}
{\sc {Schekochihin}, A.~A., {Cowley}, S.~C., {Hammett}, G.~W., {Maron}, J.~L.
  \& {McWilliams}, J.~C.} 2002{\natexlab{{\em b\/}}} {A model of nonlinear
  evolution and saturation of the turbulent MHD dynamo}. {\em New Journal of
  Physics\/} {\bf 4}, 84--+.

\bibitem[{Schekochihin} {\em et~al.\/}(2004){Schekochihin}, {Cowley}, {Maron}
  \& {McWilliams}]{schekochihietal:2004c}
{\sc {Schekochihin}, A.~A., {Cowley}, S.~C., {Maron}, J.~L. \& {McWilliams},
  J.~C.} 2004 {Critical Magnetic Prandtl Number for Small-Scale Dynamo}. {\em
  Physical Review Letters\/} {\bf 92}~(5), 054502--+.

\bibitem[{Schekochihin} {\em et~al.\/}(2005){Schekochihin}, {Haugen},
  {Brandenburg}, {Cowley}, {Maron} \& {McWilliams}]{schekochihinetal:2005c}
{\sc {Schekochihin}, A.~A., {Haugen}, N.~E.~L., {Brandenburg}, A., {Cowley},
  S.~C., {Maron}, J.~L. \& {McWilliams}, J.~C.} 2005 {The Onset of a
  Small-Scale Turbulent Dynamo at Low Magnetic Prandtl Numbers}. {\em \apjl\/}
  {\bf 625}, L115--L118.

\bibitem[{Schekochihin} {\em et~al.\/}(2007){Schekochihin}, {Iskakov},
  {Cowley}, {McWilliams}, {Proctor} \& {Yousef}]{scheketal:2007}
{\sc {Schekochihin}, A.~A., {Iskakov}, A.~B., {Cowley}, S.~C., {McWilliams},
  J.~C., {Proctor}, M.~R.~E. \& {Yousef}, T.~A.} 2007 {Fluctuation dynamo and
  turbulent induction at low magnetic Prandtl numbers}. {\em New Journal of
  Physics\/} {\bf 9}, 300--+.

\bibitem[{Schrinner} {\em et~al.\/}(2005){Schrinner}, {R{\"a}dler}, {Schmitt},
  {Rheinhardt} \& {Christensen}]{schrinner2005}
{\sc {Schrinner}, M., {R{\"a}dler}, K.-H., {Schmitt}, D., {Rheinhardt}, M. \&
  {Christensen}, U.} 2005 {Mean-field view on rotating magnetoconvection and a
  geodynamo model}. {\em Astronomische Nachrichten\/} {\bf 326}, 245--249.

\bibitem[{Schubert} \& {Soderlund}(2011)]{schsoder:2011}
{\sc {Schubert}, G. \& {Soderlund}, K.~M.} 2011 {Planetary magnetic fields:
  Observations and models}. {\em Physics of the Earth and Planetary
  Interiors\/} {\bf 187}, 92--108.

\bibitem[Schwaiger {\em et~al.\/}(2019)Schwaiger, Gastine \& Aubert]{sga19}
{\sc Schwaiger, T, Gastine, T \& Aubert, J} 2019 {Force balance in numerical
  geodynamo simulations: a systematic study}. {\em Geophysical Journal
  International\/} .

\bibitem[{See} {\em et~al.\/}(2016){See}, {Jardine}, {Vidotto}, {Donati}, {Boro
  Saikia}, {Bouvier}, {Fares}, {Folsom}, {Gregory}, {Hussain}, {Jeffers},
  {Marsden}, {Morin}, {Moutou}, {do Nascimento}, {Petit} \&
  {Waite}]{seeetal:2016}
{\sc {See}, V., {Jardine}, M., {Vidotto}, A.~A., {Donati}, J.-F., {Boro
  Saikia}, S., {Bouvier}, J., {Fares}, R., {Folsom}, C.~P., {Gregory}, S.~G.,
  {Hussain}, G., {Jeffers}, S.~V., {Marsden}, S.~C., {Morin}, J., {Moutou}, C.,
  {do Nascimento}, J.~D., {Petit}, P. \& {Waite}, I.~A.} 2016 {The connection
  between stellar activity cycles and magnetic field topology}. {\em Monthly
  Notices of the Royal Astronomical Society\/} {\bf 462}, 4442--4450.

\bibitem[{Seshasayanan} {\em et~al.\/}(2017){Seshasayanan}, {Dallas} \&
  {Alexakis}]{sesh:2017}
{\sc {Seshasayanan}, K., {Dallas}, V. \& {Alexakis}, A.} 2017 {The onset of
  turbulent rotating dynamos at the low magnetic Prandtl number limit}. {\em
  Journal of Fluid Mechanics\/} {\bf 822}, R3.

\bibitem[{Shkolnik} {\em et~al.\/}(2008){Shkolnik}, {Bohlender}, {Walker} \&
  {Collier Cameron}]{shkolnik:2008}
{\sc {Shkolnik}, E., {Bohlender}, D.~A., {Walker}, G.~A.~H. \& {Collier
  Cameron}, A.} 2008 {The On/Off Nature of Star-Planet Interactions}. {\em
  Astrophysical Journal\/} {\bf 676}, 628--638.

\bibitem[{Soward}(1987)]{Sow87}
{\sc {Soward}, A.~M.} 1987 {Fast dynamo action in a steady flow}. {\em Journal
  of Fluid Mechanics\/} {\bf 180}, 267--295.

\bibitem[{Squire} \& {Bhattacharjee}(2016)]{sb2016}
{\sc {Squire}, J. \& {Bhattacharjee}, A.} 2016 {The magnetic shear-current
  effect: generation of large-scale magnetic fields by the small-scale dynamo}.
  {\em Journal of Plasma Physics\/} {\bf 82}, 535820201.

\bibitem[{Sridhar} \& {Singh}(2010)]{srisin:2010}
{\sc {Sridhar}, S. \& {Singh}, N.~K.} 2010 {The shear dynamo problem for small
  magnetic Reynolds numbers}. {\em Journal of Fluid Mechanics\/} {\bf 664},
  265--285.

\bibitem[{Stellmach} \& {Hansen}(2004)]{stellhans:04}
{\sc {Stellmach}, S. \& {Hansen}, U.} 2004 {Cartesian convection driven dynamos
  at low Ekman number}. {\em Phys. Rev. E.\/} {\bf 70}~(5), 056312--+.

\bibitem[{Stello} {\em et~al.\/}(2016){Stello}, {Cantiello}, {Fuller}, {Huber},
  {Garc{\'{\i}}a}, {Bedding}, {Bildsten} \& {Silva Aguirre}]{stelloetal:2016}
{\sc {Stello}, D., {Cantiello}, M., {Fuller}, J., {Huber}, D., {Garc{\'{\i}}a},
  R.~A., {Bedding}, T.~R., {Bildsten}, L. \& {Silva Aguirre}, V.} 2016 {A
  prevalence of dynamo-generated magnetic fields in the cores of
  intermediate-mass stars}. {\em Nature\/} {\bf 529}, 364--367.

\bibitem[{Stieglitz} \& {M{\"u}ller}(2001)]{stmu:2001}
{\sc {Stieglitz}, R. \& {M{\"u}ller}, U.} 2001 {Experimental demonstration of a
  homogeneous two-scale dynamo}. {\em Physics of Fluids\/} {\bf 13}, 561--564.

\bibitem[{Strassmeier}(2009)]{strassmeier:2009}
{\sc {Strassmeier}, K.~G.} 2009 {Starspots}. {\em Astronomy \& Astrophysics
  Reviews\/} {\bf 17}, 251--308.

\bibitem[{Subramanian}(1999)]{sub:1999}
{\sc {Subramanian}, K.} 1999 {Unified Treatment of Small- and Large-Scale
  Dynamos in Helical Turbulence}. {\em Physical Review Letters\/} {\bf 83},
  2957--2960.

\bibitem[{Subramanian}(2003)]{sub:2003}
{\sc {Subramanian}, K.} 2003 {Hyperdiffusion in Nonlinear Large- and
  Small-Scale Turbulent Dynamos}. {\em Physical Review Letters\/} {\bf
  90}~(24), 245003--+.

\bibitem[{Tobias} \& {Weiss}(2007)]{wt:2007}
{\sc {Tobias}, S. \& {Weiss}, N.} 2007 {The solar dynamo and the tachocline}.
  In {\em The Solar Tachocline\/} (ed. D.~W. {Hughes}, R.~{Rosner} \& N.~O.
  {Weiss}), p. 319.

\bibitem[{Tobias} \& {Cattaneo}(2008{\natexlab{{\em a\/}}})]{tobcatt:2008a}
{\sc {Tobias}, S.~M. \& {Cattaneo}, F.} 2008{\natexlab{{\em a\/}}} Dynamo
  action in complex flows: the quick and the fast. {\em Journal of Fluid
  Mechanics\/} {\bf 601}~(-1), 101--122.

\bibitem[{Tobias} \& {Cattaneo}(2008{\natexlab{{\em b\/}}})]{tobcatt:2008b}
{\sc {Tobias}, S.~M. \& {Cattaneo}, F.} 2008{\natexlab{{\em b\/}}} {Limited
  Role of Spectra in Dynamo Theory: Coherent versus Random Dynamos}. {\em
  Physical Review Letters\/} {\bf 101}~(12), 125003--+.

\bibitem[Tobias \& Cattaneo(2013)]{tc2013}
{\sc Tobias, S.~M. \& Cattaneo, F.} 2013 On the measurement of the turbulent
  diffusivity of a large-scale magnetic field. {\em Journal of Fluid
  Mechanics\/} {\bf 717}, 347–360.

\bibitem[{Tobias} \& {Cattaneo}(2013)]{tc:2013}
{\sc {Tobias}, S.~M. \& {Cattaneo}, F.} 2013 {Shear-driven dynamo waves at high
  magnetic Reynolds number}. {\em \nat\/} {\bf 497}, 463--465.

\bibitem[{Tobias} {\em et~al.\/}(2011{\natexlab{{\em a\/}}}){Tobias},
  {Cattaneo} \& {Boldyrev}]{tcb11}
{\sc {Tobias}, S.~M., {Cattaneo}, F. \& {Boldyrev}, S.B.} 2011{\natexlab{{\em
  a\/}}} {MHD Dynamos and Turbulence}. In {\em Ten Chapters in Turbulence\/}
  (ed. P.~A. {Davidson}, Y.~{Kaneda} \& K.R. {Sreenivasan}).

\bibitem[{Tobias} {\em et~al.\/}(2011{\natexlab{{\em b\/}}}){Tobias},
  {Cattaneo} \& {Brummell}]{tcb:2011}
{\sc {Tobias}, S.~M., {Cattaneo}, F. \& {Brummell}, N.~H.} 2011{\natexlab{{\em
  b\/}}} {On the Generation of Organized Magnetic Fields}. {\em Astrophysical
  Journal\/} {\bf 728}, 153.

\bibitem[{Tobias} {\em et~al.\/}(2011{\natexlab{{\em c\/}}}){Tobias}, {Dagon}
  \& {Marston}]{tdm2011}
{\sc {Tobias}, S.~M., {Dagon}, K. \& {Marston}, J.~B.} 2011{\natexlab{{\em
  c\/}}} {Astrophysical Fluid Dynamics via Direct Statistical Simulation}. {\em
  \apj\/} {\bf 727}~(2), 127.

\bibitem[{Tobias} {\em et~al.\/}(1997){Tobias}, {Proctor} \&
  {Knobloch}]{tpk:1997}
{\sc {Tobias}, S.~M., {Proctor}, M.~R.~E. \& {Knobloch}, E.} 1997 {The Role of
  absolute instability in the solar dynamo.} {\em \aap\/} {\bf 318}, L55--L58.

\bibitem[{Usoskin}(2017)]{usoskin:2017}
{\sc {Usoskin}, I.~G.} 2017 {A history of solar activity over millennia}. {\em
  Living Reviews in Solar Physics\/} {\bf 14}, 3.

\bibitem[{Vainshtein} \& {Cattaneo}(1992)]{VainCatt:1992}
{\sc {Vainshtein}, S.~I. \& {Cattaneo}, F.} 1992 {Nonlinear restrictions on
  dynamo action}. {\em Astrophys. J.\/} {\bf 393}, 165--171.

\bibitem[{Vainshtein} \& {Kichatinov}(1986)]{vainkit:1986}
{\sc {Vainshtein}, S.~I. \& {Kichatinov}, L.~L.} 1986 {The dynamics of magnetic
  fields in a highly conducting turbulent medium and the generalized
  Kolmogorov-Fokker-Planck equations}. {\em Journal of Fluid Mechanics\/} {\bf
  168}, 73--87.

\bibitem[{Vallis}(2006)]{vallis:2006}
{\sc {Vallis}, G.~K.} 2006 {\em {Atmospheric and Oceanic Fluid Dynamics}\/}.

\bibitem[Venturi(2018)]{venturi2018}
{\sc Venturi, Daniele} 2018 The numerical approximation of nonlinear
  functionals and functional differential equations. {\em Physics Reports\/}
  {\bf 732}, 1 -- 102.

\bibitem[{Waleffe}(1997)]{waleffe:1997}
{\sc {Waleffe}, F.} 1997 {On a self-sustaining process in shear flows}. {\em
  Physics of Fluids\/} {\bf 9}, 883--900.

\bibitem[{Weiss} \& {Tobias}(2016)]{wt:2016}
{\sc {Weiss}, N.~O. \& {Tobias}, S.~M.} 2016 {Supermodulation of the Sun's
  magnetic activity: the effects of symmetry changes}. {\em Monthly Notices of
  the Royal Astronomical Society\/} {\bf 456}, 2654--2661.

\bibitem[{Willis}(2012)]{willis:2012}
{\sc {Willis}, A.~P.} 2012 {Optimization of the Magnetic Dynamo}. {\em Physical
  Review Letters\/} {\bf 109}~(25), 251101.

\bibitem[{Wu} \& {Roberts}(2015)]{wurob2015}
{\sc {Wu}, Cheng-Chin \& {Roberts}, Paul~H.} 2015 {On magnetostrophic
  mean-field solutions of the geodynamo equations}. {\em Geophysical and
  Astrophysical Fluid Dynamics\/} {\bf 109}, 84--110.

\bibitem[{Yokoi}(2013)]{yokoi:2013}
{\sc {Yokoi}, N.} 2013 {Cross helicity and related dynamo}. {\em Geophysical
  and Astrophysical Fluid Dynamics\/} {\bf 107}, 114--184.

\bibitem[Yokoi(2018)]{yokoi_2018}
{\sc Yokoi, Nobumitsu} 2018 Electromotive force in strongly compressible
  magnetohydrodynamic turbulence. {\em Journal of Plasma Physics\/} {\bf
  84}~(5), 735840501.

\bibitem[{Yokoi}(2019)]{yokoi19}
{\sc {Yokoi}, N.} 2019 {Turbulence, transport and reconnection}. In {\em CISM
  Courses and Lectures: Advanced Topics in MHD\/}.

\bibitem[{Yousef} {\em et~al.\/}(2003){Yousef}, {Brandenburg} \&
  {R{\"u}diger}]{yousefetal:2003}
{\sc {Yousef}, T.~A., {Brandenburg}, A. \& {R{\"u}diger}, G.} 2003 {Turbulent
  magnetic Prandtl number and magnetic diffusivity quenching from simulations}.
  {\em Astronomy \& Astrophysics\/} {\bf 411}, 321--327.

\bibitem[{Yousef} {\em et~al.\/}(2008){Yousef}, {Heinemann}, {Schekochihin},
  {Kleeorin}, {Rogachevskii}, {Iskakov}, {Cowley} \& {McWilliams}]{Yetal:2008}
{\sc {Yousef}, T.~A., {Heinemann}, T., {Schekochihin}, A.~A., {Kleeorin}, N.,
  {Rogachevskii}, I., {Iskakov}, A.~B., {Cowley}, S.~C. \& {McWilliams}, J.~C.}
  2008 {Generation of Magnetic Field by Combined Action of Turbulence and
  Shear}. {\em Physical Review Letters\/} {\bf 100}~(18), 184501.

\bibitem[Zel'dovich(1957)]{zel:1957}
{\sc Zel'dovich, Ya~B.} 1957 The magnetic field in the two-dimensional motion
  of a conducting turbulent liquid. {\em JETP\/} {\bf 4}~(3), 460.

\bibitem[{Zeldovich} {\em et~al.\/}(1990){Zeldovich}, {Ruzmaikin} \&
  {Sokoloff}]{zelruzsok:1990}
{\sc {Zeldovich}, Y.~B., {Ruzmaikin}, A.~A. \& {Sokoloff}, D.~D.} 1990 {\em
  {The almighty chance}\/}. World Scientific Lecture Notes in Physics,
  Singapore: World Scientific Publication, 1990.

\end{thebibliography}
\end{document}